%% file: main.tex
\documentclass{jfm}

\usepackage{graphicx}
\usepackage{newtxtext}
\usepackage{newtxmath}
\usepackage{bm}
\usepackage{setspace}
\usepackage{natbib}
\usepackage[table]{xcolor}

\usepackage{longtable}
\usepackage{romannum}
\usepackage{lipsum}
\usepackage{hyperref}
\hypersetup{
    colorlinks = true,
    urlcolor   = blue,
    citecolor  = blue,
    linkcolor=blue,
}
\usepackage{xpatch}

\makeatletter
\xpatchcmd\NAT@citex
 {%
  \@citea\NAT@hyper@{%
    \NAT@nmfmt{\NAT@nm}%
    \hyper@natlinkbreak{\NAT@aysep\NAT@spacechar}{\@citeb\@extra@b@citeb}%
    \NAT@date
  }%
 }
 {%
  \@citea
  \NAT@nmfmt{\NAT@nm}%
  \NAT@aysep\NAT@spacechar
  \NAT@hyper@{\NAT@date}%
 }
 {}{}
\xpatchcmd\NAT@citex
 {%
  \@citea\NAT@hyper@{%
    \NAT@nmfmt{\NAT@nm}%
    \hyper@natlinkbreak{\NAT@spacechar\NAT@@open\if*#1*\else#1\NAT@spacechar\fi}%
    {\@citeb\@extra@b@citeb}%
    \NAT@date
  }%
 }
 {
  \@citea
    \NAT@nmfmt{\NAT@nm}%
    \NAT@spacechar\NAT@@open\if*#1*\else#1\NAT@spacechar\fi
    \NAT@hyper@{\NAT@date}%
 }
 {}{}
\makeatother

\usepackage{cleveref}

\newcommand{\RomanNumeralCaps}[1]
\linenumbers
\usepackage{multirow}

\usepackage{array}
         
\usepackage{lipsum}

\usepackage{tikz,pgfplots}
\usepgfplotslibrary{groupplots}
\usepackage{pgfplotstable}
\usepgfplotslibrary{external}
\usetikzlibrary{positioning,arrows.meta,shapes}
\pgfplotsset{compat=1.15}

\tikzstyle{text} = [text centered, draw=black, fill=none,align=left]
\tikzstyle{theme} = [rectangle, rounded corners,minimum width=1.5cm, minimum height=1cm, text ragged,  align=left, draw=black, fill=gray!15,line width=0.5mm]

\tikzstyle{topic} = [rectangle, rounded corners, minimum width=3cm, minimum height=0.8cm, text ragged, draw=black, fill=white!30,  align=left]

\DeclareMathOperator\erf{erf}
\makeatletter
\newcommand{\Vast}{\bBigg@{5.5}}
\newcommand{\vast}{\bBigg@{5}}
\makeatother
\usepackage{comment}
\usepackage{tabularx}
\usepackage{booktabs}

\title{Linking turbulent waves and bubble diffusion in self-aerated open-channel flows: Two-state air concentration}

\author{Matthias Kramer\aff{1}
  \corresp{\email{m.kramer@adfa.edu.au}},
  Daniel Valero \aff{2,3}}

\affiliation{\aff{1}UNSW Canberra, School of Engineering and Information Technology (SEIT), Canberra,
ACT 2610, Australia
\aff{2}Karlsruhe Institute of Technology (KIT), Institute for Water and River Basin Management (IWG), Karlsruhe, Germany
\aff{3}IHE Delft, Water Resources and Ecosystems dept., Delft, the Netherlands}
\begin{document}
\pagenumbering{arabic}
\maketitle

\begin{abstract}
High Froude-number flows become self-aerated when the destabilizing effect of turbulence overcomes gravity and surface tension forces. Traditionally, the resulting air concentration profile has been explained using single-layer approaches that invoke solutions of the advection-diffusion equation for air in water, i.e., bubbles' dispersion. Based on a wide range of experimental evidences, we argue that the complete air concentration profile shall be explained through the weak interaction of different canonical turbulent flows, namely a Turbulent Boundary Layer (TBL) and a Turbulent Wavy Layer (TWL). Motivated by a decomposition of the streamwise velocity into a pure wall flow and a free-stream flow [Krug et al., J. Fluid Mech. vol. 811, 2017, pp. 421--435], we present a physically consistent two-state formulation of the structure of a self-aerated flow. The air concentration is mathematically built upon a modified Rouse profile and a Gaussian error function, resembling vertical mass transport in the TBL and the TWL. We apply our air concentration theory to over 500 profiles from different data sets, featuring excellent agreement. Finally, we show that the turbulent Schmidt number, characterizing the momentum-mass transfer, ranges between 0.2 to 1, which is consistent with previous mass-transfer experiments in TBLs. Altogether, the proposed flow conceptualization sets the scene for more physically-based numerical modelling of turbulent mass diffusion in self-aerated flows.   
\end{abstract}

\begin{keywords}
channel flow, bubble dynamics
\end{keywords}

\newpage 
\section{Introduction}
\label{sec:intro}
\begin{figure}
\centering   
\input{Figures/fig1a.tex}
 \input{Figures/fig1b.tex}
\caption{Characteristic regions and modelling approaches of self-aerated open-channel flows (\textit{a}) snapshot of the self-aerated flow down a stepped chute (The University of Queensland) with clear distinction of the bubbly flow region and the wavy free-surface region; 
specific water discharge $q = 0.143$ m$^2$/s, chute angle $\theta = 45^\circ$; step edges 5 to 7 (\textit{b}) schools of thought in the modelling of air concentration distributions in self-aerated flows}\label{Fig1}
\end{figure}
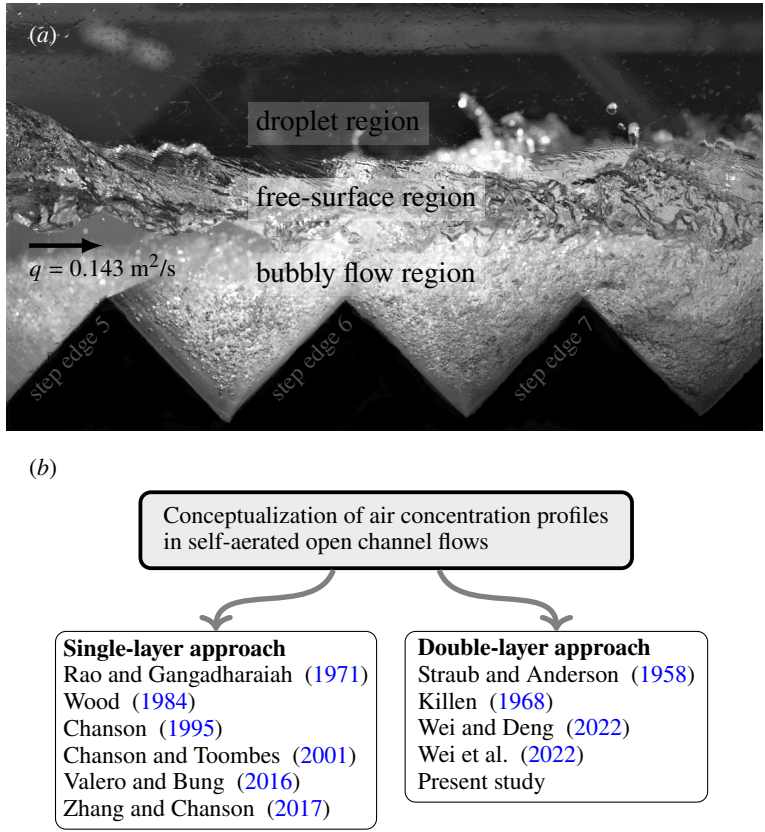

In a supercritical open-channel flow, turbulent stresses next to the free-surface can be large enough to overcome surface tension and gravity effects, thus leading to air bubble entrainment \citep{Brocchini2001JFM1,Valero2018} and their subsequent breakdown \citep{Deane2002,Chan2021}. This process is called self-aeration (Fig. \ref{Fig1}\textit{a}), and the location of the so-called inception point of air entrainment has been associated with the interaction of a developing boundary layer with the free-surface \citep{Lane39, Straub1958, Wood1984}, as well as with an unstable state of free-surface perturbations \citep{Brocchini2001JFM1, Valero2018}. Air--water multiphase flows are of key interest because entrained air affects flow properties, thereby leading to (i) flow bulking, which may compromise overtopping safety of spillways \citep{Straub1958, Hager91, Boes2000}, (ii) drag reduction, which can lead to flow velocities of twice or thrice the counter-part single-phase flow \citep{Wood1984,Chanson1994,Kramer2021}, (iii) cavitation protection of solid surfaces \citep{Falvey1990, Frizell2013cavitation}, (iv) enhanced gas transfer \citep{Gulliver1990, Bung2009}, and (v) total dissolved gas super-saturation that can mortally affect fish \citep{Pleizier2020}. Therefore, the accurate description of the air concentration distribution has been a topic of sustained research interest since the second half of the twentieth century, and two different schools of thought can be distinguished \mbox{(Fig. \ref{Fig1}\textit{b}).}

The first group of researchers conceptualized the air concentration distribution using a single-layer approach,  thus assuming a ``homogeneous'' mixing process between the channel bottom and $y_{90}$, except \citet{Valero2016}, who described air concentrations within a turbulent-wavy region. Here, $y_{90}$ is the flow depth where the time-averaged  air concentration is $\overline{c} = 0.9$\textcolor{black}{, with $\overline{c}$ being defined as volume of air per volume of air-water mixture}. \citet{Rao1971} and \citet{Wood1984} derived expressions for the air concentration distribution based on mass conservation considerations, while \citet{Chanson1995}, \citet{Chanson2001DIFF}, and \citet{Zhang2017} presented solutions of the advection-diffusion equation for air in water under the assumption of variable turbulent diffusivity across the water depth (up to $y_{90}$). We argue that the assumption of a  ``homogeneous'' bubbly air-water mixture does not contemplate the real structure of a self-aerated flow, as depicted in Fig. \ref{Fig1}\textit{a}, although we acknowledge that single-layer approaches can show a \textcolor{black}{good data-driven} agreement with typical \textit{S}-shaped concentration profiles, which is however at the expense of empirically fitted coefficients.

Based on flow visualization, \citet{Killen1968} discerned several distinct flow regions of a self-aerated flow (Fig. \ref{Fig1}\textit{a}), comprising:

\begin{enumerate}
    \item a single-phase (water) region next to the channel bottom (not always present),
    \item a bubbly flow region,
    \item a free-surface region, characterised by free-surface perturbations/waves,
    \item and a spray/droplet region.
\end{enumerate}

\citet{Wilhelms2005} identified that measured air concentrations typically comprise entrained air in the form of bubbles and entrapped air between surface roughness/waves, corresponding to regions (ii) and (iii), while very fine droplets, forming region (iv), have only been observed at prototype scale and in near-full-scale facilities \citep[see Table 1 of ][]{Hohermuth2021Prototype, Bai2022}. \textcolor{black}{Further, it is known} that the single-phase region (i) vanishes for depth-averaged (mean) air concentrations $\langle \overline{c} \rangle \gtrapprox 0.25$, which is because the bubbly flow layer protrudes to the channel bottom \citep{Straub1958, Hager91, Wei2022b}. Here, the mean air concentration is defined as:
\begin{equation}
\langle \overline{c} \rangle = \frac{1}{y_{90}} \int_{y = 0}^{y_{90}} \overline{c} \, \text{d}y
\label{eq:cmean}
\end{equation}
where $y$ is the bed-normal coordinate with $y$ = 0 at the channel invert.

The second group of researchers differentiated between the aforementioned regions using multi-layer approaches, mostly in the form of double-layer models (Fig. \ref{Fig1}\textit{b}), such as those from \citet{Straub1958} and \citet{Killen1968}, consisting of a lower layer where air bubbles are transported by turbulence throughout the flow, and an upper layer with a heterogeneous mixture of water droplets ejected from the flowing stream. The transition point between the two regions, defined by the depth $y_\star$ with corresponding air concentration $\overline{c}_\star$, was determined based on the maximum gradient $\left(\text{d}\overline{c}/\text{d}y\right)_\text{max}$ by \cite{Straub1958}. In contrast, \citet{Wei2022a} argued that the flow depth $y_{50}$, i.e., the depth where $\overline{c} = 0.5$, can be used as interior transition depth, and \citet{Wei2022b} proposed some additive concepts to model the air concentration distribution. Generally,  a double-layer (or multi-layer) approach better reflects the physical nature of an air-water flow, but an in-depth understanding of the flow transition is currently missing. Furthermore, none of the double-layer models has established a link between the air concentration and the velocity distribution, i.e., a coupling of mass and momentum transfer. 

\begin{figure*}
\centering
\input{Figures/fig2}
\caption{Two-state model for air concentration distribution (\textit{a}) representation of the TBL air concentration [\textcolor{black}{teal} line, Eq. (\ref{eq:voidfraction1})], TWL air concentration [blue line, Eq. (\ref{eq:voidfraction2})], interface position $y_i$, and \textcolor{black}{two-state model $\hat{c}$} (\textit{b}) probability distribution of interface position, centered around $y_\star$  (\textit{c}) comparison of the convoluted profile (convolution operation indicated by the $\bf{*}$ symbol) with data from \citet[ specific discharge $q=0.322$ m$^2$/s; streamwise position $x = 13.88$ m]{Straub1958}}
\label{fig:two-state}
\end{figure*}
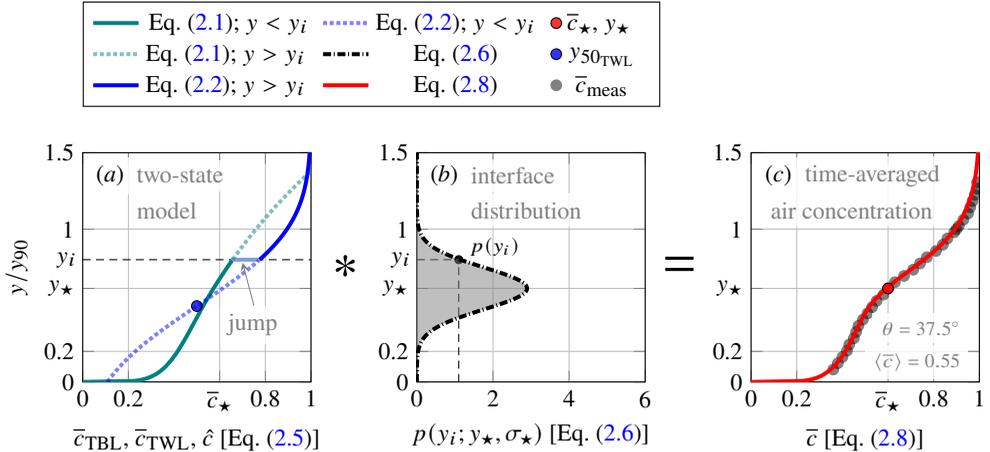

\newpage
Here, we propose a novel double-layer \textcolor{black}{conceptualization of the air transport}, which builds on two canonical flow layers of momentum, namely a Turbulent Boundary Layer (TBL) and a Turbulent Wavy Layer (TWL). We note that the TBL and the TWL conceptually feature a log-law and a constant free-stream velocity distribution, respectively. The air concentration of these layers is modelled using a modified Rouse profile (TBL, corresponds to regions i and ii) and a Gaussian error function (TWL, corresponds to region iii), both layers being convoluted using a two-state-principle, which is detailed in $\S$ \ref{sec:methods}. Through analysis of concentration profiles from literature, we show that the transition between TBL and TWL is closely linked to the boundary layer thickness ($\delta$), and that other model parameters, such as the Rouse number \citep{Rouse1961, Dey2014}, are unequivocally determined by the mean air concentration \mbox{($\S$ \ref{sec:results})}. For $\langle \overline{c} \rangle \lessapprox 0.25$, the Gaussian error function alone is able to predict the measured air concentration distributions, which suggests that entrapped air (within wave troughs) is prevalent in those flows. In $\S$ \ref{sec:discussion}, we quantify the turbulent Schmidt number for aerated flows, followed by a discussion on model limitations.

\section{Methodology: Two-state air concentration} \label{sec:methods}
The general \textcolor{black}{principle} of decomposing the streamwise velocity profile (and other flow statistics) into a pure wall flow state and a free-stream state was first introduced by \cite{Krug2017}. Here, we hypothesize that the two-state concept can be \textcolor{black}{extended to describe turbulent mass transport, such as} 
the air concentration distribution in supercritical open-channel flows (Fig. \ref{fig:two-state}). In the following, we seek expressions for the air concentration distribution within the TBL and the TWL.

\subsection{Air concentration within the TBL}
We note that the TBL features a log-law velocity distribution and turbulent air bubble diffusion. Assuming a parabolic distribution of turbulent diffusivity ($D_{t,y}$) up to half the boundary layer thickness, as well as a constant turbulent diffusivity for $y > \delta/2$ (\textcolor{black}{also known as parabolic-constant diffusivity distribution}), we obtain a modified Rouse equation by solving the advection-diffusion equation for air in water. The expression for the air concentration $\overline{c}_\text{TBL}$ within the TBL reads (Appendix \ref{app:bubblyflow}, Fig. \ref{fig:two-state}\textit{a}):

\begin{equation}
\overline{c}_\text{TBL} = 
\begin{cases}
\textcolor{black}{\overline{c}_{\delta/2}}  \left(\textcolor{black}{\frac{y}{\delta-y}} \right)^{\beta}, & y \leq \delta/2
\label{eq:voidfraction1}\\
\vphantom{\left(\frac{\frac{\delta}{y_\star}-1}{\frac{\delta}{y}-1} \right)^{\frac{\overline{v}_r S_c}{\kappa u_*}}}
\overline{c}_{\delta/2} \, \exp \left(\frac{4\beta}{\delta} \left(y - \frac{\delta}{2} \right)   \right), \quad & y > \delta/2 \\
\end{cases}
\end{equation}
\textcolor{black}{where $\overline{c}_{\delta/2}$ is the air concentration at half the boundary layer thickness}, $\overline{v}_r$ is the \textcolor{black}{bed-normal} bubble rise velocity, $\kappa$ is the van  K\a'{a}rm\a'{a}n constant,  $u_*$ is the friction velocity, and $\beta =  \overline{v}_r S_c /\kappa u_*$ is \textcolor{black}{a modified} Rouse number, which \textcolor{black}{encapsulates} the  turbulent Schmidt number $S_c$, \textcolor{black}{the latter} defined as the ratio of eddy viscosity and turbulent mass diffusivity. Here, \textcolor{black}{we adopt the classical definition of the boundary layer thickness as the bed-normal distance at which 99\% of the free-stream velocity is attained, as well as} a constant value of $\kappa = 0.41$, while we acknowledge \textcolor{black}{that slightly different values of the von {K}ármán constant have been discussed 
\citep{Nagib2008,Marusic2010,Morill2017}.}

\subsection{Air concentration within the TWL}
We emphasize that the air concentration of the TWL comprises surface waves/perturbations (entrapped air) as well as some entrained air bubbles. \textcolor{black}{An analytical solution for the air concentration $\overline{c}_\text{TWL}$ involves}  the Gaussian error function \citep{Valero2016}; (Fig. \ref{fig:two-state}\textit{a}):
\begin{equation}
\overline{c}_{\text{TWL}} = \frac{1}{2} \, \left( 1 + \erf  \left( \frac{y - y_{50_\text{TWL}}}{ \sqrt{2} \, \mathcal{H} } \right) \right)
\label{eq:voidfraction2}
\end{equation}
where $y_{50_\text{TWL}}$ is the mixture flow depth where the free-surface air concentration is $\overline{c}_\text{TWL} = 0.5$, $\mathcal{H}$ is a characteristic length-scale that describes the thickness/height of the aerated wavy layer, and $\erf$ is the Gaussian error function. \textcolor{black}{In  single-phase flows, $\mathcal{H}$ is defined as the} the root mean square wave height \citep{Valero2016}. \textcolor{black}{The air concentration of the TWL results from a superposition of entrapped air, transported between wave crests and troughs, and entrained air in the form of air bubbles travelling within the waves \citep{Killen1968, Wilhelms2005}. Because the volume of entrapped air within the TWL is typically much larger than the volume of entrained air,  $\mathcal{H}$ still provides a clear indication of the root mean square wave height. However, for the sake of accuracy, we hereafter refer to $\mathcal{H}$ as length-scale of the TWL.}

\subsection{Two-state convolution}
\textcolor{black}{Consistent with} the two-state formulation of \citet{Krug2017}, \textcolor{black}{we introduce} the concept of an interface position $y_i$ separating TBL and TWL. The air concentration profile $\hat{c}$ is defined for every position of $y_i$:
\begin{equation}
\hat{c} = 
\begin{cases}
\overline{c}_{\text{TBL}} \, \, \text{(Eq. \ref{eq:voidfraction1})}, \quad y \leq y_i\\
\overline{c}_{\text{TWL}}  \, \, \text{(Eq. \ref{eq:voidfraction2})}, \quad y > y_i\\
\end{cases}
\label{eq:chat1}
\end{equation}
\textcolor{black}{Using the Heaviside step function:
\begin{equation}
H(y-y_i) = 
\begin{cases}
0, \quad  y - y_i \leq 0\\
1, \quad  y - y_i > 0 \\
\end{cases}
\label{eq:chat2}
\end{equation}
we can write Eq. (\ref{eq:chat1}) as follows:
\begin{equation}
\hat{c} = \overline{c}_{\text{TBL}} (1-H) + \overline{c}_{\text{TWL}}H
\label{eq:chat2}
\end{equation}}

\textcolor{black}{
Equations (\ref{eq:chat1}) and (\ref{eq:chat2}) correspond to the discontinuous profile shown in Fig. \ref{fig:two-state}\textit{a}, \textcolor{black}{i.e., the two-state model}. It is implied that the flow below the interface level $y_i$ is fully explained by turbulent air bubble diffusion, whereas turbulent waves describe the air concentration above $y_i$. The interface position is now assumed to follow a random independent process, which is governed by a Gaussian probability distribution, shown in Fig. \ref{fig:two-state}\textit{b}:
\begin{equation}
p(y_i; y_\star, \sigma_\star) = \frac{1}{\sigma_\star \sqrt{2\pi}} \exp \left( \frac{-(y_i - y_\star)^2}{2\sigma_\star^2}  \right)
\label{eq:normdist}
\end{equation}
where the transition depth $y_\star$ can be interpreted as the time-averaged $y$-location of the interface, and $\sigma_\star$ describes the standard deviation of $y_i$. To obtain a complete, time-averaged expression for the double-layer air concentration (Fig. \ref{fig:two-state}\textit{c}), we convolute $\hat{c}$ [Eq. (\ref{eq:chat2})] with the interface probability $p$ [Eq. (\ref{eq:normdist})], which leads to \citep{Krug2017}:
\begin{equation}
\overline{c}(y) =  \int_{-\infty}^{\infty} \hat{c} p \, \text{d}y_i = \overline{c}_{\text{TBL}} \int_{-\infty}^{\infty}
(1-H)  p \, \text{d}y_i 
+ \overline{c}_{\text{TWL}} \int_{-\infty}^{\infty}
H p \, \text{d}y_i
\label{eq:voidfractionfinal0}
\end{equation}
where $\overline{c}_\text{TBL}$ and $\overline{c}_\text{TWL}$ are independent of $y_i$, thus allowing to simplify:
\begin{equation}
\overline{c}(y) = \overline{c}_{\text{TBL}}(1-\Gamma) +  \overline{c}_{\text{TWL}} \Gamma
\label{eq:voidfractionfinal}
\end{equation}
with:}
\begin{equation}
\Gamma(y; y_\star, \sigma_\star) = \frac{1}{2} \left( 1+\erf \left(\frac{y - y_\star }{ \sqrt{2} \sigma_\star}  \right)  \right)
\label{eq:gaussianerr}
\end{equation}

We note that the lower limit of the integral in Eq. (\ref{eq:voidfractionfinal0}) was extended to $- \infty$, which however did not affect the results as $p(y_i< 0) \ll 1$. From a physical point of view, the convolution can be interpreted as a weighted-averaging operation, which lumps concentration discontinuities between TBL and TWL, such as the jump shown in Fig. \ref{fig:two-state}\textit{a}, into a smooth, continuous profile. At last, the interface distribution of the two-state model is not expected to be the same as the turbulent/non-turbulent interface (TNTI) in turbulent boundary layers, see discussion in \citet{Krug2017}. 

\subsection{Determination of model parameters}
\label{sec:fitting}
\textcolor{black}{The \textcolor{black}{convoluted} two-state air concentration model [Eq. (\ref{eq:voidfractionfinal})] has four free physical parameters, including the Rouse number  ($\beta$), the length-scale of the TWL ($\mathcal{H}$), and two transition/interface parameters ($y_\star$, $\sigma_\star$).} \textcolor{black}{We note that the parameters $\delta$, $\overline{c}_{\delta/2}$, and $y_{50_\text{TWL}}$ are regarded as fixed, as they are directly available from measurements.} 
\textcolor{black}{The free model parameters were derived through a two-step fitting procedure to an extensive experimental dataset of air concentration profiles. It is noted that both layers (TBL and TWL) allowed for an independent (simultaneous) fit away from the mean interface position $y_\star$, thus preserving the physical significance of their parameters.} 

\newpage
\textcolor{black}{In a first step, the Rouse number $\beta$ was obtained by minimizing the sum of squared differences between measurements up to $\delta/2$ and Eq. (\ref{eq:voidfraction1}). Here, the boundary layer thickness was computed from measured velocities; if such velocity measurements were not available, we assumed $\delta = 1.25 \, y_\star$ (see $\S$ \ref{sec:transition}).}
\textcolor{black}{At the same time, the length-scale $\mathcal{H}$ was obtained by minimizing the sum of squared differences between measurements and modelled air concentrations within the upper flow region. We note that $y_{50_\text{TWL}}$ corresponds to $y_{50}$, which is a physical quantity that can be directly measured. However, this was not the case for 46 out of 571 re-analysed profiles with $\langle \overline{c} \rangle \gtrapprox 0.5$, where $y_{50_\text{TWL}}$ was also obtained through fitting; in this case, the number of free parameters increased by one.} \textcolor{black}{In a second step,} the mean interface position $y_\star$ was determined at the location where Eq. (\ref{eq:voidfraction2}) departed from \textcolor{black}{the} measured air concentrations, which yielded better results when compared to using the maximum gradient $\left(\text{d}\overline{c}/\text{d}y\right)_\text{max}$. Subsequently, the standard deviation $\sigma_\star$ of the interface position was determined using a best-fit approach.

\section{Results}
\label{sec:results}

\textcolor{black}{
We apply our air concentration model} to \textcolor{black}{571} concentration profiles from different data sets, \textcolor{black}{as presented in the supplementary material} (available at  \url{https://doi.org/10.1017/jfm.2023.440}), comprising smooth chute data from \citet[74 profiles]{Straub1958}, \citet[17 profiles]{Killen1968}, \citet[28 profiles]{Bung2009}, \citet[261 profiles]{Severi2018}, and stepped chute data from \citet[151 profiles]{Bung2009}, \citet[6 profiles]{Zhang2017DISS}, and \citet[34 profiles]{Kramer2018Transiton}. The terms ``smooth'' and ``stepped'' chute are commonly used in accordance with different roughness heights ($k_s$) of chute inverts, see Fig. \ref{Fig1} for an example of a stepped macro-roughness. The category ``smooth'' also comprises micro-rough inverts, and laboratory spillways are often considered as micro-rough for $k_s \gtrapprox 0.1$ mm \citep{Felder2023}. We note that this description slightly differs from the classic smooth/rough classification of wall flows \citep[Ch. 7]{Pope2000}. 

\subsection{Representative application}
\textcolor{black}{We demonstrate the application of our theory using a} seminal series of measurements by \citet{Straub1958}, who sampled air concentrations in the uniform region of a smooth chute ($k_s$ = 6.1 mm) for chute angles from $\theta = 7.5^\circ$ to $75^\circ$, covering a wide range of flow rates from $q = 0.13$ to 0.92 m$^2$/s. These measurements remain one of the most comprehensive and complete data sets to date, allowing us to illustrate the relative importance of each of the two states considered.

\begin{figure*}
\centering
\input{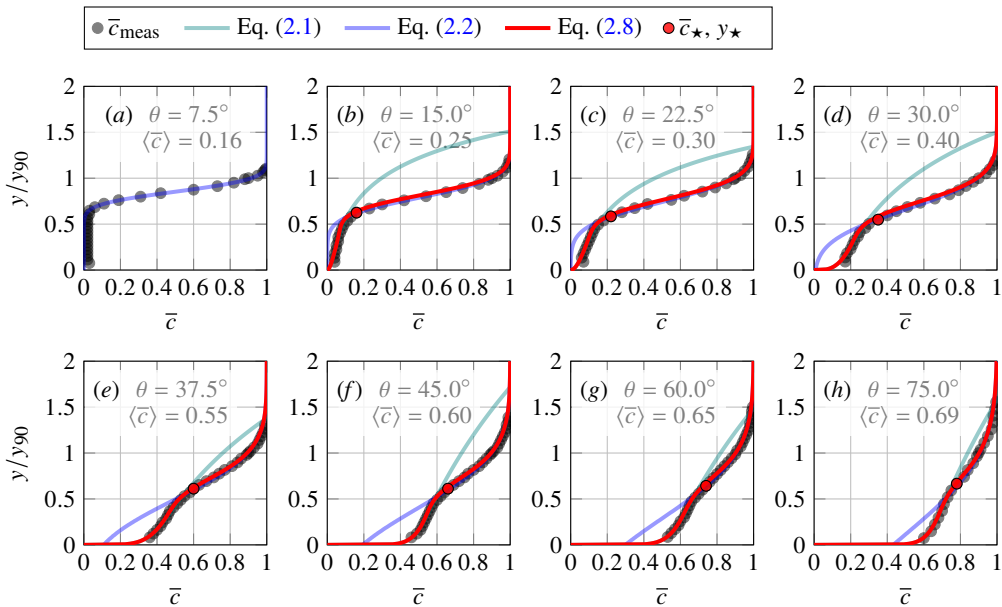}
\caption{Measured air concentration distributions in flows down a laboratory smooth chute; data from \citet[$q = 0.322$ m$^2$/s; \textcolor{black}{$x = 13.88$ m}]{Straub1958} (\textit{a} - \textit{h}) comparison of Eqns. \textcolor{black}{(\ref{eq:voidfraction1})}, 
 (\ref{eq:voidfraction2}), and (\ref{eq:voidfractionfinal}) with measurements}
\label{Fig:stepped_air}
\end{figure*}

Figure \ref{Fig:stepped_air} shows measured air concentrations \textcolor{black}{from \citet{Straub1958}} for \mbox{$q = 0.322$ m$^2$/s}, together with \textcolor{black}{the theoretical air concentration profiles} of the \textcolor{black}{TBL}, TWL, and \textcolor{black}{their convolution through the} two-state model. For mean air concentrations $\langle \overline{c} \rangle \lessapprox \textcolor{black}{0.25}$ (Fig. \ref{Fig:stepped_air}\textit{a}), entrained air bubbles did not reach the channel bottom and aeration was mostly confined to the TWL. Such flows are dominated by entrapped air (free surface perturbations and turbulent waves), and the air concentration was well described by Eq. (\ref{eq:voidfraction2}), see discussion in \citet{Felder2023}. For larger $\langle \overline{c} \rangle$ (Figs. \ref{Fig:stepped_air}\textit{\textcolor{black}{b}} - \textit{h}), the air concentration of the TWL (indicated as blue line) deviated from the measurements at the transition point. Here, the two-state model [Eq. (\ref{eq:voidfractionfinal})] excellently detailed the air concentration measurements, \textcolor{black}{and the corresponding model parameters are presented in Table \ref{tab1}.}

\input{Figures/table1}

\newpage
\subsection{Profile transition}
\label{sec:transition}
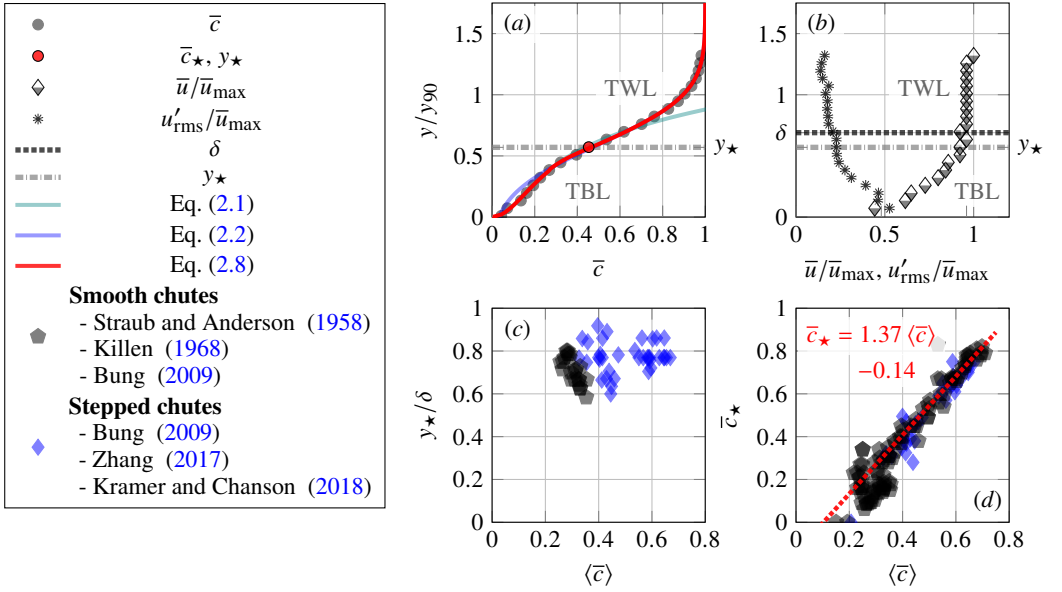
\begin{figure}
\centering
\input{Figures/fig4.tex}
\caption{Determination of the transition point parameters $y_\star$ and $c_\star$ (\textit{a}) 
exemplary air concentration profile, measured within the flow down a stepped spillway; $\theta = 45^\circ$; $q = 0.067$ m$^2$/s; step edge 11 (\textit{b}) corresponding normalised velocity profile and velocity fluctuations in bed-normal direction (\textit{c}) ratio of $y_\star$ and boundary layer thickness $\delta$ (\textit{d}) air concentration at transition point versus mean air concentration}
\label{fig:transition}
\end{figure}

\textcolor{black}{Next}, we focus our attention on the flow transition between TBL and TWL. Figures \ref{fig:transition}\textit{a,b} show a measured air concentration profile within the flow \textcolor{black}{centreline of} a stepped chute, together with the corresponding streamwise interfacial velocities ($\overline{u}$) and fluctuations from \cite{Kramer2018Transiton}. \textcolor{black}{Here}, $u'_\text{rms}$ is the root mean square of velocity fluctuations and $\overline{u}_\text{max}$ is the free-stream velocity. The chute angle was $\theta = 45^\circ$ and measurements were taken at step edge 11, at a specific flow rate of $q = 0.067$ m$^2$/s. The transition depth $y_\star$ (or mean interface position) was determined through the procedure described in $\S$ \ref{sec:fitting} and is shown in Fig. \ref{fig:transition}\textit{a}.

\newpage
A comparison of Figs. \ref{fig:transition}\textit{a,b}, confirms that the layer below $y_\star$ (i.e., mainly TBL) was characterised by high flow shearing, whereas the layer above $y_\star$ (TWL) corresponds to a uniform free-stream velocity, and, in this instance, the ratio between the transition depth and the boundary layer thickness is $y_\star/\delta = 0.83$. This finding supports our hypothesis that the air concentration distribution is intrinsically connected to the flow momentum layers, both separated by a fluctuating TBL - TWL interface. We find that $y_\star/ \delta$ ranges between 0.6 and 0.9 (Fig. \ref{fig:transition}\textit{c}), which is consistent with interface positions reported in \citet{Krug2017}. The \textcolor{black}{dimensionless} standard deviation \textcolor{black}{of the interface position remains constant at} $\sigma_\star/\delta = 0.1$ \textcolor{black}{to 0.2}, \textcolor{black}{which yields a good description of all concentration profiles. Our expression for $\sigma_\star$ assumes that the interface position $y_i$} is linked to the underlying TBL edge, while \textcolor{black}{we consider that the waves/perturbations of the TWL -- described via $\mathcal{H}$ --} are a reflection of that turbulent process. \textcolor{black}{Figure \ref{fig:transition}\textit{d}} shows that the air concentration at the transition point is linearly dependent on the mean air concentration, and can be estimated through the empirical relationship $\overline{c}_\star = 1.37 \, \langle \overline{c} \rangle - 0.14$, valid for smooth and stepped chutes.

\textcolor{black}{Finally, we acknowledge the scatter in Fig. \ref{fig:transition}\textit{c}, which likely stems from measurement uncertainties of dual-tip phase-detection probes related to air concentration and interfacial velocity \citep{Kramer2020,Hohermuth2021}. We also note that the smooth chute data of \citet{Straub1958,Killen1968,Severi2018} were not added to Fig. \ref{fig:transition}\textit{c}, which is because no velocity information was available or because flows were dominated by entrapped air, implying that no profile transition occurred.}

\newpage
\subsection{Model parameters related to $\overline{c}_\text{TBL}$ and $\overline{c}_\text{TWL}$}
\label{section:params}
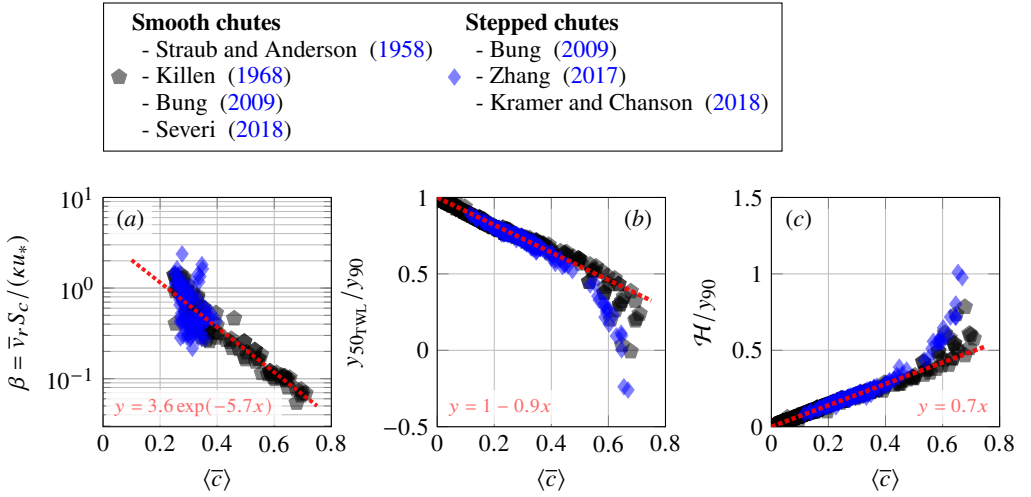
\begin{figure}
\centering
\input{Figures/fig5}
\caption{Model parameters are a function of the mean air concentration: (\textit{a}) Rouse number $\beta$ (\textit{b}) relation of TWL mixture flow depth $y_{50_\text{TWL}}$ to $y_{90}$  (\textit{c}) relation of TWL length-scale $\mathcal{H}$ to $y_{90}$}
\label{fig:modelparams}
\end{figure}

\textcolor{black}{Here, we present estimations of the} model parameters related to the air concentration within the TBL and the TWL. An inspection of Eq. (\ref{eq:voidfraction1}) shows that $\overline{c}_\text{TBL} = f(\beta, \delta, \overline{c}_{\delta/2})$, of which \textcolor{black}{$\delta$ and $\overline{c}_{\delta/2}$ were directly determined from velocity and air concentration measurements, respectively}. The Rouse number $\beta$ reflects the ratio of bubble rise velocity and the strength of turbulence (shear velocity) acting on entrained air bubbles, and consequently defines the mode of entrained air transport. For example, a small Rouse number implies dominant turbulent forces, which thereof results in a large quantity of small air bubbles being transported close to the channel bed, whereas a large Rouse number implies that large air bubbles are being transported next to the free-surface. We find that the described transport modes are dependent on the mean air concentration and that the $\beta$-parameters for stepped chutes were slightly larger than those for smooth chutes (Fig. \ref{fig:modelparams}\textit{a}), which could be associated to the entrainment of larger air bubbles. 

The air concentration of the TWL is described by two parameters $\overline{c}_\text{TWL} = f(y_{50_\text{TWL}}, \mathcal{H})$, compare Eq. (\ref{eq:voidfraction2}), \textcolor{black}{of which the mixture flow depth $y_{50_\text{TWL}}$ was extracted directly from measurements for most of the data sets.} Here, we normalise $y_{50_\text{TWL}}$ and the length-scale $\mathcal{H}$ with $y_{90}$, showing their functional dependence on the mean air concentration (Figs. \ref{fig:modelparams}\textit{b,c}). We note that the normalisation with $y_{90}$ provided the most clear relationship, which may simply be  because  $\langle \overline{c} \rangle$ is defined in terms of $y_{90}$ [Eq. (\ref{eq:cmean})]. Further, the unique dependence between model parameters and mean air concentration is not completely unexpected, as previous researchers have fitted empirical parameters that only depend on $\langle \overline{c} \rangle$, see for example \citet{Chanson1995,Chanson2001DIFF}. However, it is remarkable that model parameters of the \textcolor{black}{TWL} are similarly well behaved for smooth and stepped chutes, and data scatter is only observed beyond $\langle \overline{c} \rangle \gtrapprox 0.5$ to $0.6$. For stepped chutes, the deviation from the linear trend is likely to result from a change in flow regime, i.e., the flow changes from skimming flow to transition flow \citep{Kramer2018Transiton}.

\section{Discussion}
\label{sec:discussion}

\subsection{Turbulent Schmidt number}
\label{sec:schmidt}
In $\S$ \ref{section:params}, we have discussed the dependence of the depth-averaged parameter $\beta$ on the mean air concentration. From a Reynolds-Averaged modelling perspective, quantifying the turbulent Schmidt number is of interest since it enables a direct relationship between turbulent momentum diffusivity and turbulent mass diffusivity \citep{Pope2000, Gualtieri2017}. Hence, we adopt a Reynolds-Averaged form of the advection-diffusion equation for air in water (see also Appendix \ref{app:bubblyflow}, Eq. \ref{eq:diffusivity1}):
\begin{equation}
\overline{v}_r \, \overline{c} = D_{t,y} \frac{\partial \overline{c}}{\partial y} \end{equation}
which can be conveniently re-arranged:
\begin{equation}
D_{t,y} = \frac{\overline{v}_r \, \overline{c}}{\partial \overline{c} / \partial y}
\label{eq:Dt}
\end{equation}

\begin{figure}
\centering
\input{Figures/fig6.tex}
\caption{Measured turbulent diffusivities versus theoretical parabolic distribution for different turbulent Schmidt numbers}
\label{fig:schmidt}
\end{figure}
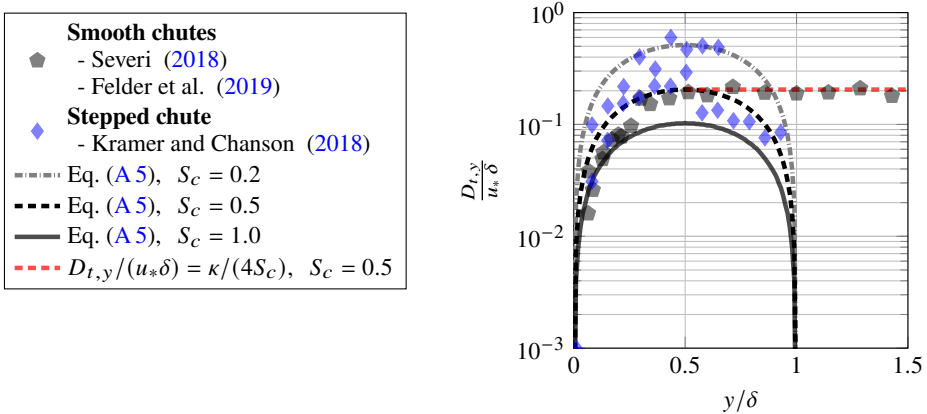

Equation (\ref{eq:Dt}) allows for a direct evaluation of $D_{t,y}$ \textcolor{black}{from measurements}, given that the local air concentrations, their gradients, and bubble rise velocities are known. For selected profiles from \citet{Severi2018}, \citet{Kramer2018Transiton}, and \citet{Felder2019}, we evaluate $\partial \overline{c}/\partial y$  using a central differences approach, and we characterize bubble sizes from intrusive phase-detection probe measurements by adopting the Sauter diameter $d = 6 \, \overline{c}/a = 1.5 \overline{u} \,  \overline{c}/F$ \citep{Ishii2011, Hohermuth2021}, \textcolor{black}{where $a = 4F/\overline{u}$ is the interfacial area per volume of air water mixture \citep[Eq. 4.6.13]{Cummings1996}}, and $F$ is the particle count rate. We estimate \textcolor{black}{still water} rise velocities ($\overline{v}_0$) for bubbles with 1 mm $< d < 10$ mm using the approach of \citet[Eq. 7-3]{Clift1978}, \textcolor{black}{which we correct for gravity slope and concentration effects after \citet{Chanson1995,Chanson1996} as $\overline{v}_r =  \overline{v}_0 \cos(\theta) \sqrt{1-\overline{c}}$}. \textcolor{black}{In a next step,} we compute turbulent diffusivities through Eq. (\ref{eq:Dt}), which are compared against the boundary layer thickness \textcolor{black}{in Fig. \ref{fig:schmidt}}.

We consider the classical parabolic turbulent diffusivity distribution (Eq. \ref{eq:parabolic}) for different $S_c$-values, which allows us to conclude that the turbulent Schmidt number for air water flows ranges between 0.2 and 1.0 \textcolor{black}{(compare Fig. \ref{fig:schmidt})}, which is well in accordance with published literature values for turbulent mass transfer in other environmental flows \citep{Gualtieri2017}. Lastly, the re-analysed smooth chute data suggests that the turbulent diffusivity becomes constant for $y > \delta/2$ (red dotted line), similar to other open-channel flows \citep{Coleman1970, Dey2014}. Based on this finding \textcolor{black}{and reasons outlined in Appendix \ref{app:bubblyflow}, we had already adopted a parabolic-constant diffusivity distribution, splitting the} expression for the air concentration within the TBL into two parts, separated by $\delta/2$ (see Eq. \ref{eq:voidfraction1}).

\subsection{Comparison with other models and limitations}
\textcolor{black}{The determination of the four (five) free physical parameters of the \textcolor{black}{convoluted} two-state air concentration profile was outlined in $\S$ \ref{sec:fitting}.} Naturally, the number of \textcolor{black}{free} parameters is larger than for commonly used single-layer models, e.g., \citet[Eq. 4.1, \textcolor{black}{two free} parameters]{Chanson2001DIFF}, and comparable to other double layer-models, e.g., \citet[\textcolor{black}{four free} parameters]{Straub1958}. However, the introduced parameters respond to physical properties of the flow, and they allow to assess the relative contribution of individual physical \textcolor{black}{momentum processes} on the air concentration profile.

Our theoretical profile [Eq. (\ref{eq:voidfraction2}) for $\langle \overline{c} \rangle \lessapprox \textcolor{black}{0.25}$ and Eq. (\ref{eq:voidfractionfinal}) for $\langle \overline{c} \rangle \gtrapprox \textcolor{black}{0.25}$] was able to characterize the air concentration distribution of all tested data sets (see Supplementary Material), including different flow rates ($q = 0.03$ to  0.92 m$^2$/s) and flow regimes over smooth and stepped chutes (transition versus skimming flow), with angles ranging from $\theta = 7.5^\circ$ and $75^\circ$. Model parameters of the TBL and TWL were between $\beta = 0.05$ to $1.2$ (Fig. \ref{fig:modelparams}\textit{a}) and $\mathcal{H}/y_{90} = 0$ to 1 (Fig. \ref{fig:modelparams}\textit{c}), both showing a unique dependence on the mean air concentration. The transition/interface parameters were determined as $y_\star/\delta$ = 0.6 to 0.9 (Fig. \ref{fig:transition}\textit{c}) and $\sigma_\star/\delta = 0.1$ to 0.2. Given a measured air concentration distribution, $y_\star$ can also be obtained from $\overline{c}_\star$, where the latter followed an empirical relationship $\overline{c}_\star = 1.37 \langle \overline{c} \rangle - 0.14$ (Fig. \ref{fig:transition}\textit{d}).

\begin{figure}
\centering
\input{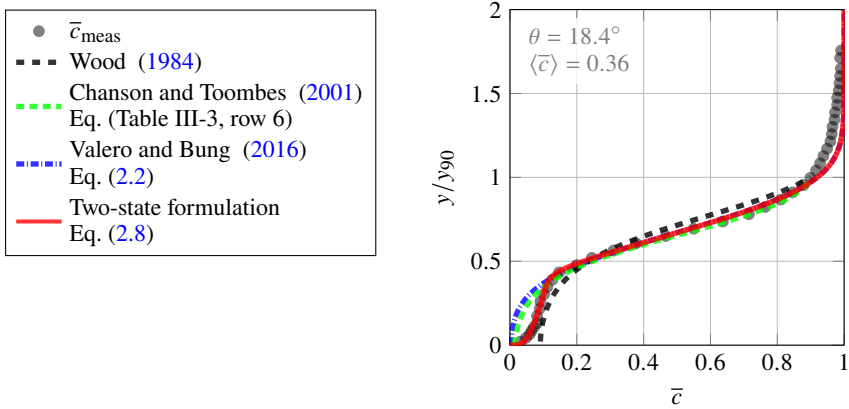}
\caption{Comparison of the proposed two-state formulation with common air concentration models for stepped chutes; data from \cite{Bung2009} with $q = 0.07$ m$^2$/s; step height $h =$ 0.06 m; measurements taken at step edge 16}
\label{fig:comp}
\end{figure}

When compared to previous models, one of the main advantages of the two-state model is its \textit{universal applicability}, together with \textit{physically interpretable} model parameters. Previous models rely more heavily on empirical fitting parameters, with application domains being limited to certain chute geometries (smooth versus stepped) or to certain flow regimes (transition versus skimming flow), see for example \citet[Table \Romannum{3}-3]{Chanson2001DIFF}. Figure \ref{fig:comp} compares the \textcolor{black}{convoluted} two-state model with common single-layer models for stepped chute data from \citet{Bung2009}. 
All models describe the air concentration reasonably well for $y/y_{90} \gtrapprox 0.4$, while the models of \citet{Wood1984} and \citet{Chanson2001DIFF} are upper bounded by $y/y_{90} = 1$. Below $y/y_{90} \lessapprox 0.4$, only the two-state model is able to capture the region which has traditionally been referred to as concentration boundary layer \citep{Chanson1996}, which is due to the novel representation of underlying physical processes.

The present two-state air concentration model has some limitations, which are herein discussed. \textcolor{black}{The parameters $\overline{c}_{\delta/2}$ and  $\delta$ were directly extracted from original measurements, i.e., no predictive formulas exist, which however does not hinder the physical interpretation of mass transport parameters in self-aerated flows.} In our derivations related to the modified Rouse profile for air bubbles in water (Appendix \ref{app:bubblyflow}), we have assumed a constant bubble rise velocity as well as uniform flow conditions. To estimate the effect of a concentration-dependent rise velocity, we derive alternative solutions of the advection-diffusion equation using $\overline{v}_r = \textcolor{black}{\overline{v}_0  \cos{(\theta)}} \sqrt{1-\overline{c}}$ \citep[\textcolor{black}{Appendix C}]{Chanson1995,Chanson1996}, where $\overline{v}_0$ is the rise velocity for clear/\textcolor{black}{still} water conditions, i.e., $\overline{c} \approx 0$ \textcolor{black}{(similar to $\S$ \ref{sec:schmidt})}. Although only marginal differences in resulting concentration profiles were observed (not shown), we cannot rule out that a more detailed assessment of bubble size distributions would help to improve our understanding of the air concentration distribution within the TBL. Related to the uniform equilibrium assumption, our data indicate that Eq. (\ref{eq:voidfraction1}) is also applicable within the gradually-varied flow region, which is in agreement with previous findings \citep{Chanson1995}.

\section{Conclusion}
In this research, we formulate a two-state model to describe air concentration distributions in self-aerated free-surface flows. The rationale behind the model is that the flow can be decomposed into a Turbulent Boundarly Layer (TBL) and a Turbulent Wavy Layer (TWL), featuring a log-law and a constant free-stream velocity, respectively. The corresponding air concentration distributions are mathematically described by a modified Rouse profile and a Gaussian error function, which conceptually implies that the bubbly flow (within the TBL) is driven by high shear and turbulent diffusion, whereas free-surface waves/perturbations of the TWL lead to large concentrations due to voids within wave troughs. 

The transition point between the two layers was previously discussed in aerated flow literature \citep{Straub1958}, but no physical reasoning was given. Here, we argue that the flow transition corresponds to a time-averaged TBL - TWL interface position that is closely related to the boundary layer thickness. From an instantaneous point of view, the flow takes either a TBL or a TWL state, and the interface position is described by a Gaussian probability density function. Subsequently, a convolution of the two states with the interface probability provides the time-averaged air concentration profile. As such, our model is the first to establish a connection between air concentration and velocity distribution based on physically explainable model parameters. 

We test our theoretical air concentration profile against more than \textcolor{black}{500} experimental concentration profiles from smooth and stepped chute literature data sets. We show that, regardless of the data set, the model is able to capture the profiles and to discern the different air concentration regions contained within. It is noted that laboratory flows with $\langle \overline{c} \rangle \lessapprox \textcolor{black}{0.25}$ are dominated by free-surface aeration/entrapped air, and the air concentration distribution is confined to the TWL. For \textcolor{black}{larger} $\langle \overline{c} \rangle$, air bubbles are diffused deeper into the water column, implying that the time-averaged air concentration is described by the convolution of the Rouse profile and the Gaussian error function with the interface probability. We characterize different transport modes of entrained air bubbles (within the TBL) based on the ratio of bubble rise velocity and shear velocity, expressed through the Rouse number $\beta$, and we show that the transport mode is a function of the mean air concentration. Model parameters related to the TWL were also explained via the mean air concentration and behaved similarly for smooth and stepped chutes. Finally, we present values of the turbulent Schmidt number in highly aerated flows, which we anticipate to be of high relevance for future numerical model applications based on Reynolds-averaged Navier-Stokes equations.

\backsection[Data availability statement]{
Supplementary material is available at \url{https://doi.org/10.1017/jfm.2023.440}
All data, models, or code that support the findings of this study are available from the corresponding author upon reasonable request.}

\backsection[Acknowledgements]{We would like to acknowledge fruitful discussions with Dr Gangfu Zhang, AECOM, Australia. 
We further would like to thank Prof Daniel Bung (FH Aachen), A/Prof Stefan Felder (UNSW Sydney), Dr Benjamin Hohermuth (ETH Zurich), and Dr Armaghan Severi (Manly Hydraulics Laboratory) for sharing their data  sets. At last, we would like to thank the Editor for handling our manuscript, and the anonymous reviewers, whose comments have helped to improve our work.}

\backsection[Funding]{This research received no specific grant from any funding agency, commercial or not-for-profit sectors.}

\backsection[Declaration of interests]{The authors report no conflict of interest.}

\backsection[Author ORCID]{M. Kramer, https://orcid.org/0000-0001-5673-2751; D. Valero, https://orcid.org/0000-0002-7127-7547}

\backsection[CRediT authorship contribution statement]{
Conceptualization: MK, DV; Data curation: MK; Formal
analysis: MK; Methodology: MK; Investigation:
MK; Visualization: MK; Software: MK; Writing – original draft: MK; Writing – review \& editing: MK, DV.
}

\begin{appendix}
\section{Modified Rouse profile for air bubbles in water}
\label{app:bubblyflow}

A governing equation for the bed-normal air concentration distribution can be written by simplifying the advection-diffusion equation for air in water. We assume a two-dimensional steady flow, where the air concentration only varies in bed-normal, but not in streamwise ($x$) or transverse ($z$) direction. Therefore, the following gradients can be neglected
$\partial (\cdot)/ \partial t = 0$, $\partial (\cdot)/ \partial x = 0$, $\partial (\cdot)/ \partial z = 0$, and the advection-diffusion equation reduces to:
\begin{equation}
 \frac{\partial (\overline{v}_r \, \overline{c})}{\partial y}  = \frac{\partial }{\partial y} \left( D_{t,y} \frac{\partial \overline{c}}{\partial y} \right)
 \label{eq:start}
\end{equation}
where $\overline{v}_r$  is the rise velocity of air bubbles, which is defined positive in bed-normal direction, $\overline{c}$ is the volumetric air concentration (volume of air/total volume), and $D_{t,y}$ is the turbulent diffusivity. We note that the full derivation of the generalized advection-diffusion equation, including Reynolds-Averaging and gradient diffusion theory, is presented, amongst others, in \citet{Dey2014}. A first integration of Eq. (\ref{eq:start}) leads to:
\begin{equation}
\int \frac{\partial}{\partial y} \left(\overline{v}_r \, \overline{c} \right) \text{d}y = \int \frac{\partial }{\partial y} \left(D_{t,y} \frac{\partial \overline{c}}{\partial y} \right) \text{d}y
\end{equation}
\begin{equation}
\overline{v}_r \, \overline{c} = D_{t,y} \frac{\partial \overline{c}}{\partial y} + \mathcal{K}_1
\label{eq:diffusivity1}
\end{equation}

To simplify Eq. (\ref{eq:diffusivity1}), the partial differential is replaced by the total differential, only the first set of solutions ($\mathcal{K}_1=0$) is considered, and a constant rise velocity is assumed. Separating variables and performing a second integration from an arbitrary elevation $y$ to $\delta/2$ yields:
\begin{equation}
\int_{\overline{c}}^{\overline{c}_{\textcolor{black}{\delta/2}}} \frac{1}{\overline{c}} \, \text{d} \overline{c}  = \overline{v}_r \, \int_{y}^{\textcolor{black}{\delta/2}} \frac{1}{D_{t,y}} \text{d} y
\label{eqA9}
\end{equation}
Now, we invoke a parabolic distribution of turbulent diffusivity \citep[likewise ][]{Rouse1961}:
\begin{equation}
D_{t,y} = \frac{\kappa u_* y}{S_c} \left(1 - \frac{y}{\delta}\right)
\label{eq:parabolic}
\end{equation}
where $\kappa$ is the von {K}ármán constant, $u_*$ is the friction velocity, and $S_c$  (= $\nu_t/D_{t,y}$) is the turbulent Schmidt number, defined as the ratio of eddy viscosity $\nu_t$ (i.e., momentum diffusivity) and turbulent mass diffusivity. Substitution and integration of Eq. (\ref{eqA9}) gives:
\begin{equation}
\frac{\overline{c}}{\overline{c}_{\textcolor{black}{\delta/2}} }=   \exp \left( \frac{\overline{v}_r S_c}{\kappa u_*} 
\ln \left[ \frac{\textcolor{black}{1}}{\frac{\delta}{y}-1}\right]
\right)
\end{equation}
which simplifies to:
\begin{equation}
\overline{c}(y) = \overline{c}_{\textcolor{black}{\delta/2}}  \left(\textcolor{black}{\frac{y}{\delta-y}} \right)^{\beta}, \quad y \leq \delta/2
\label{Rouse1}
\end{equation}
where $\beta = \overline{v}_r S_c/(\kappa u_* )$ is the Rouse number. It is noted that \textcolor{black}{$S_c$ is assumed constant and} Eq. (\ref{Rouse1}) \textcolor{black}{is similar} to the well-known Rouse equation for sediment transport but incorporating subtle differences, which are: (1) a positively defined bubble rise velocity, (2) a change of integration limits, and (3) a use of the boundary layer thickness $\delta$ instead of the water depth. The parameter $S_c$ was encapsulated within $\beta$, for convenience. 

\textcolor{black}{We note that a purely parabolic diffusivity profile (Eq. \ref{eq:parabolic}) becomes negative for $y_i/\delta > 1$, which could occasionally happen if $\sigma_*$ is large. The next appropriate choice is a parabolic-constant diffusivity distribution, which has been used for suspended sediment transport \citep{Coleman1970,Dey2014}, and which we also observe in Fig. \ref{fig:schmidt}, where the turbulent diffusivity $D_{t,y}$ becomes independent of $y$}. Adopting a constant $D_{t,y}\textcolor{black}{(\delta/2)} = \kappa u_* \delta/(4 S_c)$ for $y > \delta/2$, we integrate Eq. (\ref{eq:diffusivity1}) between \textcolor{black}{$\delta/2$} and an arbitrary elevation:
\begin{equation}
\int_{\overline{c}_{\textcolor{black}{\delta/2}}}^{\overline{c}} \frac{1}{\overline{c}} \, \text{d} \overline{c}  = \overline{v}_r \, \int_{\textcolor{black}{\delta/2}}^{y} \frac{4 S_c}{ \kappa u_* \delta} \text{d} y
\end{equation}
yielding the following exponential distribution:
\begin{equation}
\overline{c}(y) = \overline{c}_{\textcolor{black}{\delta/2}} \exp \left(\frac{4 \beta}{\delta} \left(y - \textcolor{black}{\frac{\delta}{2}} \right)   \right), \quad  y > \delta/2
\end{equation}

\end{appendix}

\bibliographystyle{jfm}

\end{document}

%% file: Figures/fig1a.tex
\begin{tikzpicture}

\node[anchor=north west] at  (-1,0)
{\includegraphics[width= 10 cm]{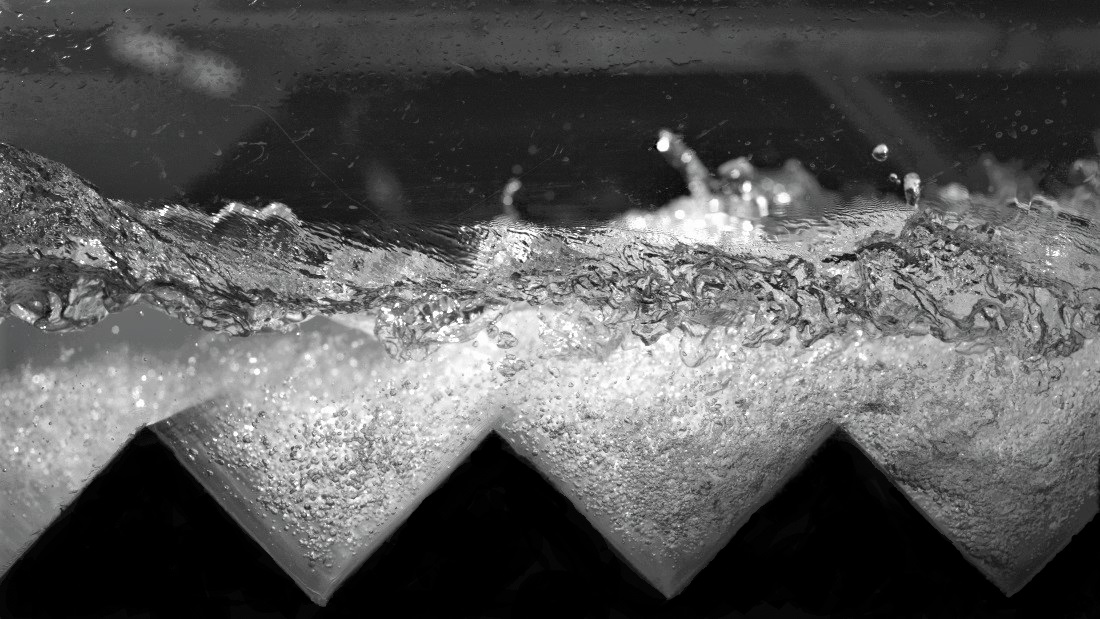} };  

\node[align=center,color=white,anchor=north west] at (-0.7cm,-0.3cm) {(\textit{a})};

\node[align=center,color=black,anchor=north west] at (-0.7cm,-3.3cm) {$q = 0.143$ m$^2$/s};

\node[align=center,color=black,anchor=north west] at (-0.7cm,-6cm) {(\textit{b})};

\node (A) at (-0.7, -3.3){};
\node (B) at (0.5, -3.3){};

\draw[-latex, line width = 1.7] (A) edge (B);

\node[align=center,color=black,anchor=north west,fill=white,fill opacity=0.2,text opacity=1] at (2.3cm,-3.4cm) {\large{bubbly flow region}};

\node[align=center,color=black,anchor=north west,fill=white,fill opacity=0.4,text opacity=1] at (2.3cm,-2.4cm) {\large{free-surface region}};

\node[align=center,color=black,anchor=north west,fill=white,fill opacity=0.2,text opacity=1] at (2.3cm,-1.4cm) {\large{droplet region}};

\node[align=center,color=white,anchor=north west,rotate=45,opacity=0.3] at (-0.8cm,-5.2cm) {step edge 5};

\node[align=center,color=white,anchor=north west,rotate=45,opacity=0.3] at (2.4cm,-5.2cm) {step edge 6};

\node[align=center,color=white,anchor=north west,rotate=45,opacity=0.3] at (5.6cm,-5.2cm) {step edge 7};

\end{tikzpicture}

%% file: Figures/fig1b.tex
\begin{tikzpicture}[node distance=2cm,every node/.style={fill=white, font=\normalsize, align=left}]

\node (theme) [theme,minimum height=1cm,minimum width=6.5cm, font=\small] {
Conceptualization of air concentration profiles\\
in self-aerated open channel flows};

\node (twolayer) [topic, below  = 0.8 cm of theme,  xshift=2.25cm ,minimum height=1cm,minimum width =4cm, font=\small] {
\textbf{Double-layer approach}\\
\cite{Straub1958}\\
\cite{Killen1968}\\
\cite{Wei2022a}\\
\cite{Wei2022b}\\
Present study
};

\node (onelayer) [topic, below  = 0.8 cm of theme,  xshift=-2.25cm ,minimum height=1cm,minimum width =4cm,font=\small] {
\textbf{Single-layer approach}\\
\cite{Rao1971}\\
\cite{Wood1984}\\
\cite{Chanson1995} \\
\cite{Chanson2001DIFF}\\
\cite{Valero2016}\\
\cite{Zhang2017}
};

\draw [->,line width=0.7mm,color=gray]  ($(theme.south)+(0.7,0)$)  to [out=290,in=90]  ($(twolayer.north)-(0,0)$); 

\draw [->,line width=0.7mm,color=gray]  ($(theme.south)-(0.7,0)$)  to [out=250,in=90]   ($(onelayer.north)-(0,0)$); 

\end{tikzpicture}

%% file: Figures/fig2.tex
\begin{tikzpicture}
\definecolor{color1}{rgb}{1,0.7,0.1}
\definecolor{color2}{rgb}{0.15,0.45,0.55}

 \begin{groupplot}[
     group style = {group size = 3 by 2,horizontal sep=1.4cm,vertical sep=1.4cm},
     width = 1\textwidth]

\nextgroupplot[height = 4.6cm,
     width = 4.6cm,
     ylabel={$y/y_{90}$},
     xlabel={$\overline{c}_\text{TBL}$, $\overline{c}_\text{TWL}$, $\hat{c}$ [Eq. (\ref{eq:chat2})] },
      xtick={0, 0.2,  0.60, 0.8,  1},
     xticklabels={0,  0.2,  \textcolor{black}{$\overline{c}_\star$}, 0.8, 1},
     ytick={0, 0.2, 0.6122, 1, 1.5},
     yticklabels={0, 0.2, \textcolor{black}{$y_\star$}, 1, 1.5},
     grid=both,
     xmin=0,
     xmax=1,
     ymin=0,
     ymax=1.5,
    legend columns=3,
    transpose legend,
   legend style={cells={align=left},anchor = south west,at={(0,1.2)},font=\small},
   clip mode=individual
   ]

\node[align=center,fill=white,fill opacity=0.8,text opacity=1] at (rel axis cs:0.12,0.9) {(\textit{a})};

\node[align=left,anchor=west,align=left,anchor=west,fill=white,fill opacity=0.8,text opacity=1] at (rel axis cs:0.2,0.9) {\textcolor{gray}{\footnotesize{two-state}}};

\node[align=left,anchor=west,align=left,anchor=west,fill=white,fill opacity=0.8,text opacity=1] at (rel axis cs:0.6,0.25) {\textcolor{gray}{\footnotesize{jump}}};

\node[align=left,anchor=west,align=left,anchor=west,fill=white,fill opacity=0.8,text opacity=1] at (rel axis cs:0.2,0.75) {\textcolor{gray}{\footnotesize{model}}};

\addplot[color=teal,mark=none,line width = 1.5,draw opacity=1] 
table[row sep=crcr]{%
0 -1	\\
}; \addlegendentry{Eq. (\ref{eq:voidfraction1}); $y < y_i  \,$};

\addplot[densely dotted,color=teal,mark=none,line width = 1.5,draw opacity=0.5] 
table[row sep=crcr]{%
0 -1	\\
}; \addlegendentry{Eq. (\ref{eq:voidfraction1}); $y > y_i  \,$  };

\addplot[color=blue,mark=none,line width = 1.5,draw opacity=1] 
table[row sep=crcr]{%
0 -1	\\
}; \addlegendentry{Eq. (\ref{eq:voidfraction2}); $y > y_i \,$ };

\addplot[densely dotted,color=blue,mark=none,line width = 1.5,draw opacity=0.5] 
table[row sep=crcr]{%
0 -1	\\
}; \addlegendentry{Eq. (\ref{eq:voidfraction2}); $y < y_i  \,$ };

\addplot[densely dashdotted,color=black,mark=none,line width = 1.5,draw opacity=1] 
table[row sep=crcr]{%
0 -1	\\
}; \addlegendentry{Eq. (\ref{eq:normdist})  \,};

\addplot[color=red,mark=none,line width = 1.5,draw opacity=1] 
table[row sep=crcr]{%
0 -1	\\
}; \addlegendentry{Eq. (\ref{eq:voidfractionfinal})  \,};

\addplot [fill=red,mark=*,only marks,fill opacity=0.8,draw opacity=1, mark size=2,clip mode=individual]
table[row sep=crcr]{%
0 -1	\\
}; \addlegendentry{$\overline{c}_\star$, $y_\star$ \, };

\addplot [fill=blue,mark=*,only marks,fill opacity=0.8,draw opacity=1, mark size=2,clip mode=individual]
table[row sep=crcr]{%
0 -1	\\
}; \addlegendentry{$y_{50_\text{TWL}}  \,$  };

\addplot [fill=black,mark=*,only marks,fill opacity=0.5,draw opacity=0.5, mark size=2,clip mode=individual]
table[row sep=crcr]{%
0 -1	\\
}; \addlegendentry{$\overline{c}_\text{meas} \,$ };

\node[align=center] at (rel axis cs:1.15,0.5) {\huge{$\bf{*}$}};

\addplot [densely dashed,color=black,mark=none,draw opacity=1,clip mode=individual]
table[row sep=crcr]{%
0	0.8	\\
1	0.8	\\
};

\node[align=center] at (-0.07,0.8) {\footnotesize{$y_i$}};  

\draw[line width=1.5,color=teal!50!blue!50,opacity=1] (0.65,0.8) -- (0.78,0.8);

\addplot[color=teal,mark=none,line width = 1.5,draw opacity=1] 
table[row sep=crcr]{%
0	0	\\
0.226299661	0.006617433	\\
0.254285459	0.013234866	\\
0.272373638	0.019852299	\\
2.86E-01	0.026469732	\\
2.97E-01	0.033087165	\\
3.07E-01	0.039704598	\\
3.15E-01	0.046322031	\\
3.23E-01	0.052939464	\\
3.30E-01	0.059556897	\\
3.36E-01	0.06617433	\\
3.42E-01	0.072791763	\\
3.47E-01	0.079409196	\\
3.52E-01	0.086026629	\\
3.57E-01	0.092644062	\\
3.62E-01	0.099261494	\\
3.66E-01	0.105878927	\\
3.70E-01	0.11249636	\\
3.74E-01	0.119113793	\\
3.78E-01	0.125731226	\\
3.82E-01	0.132348659	\\
3.86E-01	0.138966092	\\
3.89E-01	0.145583525	\\
3.93E-01	0.152200958	\\
3.96E-01	0.158818391	\\
3.99E-01	0.165435824	\\
4.02E-01	0.172053257	\\
4.06E-01	0.17867069	\\
4.09E-01	0.185288123	\\
4.12E-01	0.191905556	\\
4.15E-01	0.198522989	\\
4.18E-01	0.205140422	\\
4.20E-01	0.211757855	\\
4.23E-01	0.218375288	\\
4.26E-01	0.224992721	\\
4.29E-01	0.231610154	\\
4.31E-01	0.238227587	\\
4.34E-01	0.24484502	\\
4.37E-01	0.251462453	\\
4.39E-01	0.258079886	\\
4.42E-01	0.264697319	\\
4.45E-01	0.271314752	\\
4.47E-01	0.277932185	\\
4.50E-01	0.284549618	\\
4.52E-01	0.29116705	\\
4.55E-01	0.297784483	\\
4.57E-01	0.304401916	\\
4.60E-01	0.311019349	\\
4.62E-01	0.317636782	\\
4.65E-01	0.324254215	\\
4.67E-01	0.330871648	\\
4.69E-01	0.337489081	\\
4.72E-01	0.344106514	\\
4.74E-01	0.350723947	\\
4.77E-01	0.35734138	\\
4.79E-01	0.363958813	\\
4.81E-01	0.370576246	\\
4.84E-01	0.377193679	\\
4.86E-01	0.383811112	\\
0.488618475	0.390428545	\\
0.491017373	0.397045978	\\
0.493418464	0.403663411	\\
0.495822756	0.410280844	\\
0.498231256	0.416898277	\\
0.500644968	0.42351571	\\
0.503064902	0.430133143	\\
0.505492075	0.436750576	\\
0.507927509	0.443368009	\\
0.51037224	0.449985442	\\
0.512827315	0.456602875	\\
0.514330433	0.463220308	\\
0.516804098	0.469837741	\\
0.51928966	0.476455174	\\
0.521787177	0.483072606	\\
0.524296705	0.489690039	\\
0.526818302	0.496307472	\\
0.529352028	0.502924905	\\
0.531897939	0.509542338	\\
0.534456095	0.516159771	\\
0.537026554	0.522777204	\\
0.539609375	0.529394637	\\
0.542204619	0.53601207	\\
0.544812345	0.542629503	\\
0.547432612	0.549246936	\\
0.550065481	0.555864369	\\
0.552711014	0.562481802	\\
0.55536927	0.569099235	\\
0.55804031	0.575716668	\\
0.560724197	0.582334101	\\
0.563420993	0.588951534	\\
0.566130758	0.595568967	\\
0.568853556	0.6021864	\\
0.571589449	0.608803833	\\
0.574338501	0.615421266	\\
0.577100774	0.622038699	\\
0.579876332	0.628656132	\\
0.582665239	0.635273565	\\
0.585467559	0.641890998	\\
0.588283357	0.648508431	\\
0.591112698	0.655125864	\\
0.593955646	0.661743297	\\
0.596812267	0.66836073	\\
0.599682627	0.674978162	\\
0.602566793	0.681595595	\\
0.605464829	0.688213028	\\
0.608376804	0.694830461	\\
0.611302784	0.701447894	\\
0.614242836	0.708065327	\\
0.617197028	0.71468276	\\
0.620165429	0.721300193	\\
0.623148105	0.727917626	\\
0.626145128	0.734535059	\\
0.629156564	0.741152492	\\
0.632182483	0.747769925	\\
0.635222956	0.754387358	\\
0.638278052	0.761004791	\\
0.641347842	0.767622224	\\
0.644432395	0.774239657	\\
0.647531784	0.78085709	\\
0.650646079	0.787474523	\\
0.653775352	0.794091956	\\
};

\addplot[densely dotted,color=teal,mark=none,line width = 1.5,draw opacity=0.5] 
table[row sep=crcr]{%
0.656919675	0.800709389	\\
0.660079121	0.807326822	\\
0.663253763	0.813944255	\\
0.666443672	0.820561688	\\
0.669648924	0.827179121	\\
0.672869591	0.833796554	\\
0.676105748	0.840413987	\\
0.679357469	0.84703142	\\
0.682624829	0.853648853	\\
0.685907904	0.860266286	\\
0.689206768	0.866883718	\\
0.692521498	0.873501151	\\
0.695852171	0.880118584	\\
0.699198862	0.886736017	\\
0.702561649	0.89335345	\\
0.70594061	0.899970883	\\
0.709335821	0.906588316	\\
0.712747362	0.913205749	\\
0.71617531	0.919823182	\\
0.719619746	0.926440615	\\
0.723080747	0.933058048	\\
0.726558393	0.939675481	\\
0.730052766	0.946292914	\\
0.733563944	0.952910347	\\
0.73709201	0.95952778	\\
0.740637044	0.966145213	\\
0.744199127	0.972762646	\\
0.747778342	0.979380079	\\
0.751374772	0.985997512	\\
0.754988498	0.992614945	\\
0.758619605	0.999232378	\\
0.762268175	1.005849811	\\
0.765934293	1.012467244	\\
0.769618044	1.019084677	\\
0.773319511	1.02570211	\\
0.77703878	1.032319543	\\
0.780775937	1.038936976	\\
0.784531068	1.045554409	\\
0.788304259	1.052171841	\\
0.792095598	1.058789274	\\
0.79590517	1.065406707	\\
0.799733065	1.07202414	\\
0.80357937	1.078641573	\\
0.807444174	1.085259006	\\
0.811327565	1.091876439	\\
0.815229634	1.098493872	\\
0.819150469	1.105111305	\\
0.823090162	1.111728738	\\
0.827048803	1.118346171	\\
0.831026482	1.124963604	\\
0.835023293	1.131581037	\\
0.839039325	1.13819847	\\
0.843074673	1.144815903	\\
0.847129429	1.151433336	\\
0.851203686	1.158050769	\\
0.855297539	1.164668202	\\
0.85941108	1.171285635	\\
0.863544406	1.177903068	\\
0.86769761	1.184520501	\\
0.87187079	1.191137934	\\
0.87606404	1.197755367	\\
0.880277458	1.2043728	\\
0.88451114	1.210990233	\\
0.888765184	1.217607666	\\
0.893039688	1.224225099	\\
0.89733475	1.230842532	\\
0.901650469	1.237459965	\\
0.905986944	1.244077397	\\
0.910344275	1.25069483	\\
0.914722563	1.257312263	\\
0.919121909	1.263929696	\\
0.923542413	1.270547129	\\
0.927984177	1.277164562	\\
0.932447304	1.283781995	\\
0.936931896	1.290399428	\\
0.941438057	1.297016861	\\
0.94596589	1.303634294	\\
0.9505155	1.310251727	\\
0.955086991	1.31686916	\\
0.959680468	1.323486593	\\
0.964296038	1.330104026	\\
0.968933807	1.336721459	\\
0.97359388	1.343338892	\\
0.978276366	1.349956325	\\
0.982981373	1.356573758	\\
0.987709008	1.363191191	\\
0.992459381	1.369808624	\\
0.9972326	1.376426057	\\
1.002028776	1.38304349	\\
1.00684802	1.389660923	\\
1.011690441	1.396278356	\\
1.016556152	1.402895789	\\
1.021445265	1.409513222	\\
1.026357891	1.416130655	\\
1.031294145	1.422748088	\\
1.03625414	1.429365521	\\
1.04123799	1.435982953	\\
1.046245809	1.442600386	\\
1.051277714	1.449217819	\\
1.056333819	1.455835252	\\
1.061414242	1.462452685	\\
1.066519099	1.469070118	\\
1.071648508	1.475687551	\\
1.076802586	1.482304984	\\
1.081981453	1.488922417	\\
1.087185228	1.49553985	\\
1.09241403	1.502157283	\\
1.09766798	1.508774716	\\
1.102947199	1.515392149	\\
1.108251808	1.522009582	\\
1.113581929	1.528627015	\\
1.118937686	1.535244448	\\
1.124319201	1.541861881	\\
1.129726599	1.548479314	\\
1.135160003	1.555096747	\\
1.140619539	1.56171418	\\
1.146105333	1.568331613	\\
1.15161751	1.574949046	\\
1.157156199	1.581566479	\\
1.162721525	1.588183912	\\
1.168313618	1.594801345	\\
1.173932606	1.601418778	\\
1.179578619	1.608036211	\\
1.185251785	1.614653644	\\
1.190952237	1.621271077	\\
1.196680105	1.627888509	\\
1.202435522	1.634505942	\\
1.208218618	1.641123375	\\
1.214029529	1.647740808	\\
1.219868387	1.654358241	\\
1.225735327	1.660975674	\\
1.231630484	1.667593107	\\
1.237553993	1.67421054	\\
1.243505992	1.680827973	\\
1.249486617	1.687445406	\\
1.255496005	1.694062839	\\
1.261534295	1.700680272	\\
1.267601627	1.707297705	\\
1.273698139	1.713915138	\\
1.279823973	1.720532571	\\
1.285979268	1.727150004	\\
1.292164167	1.733767437	\\
1.298378813	1.74038487	\\
1.304623348	1.747002303	\\
1.310897915	1.753619736	\\
1.317202661	1.760237169	\\
1.323537728	1.766854602	\\
1.329903264	1.773472035	\\
1.336299416	1.780089468	\\
1.342726329	1.786706901	\\
1.349184152	1.793324334	\\
1.355673035	1.799941767	\\
1.362193125	1.8065592	\\
1.368744574	1.813176633	\\
1.375327532	1.819794065	\\
1.38194215	1.826411498	\\
1.388588582	1.833028931	\\
1.395266979	1.839646364	\\
1.401977496	1.846263797	\\
1.408720287	1.85288123	\\
1.415495508	1.859498663	\\
1.422303314	1.866116096	\\
1.429143861	1.872733529	\\
1.436017309	1.879350962	\\
1.442923814	1.885968395	\\
1.449863536	1.892585828	\\
1.456836634	1.899203261	\\
1.46384327	1.905820694	\\
1.470883603	1.912438127	\\
1.477957797	1.91905556	\\
1.485066015	1.925672993	\\
1.492208419	1.932290426	\\
1.499385175	1.938907859	\\
1.506596447	1.945525292	\\
1.513842401	1.952142725	\\
1.521123205	1.958760158	\\
1.528439026	1.965377591	\\
1.535790032	1.971995024	\\
1.543176392	1.978612457	\\
1.550598278	1.98522989	\\
1.558055858	1.991847323	\\
1.565549306	1.998464756	\\
};

\addplot[densely dotted,color=blue,mark=none,line width = 1.5,draw opacity=0.5] 
table[row sep=crcr]{%
0.108340196	0	\\
0.111432633	0.006617433	\\
0.114587738	0.013234866	\\
0.117805907	0.019852299	\\
1.21E-01	0.026469732	\\
1.24E-01	0.033087165	\\
1.28E-01	0.039704598	\\
1.31E-01	0.046322031	\\
1.35E-01	0.052939464	\\
1.38E-01	0.059556897	\\
1.42E-01	0.06617433	\\
1.46E-01	0.072791763	\\
1.50E-01	0.079409196	\\
1.54E-01	0.086026629	\\
1.57E-01	0.092644062	\\
1.61E-01	0.099261494	\\
1.66E-01	0.105878927	\\
1.70E-01	0.11249636	\\
1.74E-01	0.119113793	\\
1.78E-01	0.125731226	\\
1.82E-01	0.132348659	\\
1.87E-01	0.138966092	\\
1.91E-01	0.145583525	\\
1.96E-01	0.152200958	\\
2.00E-01	0.158818391	\\
2.05E-01	0.165435824	\\
2.10E-01	0.172053257	\\
2.14E-01	0.17867069	\\
2.19E-01	0.185288123	\\
2.24E-01	0.191905556	\\
2.29E-01	0.198522989	\\
2.34E-01	0.205140422	\\
2.39E-01	0.211757855	\\
2.44E-01	0.218375288	\\
2.50E-01	0.224992721	\\
2.55E-01	0.231610154	\\
2.60E-01	0.238227587	\\
2.66E-01	0.24484502	\\
2.71E-01	0.251462453	\\
2.76E-01	0.258079886	\\
2.82E-01	0.264697319	\\
2.88E-01	0.271314752	\\
2.93E-01	0.277932185	\\
2.99E-01	0.284549618	\\
3.05E-01	0.29116705	\\
3.10E-01	0.297784483	\\
3.16E-01	0.304401916	\\
3.22E-01	0.311019349	\\
3.28E-01	0.317636782	\\
3.34E-01	0.324254215	\\
3.40E-01	0.330871648	\\
3.46E-01	0.337489081	\\
3.52E-01	0.344106514	\\
3.58E-01	0.350723947	\\
3.64E-01	0.35734138	\\
3.71E-01	0.363958813	\\
3.77E-01	0.370576246	\\
3.83E-01	0.377193679	\\
3.89E-01	0.383811112	\\
0.395764925	0.390428545	\\
0.402119207	0.397045978	\\
0.408499464	0.403663411	\\
0.414904067	0.410280844	\\
0.421331368	0.416898277	\\
0.427779701	0.42351571	\\
0.434247381	0.430133143	\\
0.44073271	0.436750576	\\
0.447233975	0.443368009	\\
0.45374945	0.449985442	\\
0.460277397	0.456602875	\\
0.466816068	0.463220308	\\
0.473363707	0.469837741	\\
0.47991855	0.476455174	\\
0.486478829	0.483072606	\\
0.493042769	0.489690039	\\
0.499608593	0.496307472	\\
0.506174523	0.502924905	\\
0.512738781	0.509542338	\\
0.519299589	0.516159771	\\
0.525855173	0.522777204	\\
0.532403764	0.529394637	\\
0.538943597	0.53601207	\\
0.545472914	0.542629503	\\
0.551989967	0.549246936	\\
0.558493016	0.555864369	\\
0.564980336	0.562481802	\\
0.57145021	0.569099235	\\
0.577900938	0.575716668	\\
0.584330834	0.582334101	\\
0.590738231	0.588951534	\\
0.597121476	0.595568967	\\
0.603478939	0.6021864	\\
0.609809009	0.608803833	\\
0.616110097	0.615421266	\\
0.622380637	0.622038699	\\
0.628619086	0.628656132	\\
0.634823928	0.635273565	\\
0.640993672	0.641890998	\\
0.647126856	0.648508431	\\
0.653222043	0.655125864	\\
0.65927783	0.661743297	\\
0.665292841	0.66836073	\\
0.671265733	0.674978162	\\
0.677195193	0.681595595	\\
0.683079944	0.688213028	\\
0.688918741	0.694830461	\\
0.694710373	0.701447894	\\
0.700453665	0.708065327	\\
0.706147479	0.71468276	\\
0.711790711	0.721300193	\\
0.717382296	0.727917626	\\
0.722921207	0.734535059	\\
0.728406452	0.741152492	\\
0.733837081	0.747769925	\\
0.739212182	0.754387358	\\
0.74453088	0.761004791	\\
0.749792344	0.767622224	\\
0.754995777	0.774239657	\\
0.760140428	0.78085709	\\
0.765225582	0.787474523	\\
0.770250566	0.794091956	\\
};

\addplot[color=blue,mark=none,line width = 1.5,draw opacity=1] 
table[row sep=crcr]{%
0.775214746	0.800709389	\\
0.780117531	0.807326822	\\
0.784958368	0.813944255	\\
0.789736745	0.820561688	\\
0.794452191	0.827179121	\\
0.799104274	0.833796554	\\
0.803692602	0.840413987	\\
0.808216824	0.84703142	\\
0.812676626	0.853648853	\\
0.817071736	0.860266286	\\
0.821401918	0.866883718	\\
0.825666976	0.873501151	\\
0.829866751	0.880118584	\\
0.834001124	0.886736017	\\
0.838070009	0.89335345	\\
0.842073361	0.899970883	\\
0.846011166	0.906588316	\\
0.84988345	0.913205749	\\
0.853690271	0.919823182	\\
0.857431722	0.926440615	\\
0.86110793	0.933058048	\\
0.864719054	0.939675481	\\
0.868265285	0.946292914	\\
0.871746847	0.952910347	\\
0.875163993	0.95952778	\\
0.878517006	0.966145213	\\
0.881806199	0.972762646	\\
0.885031913	0.979380079	\\
0.888194516	0.985997512	\\
0.891294402	0.992614945	\\
0.894331993	0.999232378	\\
0.897307735	1.005849811	\\
0.900222096	1.012467244	\\
0.90307557	1.019084677	\\
0.905868673	1.02570211	\\
0.908601941	1.032319543	\\
0.911275931	1.038936976	\\
0.913891221	1.045554409	\\
0.916448406	1.052171841	\\
0.9189481	1.058789274	\\
0.921390934	1.065406707	\\
0.923777554	1.07202414	\\
0.926108622	1.078641573	\\
0.928384816	1.085259006	\\
0.930606824	1.091876439	\\
0.932775349	1.098493872	\\
0.934891105	1.105111305	\\
0.936954817	1.111728738	\\
0.93896722	1.118346171	\\
0.940929059	1.124963604	\\
0.942841085	1.131581037	\\
0.944704059	1.13819847	\\
0.946518747	1.144815903	\\
0.948285923	1.151433336	\\
0.950006363	1.158050769	\\
0.951680851	1.164668202	\\
0.953310172	1.171285635	\\
0.954895116	1.177903068	\\
0.956436473	1.184520501	\\
0.957935036	1.191137934	\\
0.959391598	1.197755367	\\
0.960806953	1.2043728	\\
0.962181895	1.210990233	\\
0.963517216	1.217607666	\\
0.964813706	1.224225099	\\
0.966072153	1.230842532	\\
0.967293343	1.237459965	\\
0.968478057	1.244077397	\\
0.969627075	1.25069483	\\
0.97074117	1.257312263	\\
0.97182111	1.263929696	\\
0.972867659	1.270547129	\\
0.973881575	1.277164562	\\
0.97486361	1.283781995	\\
0.975814507	1.290399428	\\
0.976735006	1.297016861	\\
0.977625836	1.303634294	\\
0.978487721	1.310251727	\\
0.979321374	1.31686916	\\
0.980127503	1.323486593	\\
0.980906805	1.330104026	\\
0.981659968	1.336721459	\\
0.982387673	1.343338892	\\
0.983090588	1.349956325	\\
0.983769376	1.356573758	\\
0.984424686	1.363191191	\\
0.985057159	1.369808624	\\
0.985667426	1.376426057	\\
0.986256107	1.38304349	\\
0.986823811	1.389660923	\\
0.987371138	1.396278356	\\
0.987898676	1.402895789	\\
0.988407003	1.409513222	\\
0.988896686	1.416130655	\\
0.989368281	1.422748088	\\
0.989822334	1.429365521	\\
0.990259377	1.435982953	\\
0.990679935	1.442600386	\\
0.99108452	1.449217819	\\
0.991473632	1.455835252	\\
0.991847763	1.462452685	\\
0.992207391	1.469070118	\\
0.992552986	1.475687551	\\
0.992885005	1.482304984	\\
0.993203894	1.488922417	\\
0.99351009	1.49553985	\\
0.993804019	1.502157283	\\
0.994086096	1.508774716	\\
0.994356725	1.515392149	\\
0.9946163	1.522009582	\\
0.994865206	1.528627015	\\
0.995103817	1.535244448	\\
0.995332496	1.541861881	\\
0.995551597	1.548479314	\\
0.995761465	1.555096747	\\
0.995962434	1.56171418	\\
0.99615483	1.568331613	\\
0.996338968	1.574949046	\\
0.996515155	1.581566479	\\
0.996683689	1.588183912	\\
0.996844859	1.594801345	\\
0.996998944	1.601418778	\\
0.997146216	1.608036211	\\
0.997286938	1.614653644	\\
0.997421365	1.621271077	\\
0.997549744	1.627888509	\\
0.997672314	1.634505942	\\
0.997789306	1.641123375	\\
0.997900944	1.647740808	\\
0.998007443	1.654358241	\\
0.998109013	1.660975674	\\
0.998205856	1.667593107	\\
0.998298166	1.67421054	\\
0.998386132	1.680827973	\\
0.998469936	1.687445406	\\
0.998549753	1.694062839	\\
0.998625752	1.700680272	\\
0.998698096	1.707297705	\\
0.998766942	1.713915138	\\
0.998832442	1.720532571	\\
0.998894742	1.727150004	\\
0.998953981	1.733767437	\\
0.999010294	1.74038487	\\
0.999063813	1.747002303	\\
0.999114661	1.753619736	\\
0.999162959	1.760237169	\\
0.999208823	1.766854602	\\
0.999252363	1.773472035	\\
0.999293685	1.780089468	\\
0.999332893	1.786706901	\\
0.999370084	1.793324334	\\
0.999405352	1.799941767	\\
0.999438788	1.8065592	\\
0.999470479	1.813176633	\\
0.999500506	1.819794065	\\
0.999528951	1.826411498	\\
0.999555889	1.833028931	\\
0.999581393	1.839646364	\\
0.999605532	1.846263797	\\
0.999628374	1.85288123	\\
0.999649983	1.859498663	\\
0.999670419	1.866116096	\\
0.999689741	1.872733529	\\
0.999708004	1.879350962	\\
0.999725263	1.885968395	\\
0.999741567	1.892585828	\\
0.999756966	1.899203261	\\
0.999771506	1.905820694	\\
0.999785231	1.912438127	\\
0.999798183	1.91905556	\\
0.999810402	1.925672993	\\
0.999821927	1.932290426	\\
0.999832795	1.938907859	\\
0.999843039	1.945525292	\\
0.999852693	1.952142725	\\
0.999861789	1.958760158	\\
0.999870357	1.965377591	\\
0.999878425	1.971995024	\\
0.999886019	1.978612457	\\
0.999893167	1.98522989	\\
0.999899892	1.991847323	\\
0.999906218	1.998464756	\\
};

\addplot [fill=blue,mark=*,only marks,fill opacity=0.8,draw opacity=1, mark size=2,clip mode=individual]
table[row sep=crcr]{%
0.50	 0.4967	\\
};

\draw [-stealth, color=gray](0.75,0.5) -- (0.7,0.78);

\nextgroupplot[height = 4.6cm,
     width = 4.6cm,
     xlabel={$p(y_i; y_\star, \sigma_\star)$ [Eq. (\ref{eq:normdist})]},
     grid=both,
     xmin=0,
     xmax=6,
     ymin=0,
     ymax=1.5,
      ytick={0, 0.2,  0.6122, 1, 1.5},
     yticklabels={0, 0.2, \textcolor{black}{$y_\star$}, 1, 1.5},
       xtick={0, 2, 4, 6},
    xticklabels={0,2,4,6},
    legend columns=2,
   legend style={cells={align=left},anchor = north east,at={(1,1.4)},font=\footnotesize},
   clip mode = individual,   
   domain=0:1.5,samples=400,
   ]
   
\node[align=center] at (rel axis cs:1.15,0.5) {\huge{$\bf{=}$}};  

\node[align=center,fill=white,fill opacity=0.8,text opacity=1] at (rel axis cs:0.12,0.9) {(\textit{b})};   
\addplot [densely dashed,color=black,mark=none,draw opacity=1,clip mode=individual]
table[row sep=crcr]{%
0	0.8	\\
1.1	0.8	\\
};

\addplot [densely dashed,color=black,mark=none,draw opacity=1,clip mode=individual]
table[row sep=crcr]{%
1.1	0	\\
1.1	0.8	\\
};

\addplot [fill=black,mark=*,only marks,fill opacity=1,draw opacity=1,mark size=1.5,clip mode=individual]
table[row sep=crcr]{%
1.1	 0.8	\\
};

\node[align=left,anchor=west,align=left,anchor=west,fill=white,fill opacity=0.8,text opacity=1] at (rel axis cs:0.2,0.9) {\textcolor{gray}{\footnotesize{interface}}};

\node[align=left,anchor=west,align=left,anchor=west,fill=white,fill opacity=0.8,text opacity=1] at (rel axis cs:0.2,0.75) {\textcolor{gray}{\footnotesize{distribution}}};

\addplot+[densely dashdotted,color=black,mark=none,line width=1.5, fill=gray,fill opacity=0.5] ({1/(0.1378
*sqrt(2*pi))*1/exp(((x-0.6122)^2)/(2*0.1378
^2 ))},{x});

\node[align=center] at (-0.42,0.8) {\footnotesize{$y_i$}};  

\node[align=center] at (2.05,0.885) {\scriptsize{$p(y_i)$}};

\nextgroupplot[height = 4.6cm,
     width = 4.6cm,
     xlabel={$\overline{c}$ [Eq. (\ref{eq:voidfractionfinal})]},
     grid=both,
     xmin=0,
     xmax=1,
     ymin=0,
     ymax=1.5,
   xtick={0, 0.2,  0.6, 0.8,  1},
     xticklabels={0,  0.2,  \textcolor{black}{$\overline{c}_\star$}, 0.8, 1},
     ytick={0, 0.2, 0.6122, 1, 1.5},
     yticklabels={0, 0.2, \textcolor{black}{$y_\star$}, 1, 1.5},
    legend columns=2,
   legend style={cells={align=left},anchor = north east,at={(1,1.4)},font=\normalsize}
   ]

\node[align=center,fill=white,fill opacity=0.8,text opacity=1] at (rel axis cs:0.12,0.9) {(\textit{c})};

\node[align=left,anchor=west,fill=white,fill opacity=0.8,text opacity=1] at (rel axis cs:0.2,0.9) {\textcolor{gray}{\footnotesize{time-averaged}}};

\node[align=left,anchor=west,fill=white,fill opacity=0.8,text opacity=1] at (rel axis cs:0.05,0.75) {\textcolor{gray}{\footnotesize{air concentration}}};

\addplot [fill=black,mark=*,only marks,fill opacity=0.5,draw opacity=0.5, mark size=2,clip mode=individual]
table[row sep=crcr]{%
0.36	0.081632653	\\
0.38	0.12244898	\\
0.4	0.163265306	\\
0.42	0.204081633	\\
0.43	0.244897959	\\
0.45	0.285714286	\\
0.46	0.326530612	\\
0.47	0.367346939	\\
0.49	0.408163265	\\
0.51	0.448979592	\\
0.53	0.489795918	\\
0.55	0.530612245	\\
0.58	0.571428571	\\
0.6	0.612244898	\\
0.63	0.653061224	\\
0.67	0.693877551	\\
0.71	0.734693878	\\
0.74	0.775510204	\\
0.77	0.816326531	\\
0.81	0.857142857	\\
0.84	0.897959184	\\
0.86	0.93877551	\\
0.89	0.979591837	\\
0.9	1	\\
0.91	1.020408163	\\
0.93	1.06122449	\\
0.95	1.102040816	\\
0.96	1.142857143	\\
0.97	1.183673469	\\
0.98	1.224489796	\\
0.99	1.265306122	\\
0.99	1.306122449	\\
}; 

\addplot[color=red,mark=none,line width = 1.5,draw opacity=1] 
table[row sep=crcr]{%
4.65E-05	0	\\
0.226243596	0.006617433	\\
0.254207989	0.013234866	\\
0.272276366	0.019852299	\\
2.86E-01	0.026469732	\\
2.97E-01	0.033087165	\\
3.07E-01	0.039704598	\\
3.15E-01	0.046322031	\\
3.22E-01	0.052939464	\\
3.29E-01	0.059556897	\\
3.35E-01	0.06617433	\\
3.41E-01	0.072791763	\\
3.47E-01	0.079409196	\\
3.52E-01	0.086026629	\\
3.57E-01	0.092644062	\\
3.61E-01	0.099261494	\\
3.65E-01	0.105878927	\\
3.70E-01	0.11249636	\\
3.74E-01	0.119113793	\\
3.77E-01	0.125731226	\\
3.81E-01	0.132348659	\\
3.85E-01	0.138966092	\\
3.88E-01	0.145583525	\\
3.91E-01	0.152200958	\\
3.95E-01	0.158818391	\\
3.98E-01	0.165435824	\\
4.01E-01	0.172053257	\\
4.04E-01	0.17867069	\\
4.07E-01	0.185288123	\\
4.10E-01	0.191905556	\\
4.12E-01	0.198522989	\\
4.15E-01	0.205140422	\\
4.18E-01	0.211757855	\\
4.20E-01	0.218375288	\\
4.23E-01	0.224992721	\\
4.25E-01	0.231610154	\\
4.28E-01	0.238227587	\\
4.30E-01	0.24484502	\\
4.33E-01	0.251462453	\\
4.35E-01	0.258079886	\\
4.37E-01	0.264697319	\\
4.40E-01	0.271314752	\\
4.42E-01	0.277932185	\\
4.44E-01	0.284549618	\\
4.46E-01	0.29116705	\\
4.48E-01	0.297784483	\\
4.51E-01	0.304401916	\\
4.53E-01	0.311019349	\\
4.55E-01	0.317636782	\\
4.57E-01	0.324254215	\\
4.59E-01	0.330871648	\\
4.61E-01	0.337489081	\\
4.63E-01	0.344106514	\\
4.65E-01	0.350723947	\\
4.67E-01	0.35734138	\\
4.69E-01	0.363958813	\\
4.72E-01	0.370576246	\\
4.74E-01	0.377193679	\\
4.76E-01	0.383811112	\\
0.478071452	0.390428545	\\
0.480289933	0.397045978	\\
0.48254375	0.403663411	\\
0.484837468	0.410280844	\\
0.48717571	0.416898277	\\
0.489563142	0.42351571	\\
0.492004454	0.430133143	\\
0.494504336	0.436750576	\\
0.497067462	0.443368009	\\
0.499698465	0.449985442	\\
0.502401913	0.456602875	\\
0.504419858	0.463220308	\\
0.507287482	0.469837741	\\
0.510239717	0.476455174	\\
0.513279866	0.483072606	\\
0.516411073	0.489690039	\\
0.519636299	0.496307472	\\
0.522958296	0.502924905	\\
0.526379588	0.509542338	\\
0.529902446	0.516159771	\\
0.533528872	0.522777204	\\
0.537260575	0.529394637	\\
0.541098955	0.53601207	\\
0.545045088	0.542629503	\\
0.549099708	0.549246936	\\
0.553263197	0.555864369	\\
0.557535575	0.562481802	\\
0.561916489	0.569099235	\\
0.566405207	0.575716668	\\
0.571000619	0.582334101	\\
0.575701229	0.588951534	\\
0.58050516	0.595568967	\\
0.585410156	0.6021864	\\
0.590413591	0.608803833	\\
0.595512472	0.615421266	\\
0.600703457	0.622038699	\\
0.60598286	0.628656132	\\
0.611346674	0.635273565	\\
0.616790583	0.641890998	\\
0.622309984	0.648508431	\\
0.627900005	0.655125864	\\
0.633555533	0.661743297	\\
0.639271233	0.66836073	\\
0.645041575	0.674978162	\\
0.650860861	0.681595595	\\
0.65672325	0.688213028	\\
0.662622787	0.694830461	\\
0.668553431	0.701447894	\\
0.674509082	0.708065327	\\
0.680483607	0.71468276	\\
0.686470873	0.721300193	\\
0.692464768	0.727917626	\\
0.69845923	0.734535059	\\
0.704448271	0.741152492	\\
0.710426004	0.747769925	\\
0.716386662	0.754387358	\\
0.722324621	0.761004791	\\
0.728234421	0.767622224	\\
0.734110784	0.774239657	\\
0.739948627	0.78085709	\\
0.745743085	0.787474523	\\
0.751489516	0.794091956	\\
0.757183515	0.800709389	\\
0.762820925	0.807326822	\\
0.768397843	0.813944255	\\
0.773910625	0.820561688	\\
0.779355893	0.827179121	\\
0.784730531	0.833796554	\\
0.790031693	0.840413987	\\
0.795256797	0.84703142	\\
0.800403523	0.853648853	\\
0.805469811	0.860266286	\\
0.810453853	0.866883718	\\
0.815354088	0.873501151	\\
0.820169197	0.880118584	\\
0.824898088	0.886736017	\\
0.829539895	0.89335345	\\
0.83409396	0.899970883	\\
0.838559828	0.906588316	\\
0.842937235	0.913205749	\\
0.847226096	0.919823182	\\
0.851426491	0.926440615	\\
0.855538659	0.933058048	\\
0.859562981	0.939675481	\\
0.863499972	0.946292914	\\
0.867350268	0.952910347	\\
0.871114613	0.95952778	\\
0.874793852	0.966145213	\\
0.878388919	0.972762646	\\
0.881900823	0.979380079	\\
0.885330645	0.985997512	\\
0.888679523	0.992614945	\\
0.891948647	0.999232378	\\
0.895139249	1.005849811	\\
0.898252595	1.012467244	\\
0.90128998	1.019084677	\\
0.90425272	1.02570211	\\
0.907142145	1.032319543	\\
0.909959597	1.038936976	\\
0.91270642	1.045554409	\\
0.915383963	1.052171841	\\
0.917993568	1.058789274	\\
0.920536571	1.065406707	\\
0.923014301	1.07202414	\\
0.925428072	1.078641573	\\
0.927779185	1.085259006	\\
0.930068925	1.091876439	\\
0.93229856	1.098493872	\\
0.934469339	1.105111305	\\
0.936582491	1.111728738	\\
0.938639226	1.118346171	\\
0.940640732	1.124963604	\\
0.942588179	1.131581037	\\
0.944482715	1.13819847	\\
0.946325464	1.144815903	\\
0.948117536	1.151433336	\\
0.949860015	1.158050769	\\
0.951553969	1.164668202	\\
0.953200445	1.171285635	\\
0.954800471	1.177903068	\\
0.956355057	1.184520501	\\
0.957865197	1.191137934	\\
0.959331865	1.197755367	\\
0.960756019	1.2043728	\\
0.962138603	1.210990233	\\
0.963480543	1.217607666	\\
0.96478275	1.224225099	\\
0.966046123	1.230842532	\\
0.967271543	1.237459965	\\
0.96845988	1.244077397	\\
0.969611989	1.25069483	\\
0.970728714	1.257312263	\\
0.971810884	1.263929696	\\
0.972859317	1.270547129	\\
0.973874817	1.277164562	\\
0.974858179	1.283781995	\\
0.975810185	1.290399428	\\
0.976731602	1.297016861	\\
0.977623192	1.303634294	\\
0.978485699	1.310251727	\\
0.979319861	1.31686916	\\
0.980126401	1.323486593	\\
0.980906033	1.330104026	\\
0.981659459	1.336721459	\\
0.98238737	1.343338892	\\
0.983090446	1.349956325	\\
0.983769356	1.356573758	\\
0.984424757	1.363191191	\\
0.985057297	1.369808624	\\
0.98566761	1.376426057	\\
0.98625632	1.38304349	\\
0.986824042	1.389660923	\\
0.987371377	1.396278356	\\
0.987898916	1.402895789	\\
0.988407238	1.409513222	\\
0.988896912	1.416130655	\\
0.989368495	1.422748088	\\
0.989822534	1.429365521	\\
0.990259563	1.435982953	\\
0.990680106	1.442600386	\\
0.991084676	1.449217819	\\
0.991473774	1.455835252	\\
0.991847891	1.462452685	\\
0.992207506	1.469070118	\\
0.992553088	1.475687551	\\
0.992885096	1.482304984	\\
0.993203974	1.488922417	\\
0.993510161	1.49553985	\\
0.993804082	1.502157283	\\
0.994086151	1.508774716	\\
0.994356772	1.515392149	\\
0.994616342	1.522009582	\\
0.994865242	1.528627015	\\
0.995103848	1.535244448	\\
0.995332523	1.541861881	\\
0.99555162	1.548479314	\\
0.995761485	1.555096747	\\
0.995962451	1.56171418	\\
0.996154844	1.568331613	\\
0.99633898	1.574949046	\\
0.996515166	1.581566479	\\
0.996683698	1.588183912	\\
0.996844866	1.594801345	\\
0.99699895	1.601418778	\\
0.997146221	1.608036211	\\
0.997286942	1.614653644	\\
0.997421369	1.621271077	\\
0.997549748	1.627888509	\\
0.997672317	1.634505942	\\
0.997789308	1.641123375	\\
0.997900946	1.647740808	\\
0.998007444	1.654358241	\\
0.998109014	1.660975674	\\
0.998205857	1.667593107	\\
0.998298167	1.67421054	\\
0.998386133	1.680827973	\\
0.998469937	1.687445406	\\
0.998549753	1.694062839	\\
0.998625752	1.700680272	\\
0.998698096	1.707297705	\\
0.998766943	1.713915138	\\
0.998832442	1.720532571	\\
0.998894742	1.727150004	\\
0.998953981	1.733767437	\\
0.999010294	1.74038487	\\
0.999063813	1.747002303	\\
0.999114661	1.753619736	\\
0.999162959	1.760237169	\\
0.999208823	1.766854602	\\
0.999252363	1.773472035	\\
0.999293685	1.780089468	\\
0.999332893	1.786706901	\\
0.999370084	1.793324334	\\
0.999405352	1.799941767	\\
0.999438788	1.8065592	\\
0.999470479	1.813176633	\\
0.999500506	1.819794065	\\
0.999528951	1.826411498	\\
0.999555889	1.833028931	\\
0.999581393	1.839646364	\\
0.999605532	1.846263797	\\
0.999628374	1.85288123	\\
0.999649983	1.859498663	\\
0.999670419	1.866116096	\\
0.999689741	1.872733529	\\
0.999708004	1.879350962	\\
0.999725263	1.885968395	\\
0.999741567	1.892585828	\\
0.999756966	1.899203261	\\
0.999771506	1.905820694	\\
0.999785231	1.912438127	\\
0.999798183	1.91905556	\\
0.999810402	1.925672993	\\
0.999821927	1.932290426	\\
0.999832795	1.938907859	\\
0.999843039	1.945525292	\\
0.999852693	1.952142725	\\
0.999861789	1.958760158	\\
0.999870357	1.965377591	\\
0.999878425	1.971995024	\\
0.999886019	1.978612457	\\
0.999893167	1.98522989	\\
0.999899892	1.991847323	\\
0.999906218	1.998464756	\\
};

\addplot [fill=red,mark=*,only marks,fill opacity=0.8,draw opacity=1, mark size=2,clip mode=individual]
table[row sep=crcr]{%
0.60	0.6122	\\
};

\node[align=left,anchor=west,fill=white,fill opacity=0.8,text opacity=1] at (rel axis cs:0.53,0.235) {\textcolor{gray}{\scriptsize{$\theta = 37.5^\circ$}}};

\node[align=left,anchor=west,fill=white,fill opacity=0.8,text opacity=1] at (rel axis cs:0.5,0.085) {\textcolor{gray}{\scriptsize{$\langle \overline{c} \rangle = 0.55$}}};

\end{groupplot}

\end{tikzpicture}

%% file: Figures/table1.tex
\begin{table*}
\centering
\caption{Normalized parameters of the two-state convolution model for  profiles shown in Fig. \ref{Fig:stepped_air}; as no detailed velocity measurements were available from \cite{Straub1958}, we determined $y_\star$ first, and subsequently assumed $\delta = 1.25 y_\star$ (see $\S$ \ref{sec:transition}); measurements taken at $x = 13.88$ m from the flume inlet}
\begin{tabular}{c c c c c c c c c c c c c c}
\toprule
sub-figure & $q$ & $x$ & $ \theta$ & $\langle \overline{c} \rangle$ & $\beta$ & $\overline{c}_{\delta/2}$ & $\delta/y_\star$  & $y_{50_\text{TWL}}/y_{90}$ & $\mathcal{H}/y_{90}$ & $y_\star/y_{90}$ & $\sigma_\star/\delta$ &  & 
\\
(-) & (m$^2$/s) & (m) & ($^\circ$) & (-) & (-) & (-) & (-) & (-) & (-) & (-) & (-) \\
\midrule
(\textit{a}) & 0.32 & 13.88 & 7.5  & 0.16 & - & - & - & 0.85 & 0.11 & - & -  \\
(\textit{b}) & 0.32 & 13.88 & 15.0 & 0.25 & 0.50 & 0.06 & 1.25 & 0.79 & 0.17 & 0.63 & 0.11 \\
(\textit{c}) & 0.32 & 13.88 & 22.5 & 0.30 & 0.43 & 0.11 & 1.25 & 0.74 & 0.21 & 0.58 & 0.12 \\
(\textit{d}) & 0.32 & 13.88 & 30.0 & 0.40 & 0.27 & 0.24 & 1.25 & 0.65 & 0.28 & 0.59 & 0.13 \\
(\textit{e}) & 0.32 & 13.88 & 37.5 & 0.55 & 0.14 & 0.47 & 1.25 & 0.48 & 0.42 & 0.61 & 0.13 \\
(\textit{f}) & 0.32 & 13.88 & 45.0 & 0.60 & 0.09 & 0.56 & 1.25 & 0.39 & 0.48 & 0.61 & 0.15 \\
(\textit{g}) & 0.32 & 13.88 & 60.0 & 0.65 & 0.09 & 0.64 & 1.25 & 0.27 & 0.56 & 0.64 & 0.17 \\
(\textit{h}) & 0.32 & 13.88 & 75.0 & 0.69 & 0.07 & 0.70 & 1.25 & 0.06 & 0.72 & 0.67 & 0.2 \\
\bottomrule
\end{tabular}
\label{tab1}
\end{table*}

%% file: Figures/fig4.tex
\begin{tikzpicture}
\definecolor{color1}{rgb}{1,0.7,0.1}
\definecolor{color2}{rgb}{0.15,0.45,0.55}

 \begin{groupplot}[
     group style = {group size = 2 by 2,horizontal sep=1.2cm,vertical sep=1.2cm},
     width = 1\textwidth]

\nextgroupplot[height = 4.4cm,
     width = 4.4cm,
      ytick={0, 0.5, 1, 1.5, 2},
     yticklabels={0, 0.5 , 1, 1.5, 2},     
     ylabel={$y/y_{90}$},
      xlabel={$\overline{c}$},
        grid=both,
     xmin=0,
     xmax=1,
     ymin=0,
     ymax=1.75,
    legend columns=1,
   legend style={cells={align=left},anchor = north east,at={(-0.5,1)},font=\footnotesize},          domain=0:1.25,samples=400,
   ]

\node[align=center,fill=white,fill opacity=0.8,text opacity=1] at (rel axis cs:0.12,0.9) {(\textit{a})};  

\node[align=center,fill=white,fill opacity=0.8,text opacity=0.6] at (rel axis cs:0.65,0.6) {\footnotesize{TWL}};   

\node[align=center,fill=white,fill opacity=0.8,text opacity=0.6] at (rel axis cs:0.45,0.12) {\footnotesize{TBL}};

\addplot [fill=black,mark=*,only marks,fill opacity=0.5,draw opacity=0.5, mark size=2,clip mode=individual]
table[row sep=crcr]{%
0 -1	\\
}; \addlegendentry{$\overline{c}$ \,};

\addplot [fill=red,mark=*,only marks,fill opacity=0.8,draw opacity=1, mark size=2,clip mode=individual]
table[row sep=crcr]{%
0 -1	\\
}; \addlegendentry{$\overline{c}_\star$, $y_\star$ \, };

\addplot
[fill=black,mark=halfdiamond*,only marks,fill opacity=0.5,draw opacity=0.9, mark size=3,clip mode=individual]
table[row sep=crcr]{%
0 -1	\\
}; \addlegendentry{$\overline{u}/\overline{u}_\text{max}$ \, };

\addplot [mark=10-pointed star
,only marks,fill opacity=0.5,draw opacity=0.8, mark size=2,clip mode=individual]
table[row sep=crcr]{%
0 -1	\\
}; \addlegendentry{$u'_\text{rms}/\overline{u}_\text{max}$ \, };

 \addplot+[densely dotted,mark=none,color=black,line width=2,draw opacity=0.7] 
 table[row sep=crcr]{%
0 -1	\\
}; \addlegendentry{$\delta$ \,};

 \addplot+[densely dashdotted,mark=none,color=gray,line width=2,draw opacity=0.7] 
 table[row sep=crcr]{%
0 -1	\\
}; \addlegendentry{$y_\star$ \,};

\addplot[color=teal,mark=none,line width = 1.5,draw opacity=0.4] 
table[row sep=crcr]{%
0 -1	\\
}; \addlegendentry{Eq. (\ref{eq:voidfraction1}) \, };

\addplot[color=blue,mark=none,line width = 1.5,draw opacity=0.4] 
table[row sep=crcr]{%
0 -1	\\
}; \addlegendentry{Eq. (\ref{eq:voidfraction2}) \, };

\addplot[color=red,mark=none,line width = 1.5,draw opacity=0.8] 
table[row sep=crcr]{%
0 -1	\\
}; \addlegendentry{Eq. (\ref{eq:voidfractionfinal}) \, };

\addplot [fill=black,mark=pentagon*,only marks,fill opacity=0.5,draw opacity=0.5, mark size=3,clip mode=individual]
table[row sep=crcr]{%
0 -1	\\
}; \addlegendentry{\textbf{Smooth chutes} \\ \, -  \citet{Straub1958} \\ 
\, - \citet{Killen1968}\\
\, - \citet{Bung2009}
};

\addplot [fill=blue,mark=diamond*,only marks,fill opacity=0.5,draw opacity=0.5,draw=blue, mark size=3,clip mode=individual]
table[row sep=crcr]{%
0 -1	\\
}; \addlegendentry{\textbf{Stepped  chutes} \\ 
\, - \citet{Bung2009}\\
\, - \citet{Zhang2017DISS}\\
\, - \citet{Kramer2018Transiton}};

 \addplot+[densely dashdotted,mark=none,color=gray,line width=2,draw opacity=0.7] {0.57};

\addplot [fill=black,mark=*,only marks,fill opacity=0.5,draw opacity=0.5, mark size=2,clip mode=individual]
table[row sep=crcr]{%
0.043289444	0.0099795	\\
0.071438333	0.072351378	\\
0.135207222	0.134723256	\\
0.162609444	0.196887227	\\
0.193477222	0.259259105	\\
0.226347222	0.322046796	\\
0.268282778	0.384418673	\\
0.321564444	0.446582645	\\
0.395193889	0.509370335	\\
0.536748889	0.634321997	\\
0.618556111	0.697109687	\\
0.701577778	0.759689471	\\
0.760947778	0.822269255	\\
0.827032778	0.884433227	\\
0.872488333	0.946805105	\\
0.903671667	1.00876117	\\
0.935738333	1.070925141	\\
0.956602778	1.133504925	\\
0.968179444	1.195876803	\\
0.975837778	1.258248681	\\
0.981897778	1.320620559	\\
};

\addplot[solid,color=teal,mark=none,line width = 1.5,draw opacity=0.4] 
table[row sep=crcr]{%
0.043561655	0.020790638	\\
0.062956807	0.041581277	\\
0.078594905	0.062371915	\\
0.092434035	0.083162553	\\
0.10523737	0.103953191	\\
0.117402972	0.12474383	\\
0.1291749	0.145534468	\\
0.140720416	0.166325106	\\
0.152164557	0.187115744	\\
0.163607894	0.207906383	\\
0.175136697	0.228697021	\\
0.186829318	0.249487659	\\
0.198760615	0.270278298	\\
0.211005328	0.291068936	\\
0.223640966	0.311859574	\\
0.236750543	0.332650212	\\
0.250425444	0.353440851	\\
0.264768668	0.374231489	\\
0.274123414	0.395022127	\\
0.289783328	0.415812765	\\
0.306337851	0.436603404	\\
0.323838088	0.457394042	\\
0.342338065	0.47818468	\\
0.361894894	0.498975319	\\
0.382568951	0.519765957	\\
0.404424061	0.540556595	\\
0.427527692	0.561347233	\\
0.45195117	0.582137872	\\
0.477769894	0.60292851	\\
0.505063571	0.623719148	\\
0.53391646	0.644509786	\\
0.564417635	0.665300425	\\
0.596661258	0.686091063	\\
0.63074687	0.706881701	\\
0.666779699	0.72767234	\\
0.704870985	0.748462978	\\
0.74513832	0.769253616	\\
0.787706018	0.790044254	\\
0.83270549	0.810834893	\\
0.880275659	0.831625531	\\
0.93056338	0.852416169	\\
0.983723899	0.873206807	\\
1.039921333	0.893997446	\\
1.099329171	0.914788084	\\
1.162130815	0.935578722	\\
1.228520144	0.956369361	\\
1.298702112	0.977159999	\\
1.372893382	0.997950637	\\
1.451322995	1.018741275	\\
1.534233076	1.039531914	\\
1.621879581	1.060322552	\\
1.714533089	1.08111319	\\
1.812479636	1.101903828	\\
1.916021599	1.122694467	\\
2.02547863	1.143485105	\\
2.141188638	1.164275743	\\
2.263508841	1.185066382	\\
2.392816859	1.20585702	\\
2.529511887	1.226647658	\\
2.674015926	1.247438296	\\
2.826775082	1.268228935	\\
2.988260947	1.289019573	\\
3.158972055	1.309810211	\\
3.339435417	1.330600849	\\
3.530208153	1.351391488	\\
3.73187921	1.372182126	\\
3.945071178	1.392972764	\\
4.170442216	1.413763403	\\
4.408688079	1.434554041	\\
4.660544272	1.455344679	\\
4.926788314	1.476135317	\\
5.208242144	1.496925956	\\
5.505774655	1.517716594	\\
5.820304378	1.538507232	\\
6.152802317	1.55929787	\\
6.504294947	1.580088509	\\
6.875867381	1.600879147	\\
7.268666723	1.621669785	\\
7.683905609	1.642460424	\\
8.122865947	1.663251062	\\
8.586902879	1.6840417	\\
9.077448961	1.704832338	\\
9.596018588	1.725622977	\\
10.14421267	1.746413615	\\
10.72372356	1.767204253	\\
11.33634032	1.787994891	\\
11.98395418	1.80878553	\\
12.66856443	1.829576168	\\
13.39228459	1.850366806	\\
14.15734887	1.871157445	\\
14.96611917	1.891948083	\\
15.82109229	1.912738721	\\
16.72490767	1.933529359	\\
17.68035552	1.954319998	\\
18.69038547	1.975110636	\\
19.75811566	1.995901274	\\
};

\addplot[solid,color=blue,mark=none,line width = 1.5,draw opacity=0.4] 
table[row sep=crcr]{%
0.033207532	0	\\
0.03825495	0.020790638	\\
0.043911103	0.041581277	\\
0.050223257	0.062371915	\\
0.057238437	0.083162553	\\
0.065002778	0.103953191	\\
0.073560835	0.12474383	\\
0.082954831	0.145534468	\\
0.093223881	0.166325106	\\
0.104403186	0.187115744	\\
0.116523225	0.207906383	\\
0.129608954	0.228697021	\\
0.143679032	0.249487659	\\
0.158745095	0.270278298	\\
0.174811092	0.291068936	\\
0.191872714	0.311859574	\\
0.209916911	0.332650212	\\
0.228921546	0.353440851	\\
0.248855165	0.374231489	\\
0.269676927	0.395022127	\\
0.291336677	0.415812765	\\
0.313775191	0.436603404	\\
0.336924574	0.457394042	\\
0.360708822	0.47818468	\\
0.385044537	0.498975319	\\
0.409841784	0.519765957	\\
0.435005078	0.540556595	\\
0.460434483	0.561347233	\\
0.486026798	0.582137872	\\
0.511676814	0.60292851	\\
0.537278606	0.623719148	\\
0.562726847	0.644509786	\\
0.587918105	0.665300425	\\
0.612752101	0.686091063	\\
0.637132906	0.706881701	\\
0.660970046	0.72767234	\\
0.6841795	0.748462978	\\
0.706684571	0.769253616	\\
0.728416612	0.790044254	\\
0.749315604	0.810834893	\\
0.769330567	0.831625531	\\
0.788419822	0.852416169	\\
0.806551074	0.873206807	\\
0.823701358	0.893997446	\\
0.839856823	0.914788084	\\
0.855012396	0.935578722	\\
0.869171308	0.956369361	\\
0.882344532	0.977159999	\\
0.894550128	0.997950637	\\
0.905812517	1.018741275	\\
0.916161716	1.039531914	\\
0.925632538	1.060322552	\\
0.934263779	1.08111319	\\
0.942097417	1.101903828	\\
0.949177828	1.122694467	\\
0.955551035	1.143485105	\\
0.96126401	1.164275743	\\
0.966364024	1.185066382	\\
0.970898059	1.20585702	\\
0.974912298	1.226647658	\\
0.97845167	1.247438296	\\
0.981559476	1.268228935	\\
0.98427708	1.289019573	\\
0.986643671	1.309810211	\\
0.988696085	1.330600849	\\
0.990468687	1.351391488	\\
0.991993309	1.372182126	\\
0.993299233	1.392972764	\\
0.994413215	1.413763403	\\
0.995359547	1.434554041	\\
0.996160143	1.455344679	\\
0.996834651	1.476135317	\\
0.997400586	1.496925956	\\
0.997873466	1.517716594	\\
0.99826696	1.538507232	\\
0.998593046	1.55929787	\\
0.998862156	1.580088509	\\
0.999083329	1.600879147	\\
0.999264354	1.621669785	\\
0.999411908	1.642460424	\\
0.999531684	1.663251062	\\
0.999628509	1.6840417	\\
0.999706459	1.704832338	\\
0.999768953	1.725622977	\\
0.999818851	1.746413615	\\
0.999858526	1.767204253	\\
0.999889943	1.787994891	\\
0.999914718	1.80878553	\\
0.999934175	1.829576168	\\
0.999949392	1.850366806	\\
0.999961244	1.871157445	\\
0.999970437	1.891948083	\\
0.999977538	1.912738721	\\
0.999983001	1.933529359	\\
0.999987186	1.954319998	\\
0.999990378	1.975110636	\\
0.999992804	1.995901274	\\
};

\addplot[solid,color=red,mark=none,line width = 1.5,draw opacity=1] 
table[row sep=crcr]{%
4.34E-13	0	\\
0.043561655	0.020790638	\\
0.062956807	0.041581277	\\
0.078594905	0.062371915	\\
0.092434034	0.083162553	\\
0.105237368	0.103953191	\\
0.117402962	0.12474383	\\
0.129174855	0.145534468	\\
0.140720246	0.166325106	\\
0.152163967	0.187115744	\\
0.16360602	0.207906383	\\
0.17513125	0.228697021	\\
0.186814852	0.249487659	\\
0.198725551	0.270278298	\\
0.210927967	0.291068936	\\
0.223486303	0.311859574	\\
0.236472456	0.332650212	\\
0.249981721	0.353440851	\\
0.264156454	0.374231489	\\
0.273824393	0.395022127	\\
0.289954793	0.415812765	\\
0.30760556	0.436603404	\\
0.32708571	0.457394042	\\
0.348613091	0.47818468	\\
0.372217068	0.498975319	\\
0.397681124	0.519765957	\\
0.424559309	0.540556595	\\
0.452268133	0.561347233	\\
0.480218543	0.582137872	\\
0.507934041	0.60292851	\\
0.535112182	0.623719148	\\
0.561618473	0.644509786	\\
0.587433184	0.665300425	\\
0.612585346	0.686091063	\\
0.637101816	0.706881701	\\
0.660982464	0.72767234	\\
0.684197632	0.748462978	\\
0.706697463	0.769253616	\\
0.728423706	0.790044254	\\
0.749318923	0.810834893	\\
0.769331937	0.831625531	\\
0.788420328	0.852416169	\\
0.806551244	0.873206807	\\
0.823701409	0.893997446	\\
0.839856838	0.914788084	\\
0.8550124	0.935578722	\\
0.869171309	0.956369361	\\
0.882344532	0.977159999	\\
0.894550128	0.997950637	\\
0.905812517	1.018741275	\\
0.916161716	1.039531914	\\
0.925632538	1.060322552	\\
0.934263779	1.08111319	\\
0.942097417	1.101903828	\\
0.949177828	1.122694467	\\
0.955551035	1.143485105	\\
0.96126401	1.164275743	\\
0.966364024	1.185066382	\\
0.970898059	1.20585702	\\
0.974912298	1.226647658	\\
0.97845167	1.247438296	\\
0.981559476	1.268228935	\\
0.98427708	1.289019573	\\
0.986643671	1.309810211	\\
0.988696085	1.330600849	\\
0.990468687	1.351391488	\\
0.991993309	1.372182126	\\
0.993299233	1.392972764	\\
0.994413215	1.413763403	\\
0.995359547	1.434554041	\\
0.996160143	1.455344679	\\
0.996834651	1.476135317	\\
0.997400586	1.496925956	\\
0.997873466	1.517716594	\\
0.99826696	1.538507232	\\
0.998593046	1.55929787	\\
0.998862156	1.580088509	\\
0.999083329	1.600879147	\\
0.999264354	1.621669785	\\
0.999411908	1.642460424	\\
0.999531684	1.663251062	\\
0.999628509	1.6840417	\\
0.999706459	1.704832338	\\
0.999768953	1.725622977	\\
0.999818851	1.746413615	\\
0.999858526	1.767204253	\\
0.999889943	1.787994891	\\
0.999914718	1.80878553	\\
0.999934175	1.829576168	\\
0.999949392	1.850366806	\\
0.999961244	1.871157445	\\
0.999970437	1.891948083	\\
0.999977538	1.912738721	\\
0.999983001	1.933529359	\\
0.999987186	1.954319998	\\
0.999990378	1.975110636	\\
0.999992804	1.995901274	\\
};

\addplot [fill=red,mark=*,only marks,fill opacity=0.8,draw opacity=1, mark size=2,clip mode=individual]
table[row sep=crcr]{%
0.453340556	0.571742213	\\
};

\nextgroupplot[height = 4.4cm,
     width = 4.4cm,
  xlabel={$\overline{u}/\overline{u}_\text{max}$, $u'_\text{rms}/\overline{u}_\text{max}$ \, },
      ytick={0,  0.69, 1, 1.5, 2},
     yticklabels={0,  \textcolor{black}{$\delta$}, 1, 1.5, 2},
     grid=both,
     xmin=0,
     xmax=1.2,
     ymin=0,
     ymax=1.75,
    legend columns=1,
   legend style={cells={align=left},anchor = north west,at={(1.1,1)},font=\normalsize},
           domain=0:1.2,samples=400,
   ]

   \node[align=center,fill=white,fill opacity=0.8,text opacity=1] at (rel axis cs:0.12,0.9) {(\textit{b})};   
   
\node[align=center,fill=white,fill opacity=0.8,text opacity=0.6] at (rel axis cs:0.6,0.6) {\footnotesize{TWL}};   

\node[align=center,fill=white,fill opacity=0.8,text opacity=0.6] at (rel axis cs:0.85,0.12) {\footnotesize{TBL}};    

 \addplot+[densely dashdotted,mark=none,color=gray,line width=2,draw opacity=0.7] {0.57};

\addplot+[densely dotted,mark=none,color=black,line width=2,draw opacity=0.7] {0.69};

\addplot [mark=10-pointed star,only marks,fill opacity=0.5,draw opacity=0.8, mark size=2,clip mode=individual]
table[row sep=crcr]{%
0.526678658	0.072351378	\\
0.461441955	0.134723256	\\
0.464065038	0.196887227	\\
0.392304316	0.259259105	\\
0.311693374	0.322046796	\\
0.275475032	0.384418673	\\
0.241744277	0.446582645	\\
0.229936228	0.509370335	\\
0.226394232	0.571742213	\\
0.22772021	0.634321997	\\
0.209111693	0.697109687	\\
0.183463902	0.759689471	\\
0.180268193	0.822269255	\\
0.175762258	0.884433227	\\
0.177640817	0.946805105	\\
0.166565706	1.00876117	\\
0.184384172	1.070925141	\\
0.149362365	1.133504925	\\
0.137182611	1.195876803	\\
0.144225957	1.258248681	\\
0.161269274	1.320620559	\\
};

\addplot
[fill=black,mark=halfdiamond*,only marks,fill opacity=0.5,draw opacity=0.9, mark size=3,clip mode=individual]
table[row sep=crcr]{%
0.44317111	0.072351378	\\
0.615384615	0.134723256	\\
0.648648649	0.196887227	\\
0.727272727	0.259259105	\\
0.8	0.322046796	\\
0.842364532	0.384418673	\\
0.857142857	0.446582645	\\
0.923076923	0.509370335	\\
0.923076923	0.571742213	\\
0.96	0.634321997	\\
0.923076923	0.697109687	\\
0.96	0.759689471	\\
0.96	0.822269255	\\
0.96	0.884433227	\\
0.96	0.946805105	\\
0.96	1.00876117	\\
0.96	1.070925141	\\
0.96	1.133504925	\\
0.96	1.195876803	\\
0.96	1.258248681	\\
1	1.320620559	\\
};

\nextgroupplot[height = 4.4cm,
     width = 4.4 cm,
     ylabel={$y_\star/ \delta$},
     xlabel={$\langle \overline{c} \rangle$},
     grid=both,
     xmin=0,
     xmax=0.8,
     ymin=0,
     ymax=1,
    legend columns=2,
   legend style={cells={align=left},anchor = south west,at={(0,1.2)},font=\normalsize}
   ]

\node[align=center,fill=white,fill opacity=0.8,text opacity=1] at (rel axis cs:0.12,0.9) {(\textit{c})};

\addplot [fill=blue,mark=diamond*,only marks,fill opacity=0.5,draw opacity=0.5,draw=blue, mark size=3,clip mode=individual]
table[row sep=crcr]{%
0.668891923	0.769230769	\\
0.654241261	0.769230769	\\
0.646286855	0.860788863	\\
0.59602408	0.721059972	\\
0.646812919	0.769230769	\\
0.592229508	0.75549227	\\
0.583634226	0.769230769	\\
0.617637863	0.860865519	\\
0.567247873	0.769230769	\\
0.587487844	0.769230769	\\
0.627622101	0.769230769	\\
0.602584342	0.856811863	\\
0.644581275	0.769230769	\\
0.59101534	0.769230769	\\
0.562580361	0.860659545	\\
0.587351306	0.7043222	\\
0.535812598	0.802226588	\\
0.473876706	0.781976744	\\
0.550975071	0.769230769	\\
0.445267873	0.599444444	\\
0.458722528	0.67095186	\\
0.448352939	0.635163308	\\
0.431439309	0.890744102	\\
0.429715168	0.670498084	\\
0.326396375	0.767261443	\\
0.413837285	0.769230769	\\
0.414374751	0.859764706	\\
0.439059377	0.705128205	\\
0.396079757	0.917669584	\\
0.359066301	0.842942865	\\
0.338403374	0.858421673	\\
0.399705525	0.769230769	\\
0.411594452	0.664279767	\\
0.405918342	0.784383954	\\
0.40351446	0.76219844	\\
0.355121332	0.746979977	\\
};

\addplot [fill=black,mark=pentagon*,only marks,fill opacity=0.5,draw opacity=0.5, mark size=3,clip mode=individual]
table[row sep=crcr]{%
0.279105263	0.8	\\
0.283871391	0.8	\\
0.296059343	0.7	\\
0.330622449	0.727272727	\\
0.355041237	0.666666667	\\
0.324873936	0.666666667	\\
0.353172437	0.583333333	\\
0.256571514	0.75	\\
0.265983683	0.75	\\
0.291258065	0.769230769	\\
0.314695496	0.714285714	\\
0.296886882	0.666666667	\\
0.328058226	0.642857143	\\
0.329708621	0.625	\\
0.331049133	0.625	\\
0.271521283	0.769230769	\\
0.264347079	0.714285714	\\
0.284164688	0.785714286	\\
0.291487752	0.785714286	\\
0.310328678	0.666666667	\\
0.32638613	0.666666667	\\
0.312257143	0.705882353	\\
};

 \nextgroupplot[height = 4.4cm,
     width = 4.4 cm,
     ylabel={$\overline{c}_\star$},
     xlabel={$\langle \overline{c} \rangle$},
     grid=both,
     xmin=0,
     xmax=0.8,
     ymin=0,
     ymax=1,     
    legend columns=2,
   legend style={cells={align=left},anchor = north east,at={(1,1.4)},font=\normalsize},
    domain=0:0.75,samples=400,
   ]

\node[align=center] at (rel axis cs:0.9,0.1) {(\textit{d})};

\addplot [fill=blue,mark=diamond*,only marks,fill opacity=0.5,draw opacity=0.5,draw=blue, mark size=3,clip mode=individual]
table[row sep=crcr]{%
0.668891923	0.75	\\
0.654241261	0.74	\\
0.646286855	0.7	\\
0.59602408	0.61	\\
0.646812919	0.71	\\
0.592229508	0.67	\\
0.583634226	0.64	\\
0.617637863	0.67	\\
0.567247873	0.63	\\
0.587487844	0.75	\\
0.627622101	0.71	\\
0.602584342	0.64	\\
0.644581275	0.73	\\
0.59101534	0.68	\\
0.562580361	0.57	\\
0.587351306	0.647	\\
0.535812598	0.59	\\
0.473876706	0.49	\\
0.550975071	0.65	\\
0.445267873	0.38	\\
0.458722528	0.47	\\
0.448352939	0.46	\\
0.431439309	0.44	\\
0.429715168	0.39	\\
0.326396375	0.3	\\
0.413837285	0.47	\\
0.414374751	0.37	\\
0.439059377	0.28	\\
0.396079757	0.35	\\
0.359066301	0.31	\\
0.338403374	0.32	\\
0.206843198	0	\\
0.399705525	0.494791111	\\
0.411594452	0.324664444	\\
0.405918342	0.388656667	\\
0.40351446	0.383502222	\\
0.355121332	0.318102222	\\
};

\addplot [fill=black,mark=pentagon*,only marks,fill opacity=0.5,draw opacity=0.5, mark size=3,clip mode=individual]
table[row sep=crcr]{%
0.24615	0.09	\\
0.238217822	0.21	\\
0.240756303	0.21	\\
0.244017857	0.16	\\
0.25441989	0.23	\\
0.246363636	0.16	\\
0.245333333	0.17	\\
0.230075188	0.16	\\
0.248318584	0.27	\\
0.312333333	0.34	\\
0.303024691	0.24	\\
0.327916667	0.22	\\
0.296348315	0.28	\\
0.31390625	0.3	\\
0.306090909	0.23	\\
0.293691275	0.19	\\
0.230075188	0.23	\\
0.345597015	0.35	\\
0.323920455	0.28	\\
0.348243243	0.31	\\
0.395625	0.37	\\
0.402910448	0.43	\\
0.398157895	0.4	\\
0.395228758	0.45	\\
0.369178767	0.38	\\
0.373228481	0.37	\\
0.361706471	0.3	\\
0.349253247	0.31	\\
0.445357143	0.45	\\
0.43175	0.45	\\
0.390645161	0.4	\\
0.532142857	0.67	\\
0.532142857	0.83	\\
0.542688679	0.6	\\
0.545816327	0.67	\\
0.503518519	0.57	\\
0.496192053	0.55	\\
0.476746032	0.53	\\
0.449411765	0.48	\\
0.5	0.53	\\
0.452702703	0.45	\\
0.370298013	0.33	\\
0.5852	0.67	\\
0.5852	0.67	\\
0.588658537	0.68	\\
0.602959184	0.66	\\
0.596626506	0.67	\\
0.577931034	0.65	\\
0.556269841	0.6	\\
0.535571429	0.56	\\
0.608196721	0.68	\\
0.62469697	0.72	\\
0.6405	0.76	\\
0.647357143	0.78	\\
0.6405	0.78	\\
0.653852459	0.79	\\
0.635454545	0.74	\\
0.622183099	0.7	\\
0.6478875	0.77	\\
0.677978723	0.81	\\
0.698043478	0.82	\\
0.694259259	0.78	\\
0.709375	0.79	\\
0.69382716	0.79	\\
0.635454545	0.73	\\
0.678454545	0.76	\\
};

\addplot [fill=black,mark=pentagon*,only marks,fill opacity=0.5,draw opacity=0.5, mark size=3,clip mode=individual]
table[row sep=crcr]{%
0.241914894	0.27	\\
0.250294118	0.34	\\
0.216935484	0.23	\\
0.330142857	0.29	\\
0.150404624	0	\\
0.250294118	0.34	\\
0.242121212	0.14	\\
0.253461538	0.17	\\
0.192535211	0	\\
0.257363636	0.12	\\
0.349590164	0.19	\\
0.346639344	0.28	\\
0.459576271	0.38	\\
0.494913793	0.49	\\
0.494913793	0.49	\\
0.55377551	0.54	\\
};

\addplot [fill=black,mark=pentagon*,only marks,fill opacity=0.5,draw opacity=0.5, mark size=3,clip mode=individual]
table[row sep=crcr]{%
0.279105263	0.093	\\
0.283871391	0.095	\\
0.296059343	0.105	\\
0.330622449	0.157	\\
0.355041237	0.178	\\
0.324873936	0.16	\\
0.353172437	0.151	\\
0.256571514	0.062	\\
0.265983683	0.089	\\
0.291258065	0.142	\\
0.314695496	0.161	\\
0.296886882	0.151	\\
0.328058226	0.153	\\
0.329708621	0.186	\\
0.331049133	0.175	\\
0.271521283	0.088	\\
0.264347079	0.085	\\
0.284164688	0.118	\\
0.291487752	0.137	\\
0.310328678	0.136	\\
0.32638613	0.142	\\
0.312257143	0.177	\\
};

\addplot+[densely dotted,mark=none,color=red,line width=1.7,draw opacity=0.9] {1.3716*x + -0.1414}; 

\node[align=left,anchor=west,fill=white,fill opacity=0.8,text opacity=1] at (rel axis cs:0.01,0.87) {\textbf{\small{\textcolor{red}{$\overline{c}_\star=1.37 \, \langle \overline{c}\rangle$}}}};

\node[align=left,fill=white,fill opacity=0.8,text opacity=1] at (rel axis cs:0.44,0.71) {
\small{\textcolor{red}{$-0.14$}} }; 

\end{groupplot}

 \begin{groupplot}[
     group style = {group size = 2 by 2,horizontal sep=1.2cm,vertical sep=1.2cm},
     width = 1\textwidth]

\nextgroupplot[height = 4.4cm,
     width = 4.4cm,     
  axis x line*=none,
  x axis line style={draw opacity=0},
  y axis line style={draw opacity=0},
  grid=none,
 xmin=0,
     xmax=1.25,
     ymin=0,
     xticklabels={ },
     ymax=1.75,
     ytick={0.57},
     yticklabels={\textcolor{black}{$y_\star$}},
     yticklabel pos=right
     ]

\nextgroupplot[height = 4.4cm,
     width = 4.4cm,     
  axis x line*=none,
  x axis line style={draw opacity=0},
  y axis line style={draw opacity=0},
  grid=none,
 xmin=0,
     xmax=1.25,
     ymin=0,
     xticklabels={ },
     ymax=1.75,
     ytick={0.57},
     yticklabels={\textcolor{black}{$y_\star$}},
     yticklabel pos=right
     ]

\end{groupplot}

\end{tikzpicture}

%% file: Figures/fig5.tex
\begin{tikzpicture}
\definecolor{color1}{rgb}{1,0.7,0.1}
\definecolor{color2}{rgb}{0.15,0.45,0.55}

 \begin{groupplot}[
     group style = {group size = 3 by 2,horizontal sep=1.4cm,vertical sep=1.4cm},
     width = 1\textwidth]

\nextgroupplot[height = 4.6cm,
     width = 4.6cm,
     ylabel={$\beta = \overline{v}_r  S_c /(\kappa u_*)$},
     xlabel={$\langle \overline{c} \rangle$},
     grid=both,
     xmin=0,
    ymode=log,
     xmax=0.8,
      ymin=0,
     ymax=10,     
    legend columns=2,
  domain=0.1:0.75,samples=400,
   legend style={cells={align=left},anchor = south west,at={(0,1.2)},font=\footnotesize}
   ]

\addplot [fill=black,mark=pentagon*,only marks,fill opacity=0.5,draw opacity=0.5, mark size=3,clip mode=individual]
table[row sep=crcr]{%
1 20	\\
}; \addlegendentry{\textbf{Smooth chutes} \\ \,  - \citet{Straub1958} \, \\
\, - \citet{Killen1968} \, \\
\, - \citet{Bung2009} \, \\
\, - \citet{Severi2018} \, };

\addplot [fill=blue,mark=diamond*,only marks,fill opacity=0.5,draw opacity=0.5,draw=blue, mark size=3,clip mode=individual]
table[row sep=crcr]{%
1 20	\\
}; \addlegendentry{\textbf{Stepped chutes}  \\
\, - \citet{Bung2009}\\
\, - \citet{Zhang2017DISS} \\
\, - \citet{Kramer2018Transiton} \\};

\node[align=center,fill=white,fill opacity=0.8,text opacity=1] at (rel axis cs:0.12,0.9) {(\textit{a})};

\node[align=left,anchor=west,fill=white,fill opacity=0.8,text opacity=0.6] at (rel axis cs:0.01,0.085) {\textbf{\scriptsize{\textcolor{red}{$y= 3.6 \exp(-5.7x)$}}}};

\addplot [fill=black,mark=pentagon*,only marks,fill opacity=0.5,draw opacity=0.5, mark size=3,clip mode=individual]
table[row sep=crcr]{%
0.279105263	1.270273576	\\
0.283871391	1.304125701	\\
0.256571514	1.296101525	\\
0.296059343	1.081884249	\\
0.265983683	1.141264555	\\
0.271521283	1.419154194	\\
0.330622449	0.78307419	\\
0.291258065	0.972707939	\\
0.264347079	0.926054544	\\
0.355041237	0.589773064	\\
0.314695496	0.587991646	\\
0.284164688	0.736660245	\\
0.324873936	0.545005246	\\
0.296886882	0.479945209	\\
0.291487752	0.520304478	\\
0.353172437	0.343682478	\\
0.328058226	0.354339716	\\
0.310328678	0.440456042	\\
0.329708621	0.333997832	\\
0.32638613	0.379370335	\\
0.331049133	0.365535901	\\
0.312257143	0.380640233	\\
0.25441989	0.404422193	\\
0.312333333	0.886082758	\\
0.303024691	0.734955373	\\
0.327916667	0.659881826	\\
0.296348315	0.452591506	\\
0.31390625	0.366857427	\\
0.306090909	0.501239994	\\
0.293691275	0.565341428	\\
0.345597015	0.286249008	\\
0.323920455	0.367623908	\\
0.348243243	0.359333986	\\
0.395625	0.528318731	\\
0.402910448	0.31790398	\\
0.398157895	0.389340661	\\
0.395228758	0.283919499	\\
0.369178767	0.324983657	\\
0.373228481	0.259081885	\\
0.361706471	0.269310243	\\
0.349253247	0.357098162	\\
0.445357143	0.254193922	\\
0.43175	0.267807089	\\
0.390645161	0.323999543	\\
0.532142857	0.206111315	\\
0.532142857	0.206111315	\\
0.542688679	0.164364601	\\
0.545816327	0.166465131	\\
0.503518519	0.194866135	\\
0.496192053	0.172729817	\\
0.476746032	0.211638646	\\
0.449411765	0.274106575	\\
0.5	0.163477186	\\
0.452702703	0.217628958	\\
0.370298013	0.359881071	\\
0.5852	0.120391115	\\
0.5852	0.120391115	\\
0.588658537	0.112489565	\\
0.602959184	0.098765787	\\
0.596626506	0.103309603	\\
0.577931034	0.111781896	\\
0.556269841	0.115195308	\\
0.535571429	0.153851783	\\
0.608196721	0.115567257	\\
0.62469697	0.10813114	\\
0.6405	0.097245643	\\
0.647357143	0.099194881	\\
0.6405	0.097245643	\\
0.653852459	0.102173908	\\
0.635454545	0.106104857	\\
0.622183099	0.102325182	\\
0.6478875	0.079869379	\\
0.677978723	0.075889683	\\
0.698043478	0.068510109	\\
0.694259259	0.075702619	\\
0.709375	0.06866121	\\
0.69382716	0.065170935	\\
0.635454545	0.106104857	\\
0.678454545	0.054182438	\\
0.250294118	1.384569592	\\
0.330142857	0.823950325	\\
0.250294118	1.384569592	\\
0.253461538	0.921910953	\\
0.257363636	0.97477037	\\
0.349590164	0.614852461	\\
0.346639344	0.55798749	\\
0.459576271	0.462114635	\\
0.494913793	0.25613746	\\
0.494913793	0.25613746	\\
0.55377551	0.18359013	\\
};

\addplot [fill=blue,mark=diamond*,only marks,fill opacity=0.5,draw opacity=0.5,draw=blue, mark size=3,clip mode=individual]
table[row sep=crcr]{%
0.269230769	0.744175454	\\
0.27199504	1.120466556	\\
0.277337662	0.739291797	\\
0.254006711	1.122677307	\\
0.283075028	1.159357325	\\
0.263007692	1.386196665	\\
0.295243523	0.57107589	\\
0.271617925	0.476839417	\\
0.286456827	1.042062363	\\
0.266199819	1.091051745	\\
0.254293301	1.435387541	\\
0.250063496	1.439497753	\\
0.296914397	0.386449911	\\
0.280762774	0.299660509	\\
0.296370849	0.906666981	\\
0.289492758	0.869920089	\\
0.275348819	1.106116772	\\
0.269317618	1.048736771	\\
0.308202231	0.453891068	\\
0.2967415	0.421496171	\\
0.306833791	0.792662547	\\
0.291355882	0.928709297	\\
0.266739108	0.850542936	\\
0.269124176	0.65261223	\\
0.314384724	0.439252151	\\
0.29878866	0.385035467	\\
0.311677043	0.777082547	\\
0.298335831	0.821398301	\\
0.272856157	0.767886789	\\
0.270133917	0.425203955	\\
0.295752091	0.316618552	\\
0.300655439	0.743314957	\\
0.284751149	0.684977232	\\
0.276189647	0.436951988	\\
0.286623997	0.548769865	\\
0.276354369	0.412828554	\\
0.332604082	1.419548203	\\
0.329862018	0.783777356	\\
0.339920407	1.343875484	\\
0.310805592	0.840674054	\\
0.321660911	0.465712864	\\
0.336193227	1.174106935	\\
0.304888628	0.368576328	\\
0.347922949	0.459866535	\\
0.32862801	0.661483331	\\
0.314566453	0.311038149	\\
0.357862671	0.598669974	\\
0.347060022	0.44671373	\\
0.322733043	0.384815864	\\
0.377078177	0.533999157	\\
0.350243243	0.623094645	\\
0.334110518	0.282631029	\\
0.378274336	0.472113435	\\
0.361236328	0.514767958	\\
0.345054369	0.315156938	\\
0.39694822	0.434047349	\\
0.369400289	0.467818705	\\
0.344064013	0.37388228	\\
0.360832657	0.437519548	\\
0.358442857	0.274069286	\\
0.34539916	0.281933878	\\
0.253654938	1.130152313	\\
0.348907862	1.632217417	\\
0.277219113	2.380833007	\\
0.314868946	1.111421224	\\
0.259292507	1.631888041	\\
0.346679949	1.816078954	\\
0.320155161	0.359902518	\\
0.274911302	0.52724401	\\
0.328201525	0.447250857	\\
0.340566098	0.391356962	\\
0.271648256	0.282612715	\\
0.294260726	0.28802785	\\
0.342660976	0.451692881	\\
0.294545775	0.36469158	\\
0.304605523	0.275816458	\\
0.345035861	0.448803964	\\
0.305895228	0.465009529	\\
0.313266467	0.216548078	\\
0.357353954	0.45125639	\\
0.324228673	0.486749162	\\
0.30804983	0.292450841	\\
0.36209481	0.453666166	\\
0.335926521	0.416171167	\\
0.298830012	0.305623985	\\
0.333474333	0.417674767	\\
0.3252352	0.352851857	\\
};

 \addplot+[densely dotted,mark=none,color=red,line width=1.5,draw opacity=0.9] {3.6*exp(-5.7*x)};

\nextgroupplot[height = 4.6cm,
     width = 4.6cm,
     ylabel={$y_{50_\text{TWL}}/y_{90}$},
     xlabel={$\langle \overline{c} \rangle$},
     grid=both,
     xmin=0,
     xmax=0.8,
     ymin=-0.5,
     ymax=1,
    legend columns=2,
   legend style={cells={align=left},anchor = north east,at={(1,1.4)},font=\normalsize},
    domain=0:0.75,samples=400,
   ]

\node[align=center,fill=white,fill opacity=0.8,text opacity=1] at (rel axis cs:0.88,0.9) {(\textit{b})};   

\addplot [fill=black,mark=pentagon*,only marks,fill opacity=0.5,draw opacity=0.5, mark size=3,clip mode=individual]
table[row sep=crcr]{%
0.210304241	0.796918615	\\
0.248783264	0.760981396	\\
0.206732801	0.800179061	\\
0.279105263	0.741193119	\\
0.23847191	0.771669342	\\
0.20681844	0.800981734	\\
0.283871391	0.738491492	\\
0.256571514	0.760992734	\\
0.239573671	0.7650187	\\
0.296059343	0.728685077	\\
0.265983683	0.754677499	\\
0.271521283	0.747407044	\\
0.330622449	0.70014634	\\
0.291258065	0.737234527	\\
0.264347079	0.759099573	\\
0.355041237	0.677540236	\\
0.314695496	0.717828422	\\
0.284164688	0.743620043	\\
0.324873936	0.714855453	\\
0.296886882	0.740243386	\\
0.291487752	0.741595231	\\
0.353172437	0.690621403	\\
0.328058226	0.711544352	\\
0.310328678	0.725417043	\\
0.329708621	0.716348296	\\
0.32638613	0.711024546	\\
0.331049133	0.716930128	\\
0.312257143	0.728182553	\\
0.065649346	0.94729808	\\
0.095667944	0.907720286	\\
0.108748755	0.89181895	\\
0.122272506	0.883236816	\\
0.135834774	0.866353587	\\
0.157674889	0.84290686	\\
0.171409974	0.831693166	\\
0.070865962	0.931731401	\\
0.092232189	0.906795394	\\
0.11042946	0.893878461	\\
0.119224335	0.883319417	\\
0.134156666	0.869613698	\\
0.140477574	0.861691112	\\
0.017426543	0.984416429	\\
0.045031428	0.960470694	\\
0.073772244	0.92696475	\\
0.093441974	0.908906114	\\
0.108060034	0.896144899	\\
0.119799689	0.88129123	\\
0.122641331	0.880302399	\\
0.018916814	0.983648612	\\
0.034551307	0.965931644	\\
0.055588428	0.947224372	\\
0.069244114	0.932856729	\\
0.0918108	0.909153798	\\
0.101487076	0.899761658	\\
0.111393785	0.8890022	\\
0.018905459	0.985549065	\\
0.023320165	0.979537836	\\
0.039796359	0.960668877	\\
0.057974646	0.945086133	\\
0.080194264	0.921822157	\\
0.090094536	0.909955198	\\
0.098966069	0.904729719	\\
0.063718126	0.991309014	\\
0.013097907	0.98787031	\\
0.01917624	0.984768035	\\
0.017546715	0.984287634	\\
0.024851114	0.977572706	\\
0.042646458	0.961769532	\\
0.051559631	0.948288913	\\
0.064179413	0.941091043	\\
0.073417071	0.92839268	\\
0.02020456	0.980473737	\\
0.017733803	0.984390475	\\
0.02026311	0.983309208	\\
0.021603408	0.982666174	\\
0.021898726	0.980386564	\\
0.036833829	0.973222586	\\
0.031747988	0.96963991	\\
0.049465667	0.949996332	\\
0.054525746	0.949302275	\\
0.01928167	0.990991144	\\
0.027292583	0.981699281	\\
0.022357173	0.979077652	\\
0.021557539	0.984675561	\\
0.024718685	0.981739699	\\
0.027741851	0.974617389	\\
0.025488662	0.976182623	\\
0.03769575	0.965503606	\\
0.053810098	0.95259709	\\
0.030483786	0.965853293	\\
0.031653926	0.975866894	\\
0.024050448	0.97878263	\\
0.036601582	0.973080337	\\
0.031667916	0.970263828	\\
0.022403435	0.979506366	\\
0.041460462	0.958281264	\\
0.041059301	0.958964904	\\
0.050252268	0.948474076	\\
0.121474531	0.882956213	\\
0.160595649	0.846139317	\\
0.183001124	0.822124827	\\
0.180082565	0.822147006	\\
0.188725198	0.813025547	\\
0.189353903	0.811159547	\\
0.208807975	0.794738016	\\
0.019091864	0.983269896	\\
0.096533695	0.906278531	\\
0.134112459	0.869370657	\\
0.165133618	0.838392028	\\
0.155809306	0.847277653	\\
0.162094341	0.84063204	\\
0.167096062	0.833623621	\\
0.18257986	0.816559021	\\
0.011755388	0.99090199	\\
0.065565457	0.929044575	\\
0.113364429	0.889998047	\\
0.152130619	0.854721896	\\
0.14387939	0.859297519	\\
0.160941962	0.840049815	\\
0.153713147	0.84741885	\\
0.174125809	0.827114535	\\
0.016251599	0.985992741	\\
0.040914672	0.961356431	\\
0.098193134	0.903236437	\\
0.133661053	0.87021502	\\
0.135406061	0.86885459	\\
0.14372617	0.857472477	\\
0.145782882	0.856495456	\\
0.16354797	0.839009264	\\
0.01205209	0.990967025	\\
0.041207114	0.972339268	\\
0.079273341	0.922737368	\\
0.116208375	0.889603576	\\
0.129746394	0.87597793	\\
0.134340982	0.870781748	\\
0.142110403	0.86022971	\\
0.152651898	0.853136301	\\
0.01332882	0.990506662	\\
0.015794479	0.98769191	\\
0.040729425	0.96086237	\\
0.082688526	0.920868633	\\
0.103385424	0.895646215	\\
0.111528816	0.892117161	\\
0.128097057	0.87060614	\\
0.132158463	0.870521614	\\
0.013415536	0.990284211	\\
0.017103841	0.98453091	\\
0.029193811	0.972659753	\\
0.062426818	0.945814288	\\
0.080545833	0.922116759	\\
0.093076896	0.909397982	\\
0.103721362	0.900675836	\\
0.118694494	0.882205786	\\
0.027329081	0.970418598	\\
0.013080055	0.993139508	\\
0.025852237	0.976084765	\\
0.041529981	0.961418132	\\
0.096606857	0.938481003	\\
0.079960702	0.922485377	\\
0.092865045	0.910635474	\\
0.100496067	0.904305167	\\
0.030566641	0.968518195	\\
0.021440589	0.983045767	\\
0.022584105	0.982116338	\\
0.038108744	0.967030649	\\
0.051308911	0.951998475	\\
0.056112547	0.954862479	\\
0.077607283	0.928953128	\\
0.089633251	0.912348994	\\
0.050252268	0.948474076	\\
0.121474531	0.882956213	\\
0.160595649	0.846139317	\\
0.169874043	0.834512299	\\
0.191715974	0.813899684	\\
0.215947998	0.785512256	\\
0.23270139	0.770194332	\\
0.234023058	0.767679963	\\
0.019091864	0.983269896	\\
0.096533695	0.906278531	\\
0.134112459	0.869370657	\\
0.149246336	0.853988553	\\
0.165311407	0.835653956	\\
0.18117404	0.820921005	\\
0.197402327	0.80265135	\\
0.192437806	0.806476937	\\
0.011755388	0.99090199	\\
0.065565457	0.929044575	\\
0.113364429	0.889998047	\\
0.137388212	0.86405291	\\
0.157331822	0.845060813	\\
0.165776391	0.837836552	\\
0.174597808	0.83072157	\\
0.175913228	0.826842336	\\
0.016251599	0.985992741	\\
0.040914672	0.961356431	\\
0.098193134	0.903236437	\\
0.129136844	0.87429617	\\
0.146485428	0.855901166	\\
0.161282554	0.841886483	\\
0.165342579	0.834656387	\\
0.162922679	0.838895388	\\
0.01205209	0.990967025	\\
0.041207114	0.972339268	\\
0.079273341	0.922737368	\\
0.115670714	0.887886275	\\
0.139852978	0.862993982	\\
0.144443596	0.860419122	\\
0.157761614	0.845237863	\\
0.160808544	0.842512327	\\
0.01332882	0.990506662	\\
0.015794479	0.98769191	\\
0.040729425	0.96086237	\\
0.091743431	0.910953319	\\
0.113912272	0.889644563	\\
0.125013774	0.875639435	\\
0.139419259	0.862839455	\\
0.147817803	0.854094657	\\
0.107584869	0.895526687	\\
0.158060233	0.848920161	\\
0.248056292	0.756399709	\\
0.232724866	0.767969781	\\
0.209531297	0.792574917	\\
0.237137182	0.769558824	\\
0.24588071	0.761711221	\\
0.238421059	0.762345196	\\
0.041747219	0.958055661	\\
0.129249748	0.874324288	\\
0.213729089	0.790154029	\\
0.19005708	0.815173511	\\
0.177845779	0.824562281	\\
0.195903376	0.808157813	\\
0.214973049	0.786988412	\\
0.209102525	0.795615001	\\
0.019942857	0.981214343	\\
0.101937861	0.89586243	\\
0.19222486	0.811134097	\\
0.162438661	0.839648041	\\
0.166140847	0.83171017	\\
0.172543727	0.830320768	\\
0.177655376	0.824443073	\\
0.180049103	0.823589305	\\
0.017301045	0.987879065	\\
0.080263005	0.916529372	\\
0.163425857	0.844915584	\\
0.148309252	0.855606865	\\
0.168681693	0.834487969	\\
0.177405891	0.827170094	\\
0.170420591	0.836966886	\\
0.172729484	0.828644047	\\
0.018143071	0.987089196	\\
0.056247521	0.946633546	\\
0.134542542	0.870414477	\\
0.129169256	0.878867728	\\
0.161150393	0.844971161	\\
0.166383071	0.836676874	\\
0.160809382	0.839194055	\\
0.163639579	0.837679616	\\
0.017296077	0.984501948	\\
0.028689176	0.977146492	\\
0.080278553	0.923871068	\\
0.107966894	0.895873784	\\
0.142569685	0.860824823	\\
0.15064514	0.85173519	\\
0.151817466	0.85486981	\\
0.156854293	0.845829961	\\
0.020558209	0.982223987	\\
0.024754779	0.975717338	\\
0.052258584	0.949148912	\\
0.077749635	0.924338969	\\
0.119831382	0.880442658	\\
0.136578185	0.869402895	\\
0.138768937	0.866220848	\\
0.146403293	0.857579478	\\
0.030970027	0.975586069	\\
0.028363423	0.97431619	\\
0.045310126	0.953542377	\\
0.059620378	0.942477433	\\
0.095794242	0.906047365	\\
0.109437629	0.897149555	\\
0.120897422	0.882504042	\\
0.129725759	0.872129154	\\
0.023794583	0.981691577	\\
0.021975598	0.980114788	\\
0.036465252	0.96519178	\\
0.048360074	0.952754008	\\
0.074627385	0.929109643	\\
0.097332954	0.903884661	\\
0.110574482	0.8902358	\\
0.121469626	0.883562666	\\
0.198709677	0.814882698	\\
0.151734266	0.861964123	\\
0.164856115	0.853717026	\\
0.162418478	0.853864734	\\
0.149237288	0.86779661	\\
0.151942675	0.862322391	\\
0.146686047	0.870016611	\\
0.153562092	0.861500156	\\
0.24615	0.769736842	\\
0.238217822	0.789250354	\\
0.240756303	0.792316927	\\
0.244017857	0.788690476	\\
0.25441989	0.785548247	\\
0.246363636	0.787878788	\\
0.245333333	0.787878788	\\
0.230075188	0.80075188	\\
0.248318584	0.786504425	\\
0.312333333	0.71114582	\\
0.303024691	0.730070276	\\
0.327916667	0.707999094	\\
0.296348315	0.743998496	\\
0.31390625	0.735212853	\\
0.306090909	0.73874743	\\
0.293691275	0.746902013	\\
0.230075188	0.80075188	\\
0.345597015	0.713932661	\\
0.323920455	0.733114636	\\
0.348243243	0.715442486	\\
0.395625	0.633704434	\\
0.402910448	0.643021614	\\
0.398157895	0.652529791	\\
0.395228758	0.666927847	\\
0.369178767	0.692261906	\\
0.373228481	0.696032767	\\
0.361706471	0.701796137	\\
0.349253247	0.708222072	\\
0.445357143	0.636567843	\\
0.43175	0.646978112	\\
0.390645161	0.68969896	\\
0.532142857	0.447965918	\\
0.532142857	0.447965918	\\
0.542688679	0.518399247	\\
0.545816327	0.520889508	\\
0.503518519	0.567784203	\\
0.496192053	0.594139807	\\
0.476746032	0.611615115	\\
0.449411765	0.630672915	\\
0.5	0.594996203	\\
0.452702703	0.641531596	\\
0.370298013	0.696483224	\\
0.5852	0.375625897	\\
0.5852	0.375625897	\\
0.588658537	0.415611086	\\
0.602959184	0.433094184	\\
0.596626506	0.486636968	\\
0.577931034	0.514946693	\\
0.556269841	0.548240128	\\
0.535571429	0.560470014	\\
0.608196721	0.201443157	\\
0.62469697	0.262957727	\\
0.6405	0.30029199	\\
0.647357143	0.322682718	\\
0.6405	0.30029199	\\
0.653852459	0.349925645	\\
0.635454545	0.432359788	\\
0.622183099	0.4693114	\\
0.6478875	0.033730738	\\
0.677978723	-0.007029644	\\
0.698043478	0.207276843	\\
0.694259259	0.194805037	\\
0.709375	0.239227041	\\
0.69382716	0.323515309	\\
0.635454545	0.432359788	\\
0.678454545	0.385811288	\\
0.195912409	0.810882548	\\
0.241914894	0.784869976	\\
0.250294118	0.764664313	\\
0.216935484	0.795698925	\\
0.330142857	0.710363498	\\
0.150404624	0.857109827	\\
0.250294118	0.764664313	\\
0.242121212	0.778787879	\\
0.253461538	0.780725671	\\
0.192535211	0.809317443	\\
0.257363636	0.767193571	\\
0.349590164	0.703877155	\\
0.346639344	0.718246822	\\
0.459576271	0.612388514	\\
0.494913793	0.600079415	\\
0.494913793	0.600079415	\\
0.55377551	0.524578363	\\
};

\addplot [fill=blue,mark=diamond*,only marks,fill opacity=0.5,draw opacity=0.5,draw=blue, mark size=3,clip mode=individual]
table[row sep=crcr]{%
0.175670377	0.832951798	\\
0.193750576	0.80550729	\\
0.246121369	0.765560166	\\
0.187047319	0.820357518	\\
0.237030233	0.76638332	\\
0.198496028	0.801169462	\\
0.16253535	0.839266586	\\
0.166561644	0.836204723	\\
0.269230769	0.758268951	\\
0.247501475	0.769281194	\\
0.27199504	0.744351504	\\
0.248006782	0.756514767	\\
0.216729802	0.794248517	\\
0.215620283	0.791598894	\\
0.277337662	0.756864674	\\
0.254006711	0.773065338	\\
0.283075028	0.737799084	\\
0.263007692	0.753662785	\\
0.242949353	0.773023549	\\
0.228395141	0.786970592	\\
0.295243523	0.739651448	\\
0.271617925	0.759373087	\\
0.286456827	0.7352426	\\
0.266199819	0.755049994	\\
0.254293301	0.767727456	\\
0.250063496	0.768912299	\\
0.296914397	0.745273336	\\
0.280762774	0.754034814	\\
0.296370849	0.728549265	\\
0.289492758	0.733241125	\\
0.275348819	0.749979174	\\
0.269317618	0.75473959	\\
0.308202231	0.733650173	\\
0.2967415	0.742036399	\\
0.306833791	0.719047655	\\
0.291355882	0.735993871	\\
0.266739108	0.759840234	\\
0.269124176	0.756749214	\\
0.314384724	0.727924685	\\
0.29878866	0.74108644	\\
0.311677043	0.715676993	\\
0.298335831	0.728127292	\\
0.272856157	0.754239179	\\
0.270133917	0.759758337	\\
0.295752091	0.74469888	\\
0.300655439	0.73080902	\\
0.284751149	0.747152016	\\
0.276189647	0.754030555	\\
0.286623997	0.747023785	\\
0.276354369	0.755084878	\\
0.185164841	0.821945229	\\
0.115554348	0.890217391	\\
0.135296417	0.863602659	\\
0.332604082	0.667615751	\\
0.233385246	0.761187417	\\
0.236936455	0.763677668	\\
0.329862018	0.687020831	\\
0.339920407	0.663658892	\\
0.310805592	0.703786187	\\
0.321660911	0.701874877	\\
0.336193227	0.677462864	\\
0.304888628	0.722984627	\\
0.347922949	0.670901092	\\
0.32862801	0.693076111	\\
0.314566453	0.719714438	\\
0.357862671	0.668635119	\\
0.347060022	0.675537097	\\
0.322733043	0.713983784	\\
0.377078177	0.648887092	\\
0.350243243	0.674947524	\\
0.334110518	0.705436614	\\
0.378274336	0.654218499	\\
0.361236328	0.670059779	\\
0.345054369	0.689800011	\\
0.39694822	0.633712479	\\
0.369400289	0.663397739	\\
0.344064013	0.69038881	\\
0.360832657	0.670618421	\\
0.358442857	0.681665159	\\
0.34539916	0.698524846	\\
0.136766043	0.866939642	\\
0.162872093	0.843103985	\\
0.168528649	0.836208747	\\
0.137345244	0.858222977	\\
0.192725332	0.811850957	\\
0.153520761	0.850989027	\\
0.118892209	0.884248946	\\
0.135765556	0.868674242	\\
0.213425339	0.791947548	\\
0.194786033	0.806522236	\\
0.20602007	0.800820111	\\
0.19384556	0.812568233	\\
0.168113402	0.832391087	\\
0.169881829	0.831706188	\\
0.231882979	0.769499299	\\
0.209383721	0.793302326	\\
0.226217401	0.778915024	\\
0.205069726	0.796919603	\\
0.191594495	0.81278801	\\
0.198777078	0.805592302	\\
0.234638006	0.772089182	\\
0.219969794	0.788002177	\\
0.235642857	0.776987798	\\
0.22060394	0.788567887	\\
0.223668127	0.784853701	\\
0.209190361	0.797266679	\\
0.23714956	0.772757197	\\
0.225765841	0.782766332	\\
0.245456888	0.767283636	\\
0.234818038	0.780217318	\\
0.22492849	0.78461975	\\
0.213903246	0.797572272	\\
0.243693878	0.764870051	\\
0.237775316	0.775316456	\\
0.239357529	0.776003819	\\
0.237730813	0.778120239	\\
0.232893395	0.776882315	\\
0.217956482	0.798519974	\\
0.253654938	0.761618565	\\
0.246917143	0.766197183	\\
0.246513038	0.766738247	\\
0.24092219	0.778201833	\\
0.248899119	0.771156669	\\
0.248877705	0.767858263	\\
0.242953721	0.77337647	\\
0.236050625	0.780535714	\\
0.226339552	0.760246988	\\
0.138396564	0.859992101	\\
0.348907862	0.649975334	\\
0.277219113	0.723955149	\\
0.184910372	0.819070603	\\
0.314868946	0.695051977	\\
0.259292507	0.748400372	\\
0.346679949	0.642988229	\\
0.320155161	0.703185596	\\
0.274911302	0.74595097	\\
0.328201525	0.673622117	\\
0.340566098	0.685444149	\\
0.271648256	0.752967781	\\
0.294260726	0.718274702	\\
0.342660976	0.684445781	\\
0.294545775	0.728652601	\\
0.304605523	0.70802076	\\
0.345035861	0.683659878	\\
0.305895228	0.721894132	\\
0.313266467	0.699567671	\\
0.357353954	0.669411495	\\
0.324228673	0.702722439	\\
0.30804983	0.708438691	\\
0.36209481	0.665370531	\\
0.335926521	0.692704081	\\
0.298830012	0.726720275	\\
0.333474333	0.698117978	\\
0.3252352	0.697215062	\\
0.668891923	-0.259571596	\\
0.654241261	-0.238599752	\\
0.646286855	0.003455264	\\
0.59602408	0.27122343	\\
0.646812919	0.095910442	\\
0.592229508	0.252630459	\\
0.583634226	0.297274254	\\
0.617637863	0.231973638	\\
0.567247873	0.32925306	\\
0.587487844	0.216803703	\\
0.627622101	0.093725559	\\
0.602584342	0.171466304	\\
0.644581275	0.111782097	\\
0.59101534	0.278175786	\\
0.562580361	0.354852406	\\
0.587351306	0.304603814	\\
0.535812598	0.429409221	\\
0.473876706	0.521576572	\\
0.550975071	0.440979601	\\
0.445267873	0.560849543	\\
0.458722528	0.552356705	\\
0.448352939	0.561034625	\\
0.431439309	0.594049213	\\
0.429715168	0.599983544	\\
0.326396375	0.709854348	\\
0.413837285	0.568579639	\\
0.414374751	0.600801639	\\
0.439059377	0.548422763	\\
0.396079757	0.635669873	\\
0.359066301	0.676387714	\\
0.338403374	0.71539343	\\
0.206843198	0.798818079	\\
0.399705525	0.604974904	\\
0.411594452	0.589873814	\\
0.405918342	0.622657031	\\
0.40351446	0.631525394	\\
0.355121332	0.681259702	\\
};

 \addplot+[densely dotted,mark=none,color=red,line width=1.7,draw opacity=0.9] {1-0.9*x}; 

\node[align=left,anchor=west,fill=white,fill opacity=0.8,text opacity=0.6] at (rel axis cs:0.01,0.085) {\textbf{\scriptsize{\textcolor{red}{$y= 1-0.9x$}}}};
 





\nextgroupplot[height = 4.6cm,
     width = 4.6cm,
     ylabel={$\mathcal{H}/y_{90}$},
     xlabel={$\langle \overline{c} \rangle$},
     grid=both,
     xmin=0,
     xmax=0.8,
     ymin=0,
     ymax=1.5,
    legend columns=2,
   legend style={cells={align=left},anchor = north east,at={(1,1.4)},font=\normalsize},
       domain=0:0.75,samples=400,
   ]

\node[align=center,fill=white,fill opacity=0.8,text opacity=1] at (rel axis cs:0.12,0.9) {(\textit{c})};   

\node[align=left,anchor=east,fill=white,fill opacity=0.8,text opacity=0.6] at (rel axis cs:0.99,0.085) {\textbf{\scriptsize{\textcolor{red}{$y= 0.7x$}}}};

\addplot [fill=black,mark=pentagon*,only marks,fill opacity=0.5,draw opacity=0.5, mark size=3,clip mode=individual]
table[row sep=crcr]{%
0.210304241	0.151259543	\\
0.248783264	0.180498742	\\
0.206732801	0.147181215	\\
0.279105263	0.193159365	\\
0.23847191	0.174122592	\\
0.20681844	0.145186083	\\
0.283871391	0.195949081	\\
0.256571514	0.178742795	\\
0.239573671	0.171211472	\\
0.296059343	0.203810188	\\
0.265983683	0.182126636	\\
0.271521283	0.191657461	\\
0.330622449	0.227594002	\\
0.291258065	0.203992473	\\
0.264347079	0.183417799	\\
0.355041237	0.242815906	\\
0.314695496	0.218238864	\\
0.284164688	0.194807688	\\
0.324873936	0.227247438	\\
0.296886882	0.208388721	\\
0.291487752	0.202118816	\\
0.353172437	0.238021501	\\
0.328058226	0.226175675	\\
0.310328678	0.21349486	\\
0.329708621	0.225579261	\\
0.32638613	0.224485838	\\
0.331049133	0.222496314	\\
0.312257143	0.208189201	\\
0.065649346	0.04779477	\\
0.095667944	0.070651501	\\
0.108748755	0.079496643	\\
0.122272506	0.090724429	\\
0.135834774	0.101290297	\\
0.157674889	0.116484604	\\
0.171409974	0.127952848	\\
0.070865962	0.051483813	\\
0.092232189	0.072436954	\\
0.11042946	0.082348192	\\
0.119224335	0.088027197	\\
0.134156666	0.099998926	\\
0.140477574	0.104611924	\\
0.017426543	0.011130023	\\
0.045031428	0.036027673	\\
0.073772244	0.054700832	\\
0.093441974	0.06818538	\\
0.108060034	0.077646595	\\
0.119799689	0.088478872	\\
0.122641331	0.091572198	\\
0.018916814	0.017361945	\\
0.034551307	0.023144904	\\
0.055588428	0.036667353	\\
0.069244114	0.052970742	\\
0.0918108	0.066525148	\\
0.101487076	0.074978781	\\
0.111393785	0.082961194	\\
0.018905459	0.014137841	\\
0.023320165	0.017725335	\\
0.039796359	0.029510569	\\
0.057974646	0.042007748	\\
0.080194264	0.05783353	\\
0.090094536	0.065679737	\\
0.098966069	0.072765603	\\
0.063718126	0.012724782	\\
0.013097907	0.009770533	\\
0.01917624	0.013534465	\\
0.017546715	0.012530979	\\
0.024851114	0.016736204	\\
0.042646458	0.030318235	\\
0.051559631	0.037622952	\\
0.064179413	0.04779185	\\
0.073417071	0.053915551	\\
0.02020456	0.013201097	\\
0.017733803	0.014798363	\\
0.02026311	0.012293463	\\
0.021603408	0.013685719	\\
0.021898726	0.017472712	\\
0.036833829	0.02205923	\\
0.031747988	0.025720129	\\
0.049465667	0.035658672	\\
0.054525746	0.042768721	\\
0.01928167	0.018094713	\\
0.027292583	0.022177064	\\
0.022357173	0.015315636	\\
0.021557539	0.016217388	\\
0.024718685	0.018911379	\\
0.027741851	0.020969883	\\
0.025488662	0.019927723	\\
0.03769575	0.028029038	\\
0.053810098	0.038044104	\\
0.030483786	0.027008297	\\
0.031653926	0.015313973	\\
0.024050448	0.01825619	\\
0.036601582	0.028776756	\\
0.031667916	0.021318189	\\
0.022403435	0.017420887	\\
0.041460462	0.025932705	\\
0.041059301	0.029696196	\\
0.050252268	0.038867822	\\
0.121474531	0.087446946	\\
0.160595649	0.123296793	\\
0.183001124	0.135203995	\\
0.180082565	0.131923612	\\
0.188725198	0.138579949	\\
0.189353903	0.134247752	\\
0.208807975	0.156293245	\\
0.019091864	0.014152788	\\
0.096533695	0.070057903	\\
0.134112459	0.096671804	\\
0.165133618	0.120434932	\\
0.155809306	0.114152235	\\
0.162094341	0.120754883	\\
0.167096062	0.119561699	\\
0.18257986	0.133317611	\\
0.011755388	0.008763902	\\
0.065565457	0.044404983	\\
0.113364429	0.083038032	\\
0.152130619	0.110764563	\\
0.14387939	0.107576989	\\
0.160941962	0.112100894	\\
0.153713147	0.110056012	\\
0.174125809	0.127566673	\\
0.016251599	0.011439078	\\
0.040914672	0.031202614	\\
0.098193134	0.069818301	\\
0.133661053	0.097941067	\\
0.135406061	0.100514534	\\
0.14372617	0.102187768	\\
0.145782882	0.104514758	\\
0.16354797	0.119217823	\\
0.01205209	0.00890273	\\
0.041207114	0.032950125	\\
0.079273341	0.053620222	\\
0.116208375	0.082504078	\\
0.129746394	0.093726888	\\
0.134340982	0.097540526	\\
0.142110403	0.102421424	\\
0.152651898	0.112344082	\\
0.01332882	0.009438205	\\
0.015794479	0.012115635	\\
0.040729425	0.032946175	\\
0.082688526	0.057953611	\\
0.103385424	0.073217572	\\
0.111528816	0.082833715	\\
0.128097057	0.089748752	\\
0.132158463	0.094664439	\\
0.013415536	0.01149926	\\
0.017103841	0.012522536	\\
0.029193811	0.021420643	\\
0.062426818	0.039138532	\\
0.080545833	0.057482911	\\
0.093076896	0.067596159	\\
0.103721362	0.076469459	\\
0.118694494	0.083376843	\\
0.027329081	0.0186198	\\
0.013080055	0.013448553	\\
0.025852237	0.019004796	\\
0.041529981	0.030212024	\\
0.096606857	0.0447676	\\
0.079960702	0.05726886	\\
0.092865045	0.066206096	\\
0.100496067	0.075425907	\\
0.030566641	0.017180917	\\
0.021440589	0.01207438	\\
0.022584105	0.018108641	\\
0.038108744	0.023953256	\\
0.051308911	0.034874938	\\
0.056112547	0.060634897	\\
0.077607283	0.057882533	\\
0.089633251	0.069541045	\\
0.050252268	0.038867822	\\
0.121474531	0.087446946	\\
0.160595649	0.123296793	\\
0.169874043	0.123773047	\\
0.191715974	0.139356766	\\
0.215947998	0.158536653	\\
0.23270139	0.167191651	\\
0.234023058	0.170936173	\\
0.019091864	0.014152788	\\
0.096533695	0.070057903	\\
0.134112459	0.096671804	\\
0.149246336	0.109893662	\\
0.165311407	0.117463892	\\
0.18117404	0.132213718	\\
0.197402327	0.144663959	\\
0.192437806	0.13983058	\\
0.011755388	0.008763902	\\
0.065565457	0.044404983	\\
0.113364429	0.083038032	\\
0.137388212	0.100872771	\\
0.157331822	0.114424255	\\
0.165776391	0.11951628	\\
0.174597808	0.125298427	\\
0.175913228	0.127483469	\\
0.016251599	0.011439078	\\
0.040914672	0.031202614	\\
0.098193134	0.069818301	\\
0.129136844	0.093099455	\\
0.146485428	0.106264987	\\
0.161282554	0.110742056	\\
0.165342579	0.118283486	\\
0.162922679	0.11669329	\\
0.01205209	0.008902737	\\
0.041207114	0.032950125	\\
0.079273341	0.053620222	\\
0.115670714	0.086032037	\\
0.139852978	0.09925834	\\
0.144443596	0.102693775	\\
0.157761614	0.111640271	\\
0.160808544	0.113769748	\\
0.01332882	0.009438205	\\
0.015794479	0.012115635	\\
0.040729425	0.032946175	\\
0.091743431	0.065120189	\\
0.113912272	0.082013623	\\
0.125013774	0.09117702	\\
0.139419259	0.099671594	\\
0.147817803	0.102934666	\\
0.107584869	0.075849004	\\
0.158060233	0.119892734	\\
0.248056292	0.181194154	\\
0.232724866	0.172524673	\\
0.209531297	0.151792978	\\
0.237137182	0.174168154	\\
0.24588071	0.186524093	\\
0.238421059	0.172850184	\\
0.041747219	0.030885872	\\
0.129249748	0.097060293	\\
0.213729089	0.152892701	\\
0.19005708	0.139998004	\\
0.177845779	0.127937183	\\
0.195903376	0.143927293	\\
0.214973049	0.156759788	\\
0.209102525	0.151469019	\\
0.019942857	0.014444283	\\
0.101937861	0.077537145	\\
0.19222486	0.138071621	\\
0.162438661	0.118281235	\\
0.166140847	0.114819633	\\
0.172543727	0.124522703	\\
0.177655376	0.132237816	\\
0.180049103	0.12919643	\\
0.017301045	0.013443152	\\
0.080263005	0.057062277	\\
0.163425857	0.116899343	\\
0.148309252	0.105631868	\\
0.168681693	0.119086495	\\
0.177405891	0.125389564	\\
0.170420591	0.125067141	\\
0.172729484	0.121433517	\\
0.018143071	0.010024094	\\
0.056247521	0.042740628	\\
0.134542542	0.098245522	\\
0.129169256	0.096825381	\\
0.161150393	0.114821757	\\
0.166383071	0.115963313	\\
0.160809382	0.115888753	\\
0.163639579	0.115038808	\\
0.017296077	0.012362163	\\
0.028689176	0.021926023	\\
0.080278553	0.062119306	\\
0.107966894	0.076724635	\\
0.142569685	0.09646022	\\
0.15064514	0.108276107	\\
0.151817466	0.107300768	\\
0.156854293	0.108934067	\\
0.020558209	0.014347373	\\
0.024754779	0.019042711	\\
0.052258584	0.039170624	\\
0.077749635	0.058400802	\\
0.119831382	0.083338735	\\
0.136578185	0.092424873	\\
0.138768937	0.09671609	\\
0.146403293	0.100776338	\\
0.030970027	0.020502301	\\
0.028363423	0.02164774	\\
0.045310126	0.030067135	\\
0.059620378	0.043303558	\\
0.095794242	0.066304344	\\
0.109437629	0.080488523	\\
0.120897422	0.086139393	\\
0.129725759	0.090167323	\\
0.023794583	0.018382668	\\
0.021975598	0.01511282	\\
0.036465252	0.027126421	\\
0.048360074	0.036011007	\\
0.074627385	0.0547381	\\
0.097332954	0.065867206	\\
0.110574482	0.07833094	\\
0.121469626	0.083712175	\\
0.198709677	0.141451363	\\
0.151734266	0.102430366	\\
0.164856115	0.113603622	\\
0.162418478	0.10756734	\\
0.149237288	0.101319409	\\
0.151942675	0.102274087	\\
0.146686047	0.099785513	\\
0.153562092	0.099267205	\\
0.24615	0.172047658	\\
0.238217822	0.170220057	\\
0.240756303	0.173482178	\\
0.244017857	0.173363791	\\
0.25441989	0.168760817	\\
0.246363636	0.17523271	\\
0.245333333	0.172676159	\\
0.230075188	0.164091062	\\
0.248318584	0.181574421	\\
0.312333333	0.224858218	\\
0.303024691	0.213679441	\\
0.327916667	0.220921202	\\
0.296348315	0.199611344	\\
0.31390625	0.204100177	\\
0.306090909	0.196343589	\\
0.293691275	0.185954418	\\
0.230075188	0.164091062	\\
0.345597015	0.226564687	\\
0.323920455	0.206298246	\\
0.348243243	0.226090715	\\
0.395625	0.286542932	\\
0.402910448	0.281557595	\\
0.398157895	0.270310266	\\
0.395228758	0.258320589	\\
0.369178767	0.248978802	\\
0.373228481	0.242197518	\\
0.361706471	0.237234056	\\
0.349253247	0.228418663	\\
0.445357143	0.280169898	\\
0.43175	0.273993212	\\
0.390645161	0.235740032	\\
0.532142857	0.423810559	\\
0.532142857	0.423810559	\\
0.542688679	0.372860389	\\
0.545816327	0.375666077	\\
0.503518519	0.334648864	\\
0.496192053	0.31827612	\\
0.476746032	0.309000814	\\
0.449411765	0.287350914	\\
0.5	0.316698749	\\
0.452702703	0.280291479	\\
0.370298013	0.239782921	\\
0.5852	0.493277422	\\
0.5852	0.493277422	\\
0.588658537	0.459755011	\\
0.602959184	0.440250705	\\
0.596626506	0.400493212	\\
0.577931034	0.378537224	\\
0.556269841	0.354294742	\\
0.535571429	0.340033643	\\
0.608196721	0.612191504	\\
0.62469697	0.580530682	\\
0.6405	0.543734663	\\
0.647357143	0.515475515	\\
0.6405	0.543734663	\\
0.653852459	0.498284363	\\
0.635454545	0.446746152	\\
0.622183099	0.405192349	\\
0.6478875	0.736865873	\\
0.677978723	0.783769899	\\
0.698043478	0.600255056	\\
0.694259259	0.608824628	\\
0.709375	0.577533336	\\
0.69382716	0.514388657	\\
0.635454545	0.446746152	\\
0.678454545	0.467278485	\\
0.195912409	0.100099095	\\
0.241914894	0.170732119	\\
0.250294118	0.169041403	\\
0.216935484	0.157405683	\\
0.330142857	0.217555993	\\
0.150404624	0.106063161	\\
0.250294118	0.169041403	\\
0.242121212	0.156487403	\\
0.253461538	0.174396076	\\
0.192535211	0.125865521	\\
0.257363636	0.173560481	\\
0.349590164	0.231972568	\\
0.346639344	0.229313181	\\
0.459576271	0.300410262	\\
0.494913793	0.323749416	\\
0.494913793	0.323749416	\\
0.55377551	0.38261538	\\
};

\addplot [fill=blue,mark=diamond*,only marks,fill opacity=0.5,draw opacity=0.5,draw=blue, mark size=3,clip mode=individual]
table[row sep=crcr]{%
0.175670377	0.126193678	\\
0.193750576	0.133180006	\\
0.246121369	0.175146764	\\
0.187047319	0.136722727	\\
0.237030233	0.168654925	\\
0.198496028	0.140907932	\\
0.16253535	0.111927189	\\
0.166561644	0.116656673	\\
0.269230769	0.193998039	\\
0.247501475	0.175484406	\\
0.27199504	0.191859275	\\
0.248006782	0.176853374	\\
0.216729802	0.161543402	\\
0.215620283	0.155328451	\\
0.277337662	0.201122245	\\
0.254006711	0.181290769	\\
0.283075028	0.201480595	\\
0.263007692	0.183356173	\\
0.242949353	0.180011087	\\
0.228395141	0.171796263	\\
0.295243523	0.205414103	\\
0.271617925	0.190036781	\\
0.286456827	0.201151315	\\
0.266199819	0.188840867	\\
0.254293301	0.183107746	\\
0.250063496	0.177375729	\\
0.296914397	0.204433124	\\
0.280762774	0.190887892	\\
0.296370849	0.21073767	\\
0.289492758	0.205984226	\\
0.275348819	0.192892477	\\
0.269317618	0.189215272	\\
0.308202231	0.214540174	\\
0.2967415	0.205146088	\\
0.306833791	0.216918965	\\
0.291355882	0.203110258	\\
0.266739108	0.189877522	\\
0.269124176	0.181634771	\\
0.314384724	0.214415933	\\
0.29878866	0.205190384	\\
0.311677043	0.220066332	\\
0.298335831	0.208703941	\\
0.272856157	0.189291183	\\
0.270133917	0.186608757	\\
0.295752091	0.207768416	\\
0.300655439	0.20862668	\\
0.284751149	0.197680846	\\
0.276189647	0.193507398	\\
0.286623997	0.197390489	\\
0.276354369	0.189006227	\\
0.185164841	0.136079682	\\
0.115554348	0.083503066	\\
0.135296417	0.093694695	\\
0.332604082	0.193541882	\\
0.233385246	0.15667758	\\
0.236936455	0.162086533	\\
0.329862018	0.226005782	\\
0.339920407	0.234796983	\\
0.310805592	0.225892273	\\
0.321660911	0.214352055	\\
0.336193227	0.240683415	\\
0.304888628	0.214647434	\\
0.347922949	0.241531962	\\
0.32862801	0.23082197	\\
0.314566453	0.21914995	\\
0.357862671	0.244255806	\\
0.347060022	0.240336187	\\
0.322733043	0.222419951	\\
0.377078177	0.244642807	\\
0.350243243	0.241257593	\\
0.334110518	0.231514586	\\
0.378274336	0.250279371	\\
0.361236328	0.252339791	\\
0.345054369	0.237018614	\\
0.39694822	0.272136851	\\
0.369400289	0.265048372	\\
0.344064013	0.241857825	\\
0.360832657	0.257126311	\\
0.358442857	0.244838996	\\
0.34539916	0.239710478	\\
0.136766043	0.098111188	\\
0.162872093	0.114753284	\\
0.168528649	0.122134127	\\
0.137345244	0.098748183	\\
0.192725332	0.136621941	\\
0.153520761	0.110272647	\\
0.118892209	0.086999835	\\
0.135765556	0.098264577	\\
0.213425339	0.146586011	\\
0.194786033	0.1378322	\\
0.20602007	0.150548701	\\
0.19384556	0.137176364	\\
0.168113402	0.121634141	\\
0.169881829	0.123544068	\\
0.231882979	0.157355927	\\
0.209383721	0.144713694	\\
0.226217401	0.161105213	\\
0.205069726	0.14717947	\\
0.191594495	0.137969885	\\
0.198777078	0.141106624	\\
0.234638006	0.166751023	\\
0.219969794	0.158554596	\\
0.235642857	0.167070247	\\
0.22060394	0.160072165	\\
0.223668127	0.157570989	\\
0.209190361	0.150889522	\\
0.23714956	0.167841502	\\
0.225765841	0.161291772	\\
0.245456888	0.173570341	\\
0.234818038	0.168705567	\\
0.22492849	0.158874507	\\
0.213903246	0.156356317	\\
0.243693878	0.172428104	\\
0.237775316	0.169297764	\\
0.239357529	0.174516478	\\
0.237730813	0.170118858	\\
0.232893395	0.165658001	\\
0.217956482	0.160477134	\\
0.253654938	0.174302912	\\
0.246917143	0.17876889	\\
0.246513038	0.173401806	\\
0.24092219	0.17259767	\\
0.248899119	0.180364451	\\
0.248877705	0.177920845	\\
0.242953721	0.180696066	\\
0.236050625	0.172715264	\\
0.226339552	0.131276673	\\
0.138396564	0.092435852	\\
0.348907862	0.21141612	\\
0.277219113	0.188933226	\\
0.184910372	0.127790787	\\
0.314868946	0.219895355	\\
0.259292507	0.181431206	\\
0.346679949	0.242610088	\\
0.320155161	0.212055682	\\
0.274911302	0.178103628	\\
0.328201525	0.220997153	\\
0.340566098	0.22042113	\\
0.271648256	0.178153022	\\
0.294260726	0.194358629	\\
0.342660976	0.223404217	\\
0.294545775	0.19384353	\\
0.304605523	0.204509033	\\
0.345035861	0.229763763	\\
0.305895228	0.211881054	\\
0.313266467	0.211448641	\\
0.357353954	0.238487574	\\
0.324228673	0.215189122	\\
0.30804983	0.202919487	\\
0.36209481	0.246940476	\\
0.335926521	0.227908478	\\
0.298830012	0.203344229	\\
0.333474333	0.222446587	\\
0.3252352	0.214964942	\\
0.668891923	0.97587249	\\
0.654241261	1.009522323	\\
0.646286855	0.802695181	\\
0.59602408	0.550882345	\\
0.646812919	0.715867813	\\
0.592229508	0.577052415	\\
0.583634226	0.552664881	\\
0.617637863	0.59803274	\\
0.567247873	0.558544865	\\
0.587487844	0.570328756	\\
0.627622101	0.717544652	\\
0.602584342	0.646242666	\\
0.644581275	0.719133087	\\
0.59101534	0.553464095	\\
0.562580361	0.511013523	\\
0.587351306	0.550373125	\\
0.535812598	0.435709088	\\
0.473876706	0.370428252	\\
0.550975071	0.44983764	\\
0.445267873	0.328178263	\\
0.458722528	0.344120004	\\
0.448352939	0.389732858	\\
0.431439309	0.31100613	\\
0.429715168	0.310931484	\\
0.326396375	0.218275459	\\
0.413837285	0.301396589	\\
0.414374751	0.293715424	\\
0.439059377	0.334891167	\\
0.396079757	0.270093186	\\
0.359066301	0.256434614	\\
0.338403374	0.228253237	\\
0.206843198	0.147034599	\\
0.399705525	0.282752075	\\
0.411594452	0.279972493	\\
0.405918342	0.283052925	\\
0.40351446	0.274720442	\\
0.355121332	0.234682932	\\
};

\addplot+[densely dotted,mark=none,color=red,line width=1.7,draw opacity=0.9] {0.7*x};



\end{groupplot}

\end{tikzpicture}

%% file: Figures/fig6.tex
\begin{tikzpicture}

 \begin{groupplot}[
     group style = {group size = 1 by 1,horizontal sep=1.5cm,vertical sep=1.5cm},
     width = 1\textwidth]

\nextgroupplot[height = 6cm,
     width = 6cm,
     ylabel={$\frac{D_{t,y}}{u_* \, \delta}  $},
     xlabel={$y/\delta$},
     grid=both,
     xmin=0,
     xmax=1.5,
     ymin=0.001,
     ymode=log,
     ymax=1,
    legend columns=1,
   legend style={cells={align=left},anchor = north east,at={(-0.5,1)},font=\footnotesize},  legend cell align={left} ]
   
\addplot [fill=black,mark=pentagon*,only marks,fill opacity=0.5,draw opacity=0.5, mark size=3,clip mode=individual]
table[row sep=crcr]{%
1 20	\\
}; \addlegendentry{\textbf{Smooth chutes} \\  \, - \citet{Severi2018} \, \\
\, - \cite{Felder2019} \, 
};

\addplot [fill=blue,mark=diamond*,only marks,fill opacity=0.5,draw opacity=0.5,draw=blue, mark size=3,clip mode=individual]
table[row sep=crcr]{%
1 20	\\
}; \addlegendentry{\textbf{Stepped chute}  \\
\, - \citet{Kramer2018Transiton} };

\addplot[densely dashdotted,color=gray,mark=none,line width = 1.7,draw opacity=1] 
table[row sep=crcr]{%
1 20	\\
};\addlegendentry{Eq. (\ref{eq:parabolic}), \, $S_c=0.2$};

\addplot[densely dashed,color=black,mark=none,line width = 1.7,draw opacity=1] 
table[row sep=crcr]{%
1 20	\\
};\addlegendentry{Eq. (\ref{eq:parabolic}), \, $S_c=0.5$};

\addplot[color=black,mark=none,line width = 1.7,draw opacity=0.7] 
table[row sep=crcr]{%
1 20	\\
};\addlegendentry{Eq. (\ref{eq:parabolic}), \, $S_c=1.0$};

\addplot[densely dashed, color=red,mark=none,line width = 1.7,draw opacity=0.7] 
table[row sep=crcr]{%
1 20	\\
};\addlegendentry{$ D_{t,y}/(u_* \delta) = \kappa/(4 S_c)$, \, $S_c = 0.5$};

\addplot[densely dashdotted,color=gray,mark=none,line width = 1.7,draw opacity=1] 
table[row sep=crcr]{%
0.00010	0.00020	\\
0.010	0.020	\\
0.020	0.040	\\
0.030	0.060	\\
0.040	0.079	\\
0.050	0.097	\\
0.060	0.116	\\
0.070	0.133	\\
0.080	0.151	\\
0.090	0.168	\\
0.100	0.185	\\
0.110	0.201	\\
0.120	0.216	\\
0.130	0.232	\\
0.140	0.247	\\
0.150	0.261	\\
0.160	0.276	\\
0.170	0.289	\\
0.180	0.303	\\
0.190	0.315	\\
0.200	0.328	\\
0.210	0.340	\\
0.220	0.352	\\
0.230	0.363	\\
0.240	0.374	\\
0.250	0.384	\\
0.260	0.394	\\
0.270	0.404	\\
0.280	0.413	\\
0.290	0.422	\\
0.300	0.431	\\
0.310	0.438	\\
0.320	0.446	\\
0.330	0.453	\\
0.340	0.460	\\
0.350	0.466	\\
0.360	0.472	\\
0.370	0.478	\\
0.380	0.483	\\
0.390	0.488	\\
0.400	0.492	\\
0.410	0.496	\\
0.420	0.499	\\
0.430	0.502	\\
0.440	0.505	\\
0.450	0.507	\\
0.460	0.509	\\
0.470	0.511	\\
0.480	0.512	\\
0.490	0.512	\\
0.500	0.513	\\
0.510	0.512	\\
0.520	0.512	\\
0.530	0.511	\\
0.540	0.509	\\
0.550	0.507	\\
0.560	0.505	\\
0.570	0.502	\\
0.580	0.499	\\
0.590	0.496	\\
0.600	0.492	\\
0.610	0.488	\\
0.620	0.483	\\
0.630	0.478	\\
0.640	0.472	\\
0.650	0.466	\\
0.660	0.460	\\
0.670	0.453	\\
0.680	0.446	\\
0.690	0.438	\\
0.700	0.431	\\
0.710	0.422	\\
0.720	0.413	\\
0.730	0.404	\\
0.740	0.394	\\
0.750	0.384	\\
0.760	0.374	\\
0.770	0.363	\\
0.780	0.352	\\
0.790	0.340	\\
0.800	0.328	\\
0.810	0.315	\\
0.820	0.303	\\
0.830	0.289	\\
0.840	0.276	\\
0.850	0.261	\\
0.860	0.247	\\
0.870	0.232	\\
0.880	0.216	\\
0.890	0.201	\\
0.900	0.185	\\
0.910	0.168	\\
0.920	0.151	\\
0.930	0.133	\\
0.940	0.116	\\
0.950	0.097	\\
0.960	0.079	\\
0.970	0.060	\\
0.980	0.040	\\
0.990	0.020	\\
0.99990	0.00020	\\
};

\addplot[densely dashed,color=black,mark=none,line width = 1.7,draw opacity=1] 
table[row sep=crcr]{%
0.00010	0.00008	\\
0.010	0.008	\\
0.020	0.016	\\
0.030	0.024	\\
0.040	0.031	\\
0.050	0.039	\\
0.060	0.046	\\
0.070	0.053	\\
0.080	0.060	\\
0.090	0.067	\\
0.100	0.074	\\
0.110	0.080	\\
0.120	0.087	\\
0.130	0.093	\\
0.140	0.099	\\
0.150	0.105	\\
0.160	0.110	\\
0.170	0.116	\\
0.180	0.121	\\
0.190	0.126	\\
0.200	0.131	\\
0.210	0.136	\\
0.220	0.141	\\
0.230	0.145	\\
0.240	0.150	\\
0.250	0.154	\\
0.260	0.158	\\
0.270	0.162	\\
0.280	0.165	\\
0.290	0.169	\\
0.300	0.172	\\
0.310	0.175	\\
0.320	0.178	\\
0.330	0.181	\\
0.340	0.184	\\
0.350	0.187	\\
0.360	0.189	\\
0.370	0.191	\\
0.380	0.193	\\
0.390	0.195	\\
0.400	0.197	\\
0.410	0.198	\\
0.420	0.200	\\
0.430	0.201	\\
0.440	0.202	\\
0.450	0.203	\\
0.460	0.204	\\
0.470	0.204	\\
0.480	0.205	\\
0.490	0.205	\\
0.500	0.205	\\
0.510	0.205	\\
0.520	0.205	\\
0.530	0.204	\\
0.540	0.204	\\
0.550	0.203	\\
0.560	0.202	\\
0.570	0.201	\\
0.580	0.200	\\
0.590	0.198	\\
0.600	0.197	\\
0.610	0.195	\\
0.620	0.193	\\
0.630	0.191	\\
0.640	0.189	\\
0.650	0.187	\\
0.660	0.184	\\
0.670	0.181	\\
0.680	0.178	\\
0.690	0.175	\\
0.700	0.172	\\
0.710	0.169	\\
0.720	0.165	\\
0.730	0.162	\\
0.740	0.158	\\
0.750	0.154	\\
0.760	0.150	\\
0.770	0.145	\\
0.780	0.141	\\
0.790	0.136	\\
0.800	0.131	\\
0.810	0.126	\\
0.820	0.121	\\
0.830	0.116	\\
0.840	0.110	\\
0.850	0.105	\\
0.860	0.099	\\
0.870	0.093	\\
0.880	0.087	\\
0.890	0.080	\\
0.900	0.074	\\
0.910	0.067	\\
0.920	0.060	\\
0.930	0.053	\\
0.940	0.046	\\
0.950	0.039	\\
0.960	0.031	\\
0.970	0.024	\\
0.980	0.016	\\
0.990	0.008	\\
0.99990	0.00008	\\
};

\addplot[color=black,mark=none,line width = 1.7,draw opacity=0.7] 
table[row sep=crcr]{%
0.00010	0.00004	\\
0.010	0.004	\\
0.020	0.008	\\
0.030	0.012	\\
0.040	0.016	\\
0.050	0.019	\\
0.060	0.023	\\
0.070	0.027	\\
0.080	0.030	\\
0.090	0.034	\\
0.100	0.037	\\
0.110	0.040	\\
0.120	0.043	\\
0.130	0.046	\\
0.140	0.049	\\
0.150	0.052	\\
0.160	0.055	\\
0.170	0.058	\\
0.180	0.061	\\
0.190	0.063	\\
0.200	0.066	\\
0.210	0.068	\\
0.220	0.070	\\
0.230	0.073	\\
0.240	0.075	\\
0.250	0.077	\\
0.260	0.079	\\
0.270	0.081	\\
0.280	0.083	\\
0.290	0.084	\\
0.300	0.086	\\
0.310	0.088	\\
0.320	0.089	\\
0.330	0.091	\\
0.340	0.092	\\
0.350	0.093	\\
0.360	0.094	\\
0.370	0.096	\\
0.380	0.097	\\
0.390	0.098	\\
0.400	0.098	\\
0.410	0.099	\\
0.420	0.100	\\
0.430	0.100	\\
0.440	0.101	\\
0.450	0.101	\\
0.460	0.102	\\
0.470	0.102	\\
0.480	0.102	\\
0.490	0.102	\\
0.500	0.103	\\
0.510	0.102	\\
0.520	0.102	\\
0.530	0.102	\\
0.540	0.102	\\
0.550	0.101	\\
0.560	0.101	\\
0.570	0.100	\\
0.580	0.100	\\
0.590	0.099	\\
0.600	0.098	\\
0.610	0.098	\\
0.620	0.097	\\
0.630	0.096	\\
0.640	0.094	\\
0.650	0.093	\\
0.660	0.092	\\
0.670	0.091	\\
0.680	0.089	\\
0.690	0.088	\\
0.700	0.086	\\
0.710	0.084	\\
0.720	0.083	\\
0.730	0.081	\\
0.740	0.079	\\
0.750	0.077	\\
0.760	0.075	\\
0.770	0.073	\\
0.780	0.070	\\
0.790	0.068	\\
0.800	0.066	\\
0.810	0.063	\\
0.820	0.061	\\
0.830	0.058	\\
0.840	0.055	\\
0.850	0.052	\\
0.860	0.049	\\
0.870	0.046	\\
0.880	0.043	\\
0.890	0.040	\\
0.900	0.037	\\
0.910	0.034	\\
0.920	0.030	\\
0.930	0.027	\\
0.940	0.023	\\
0.950	0.019	\\
0.960	0.016	\\
0.970	0.012	\\
0.980	0.008	\\
0.990	0.004	\\
0.99990	0.00004	\\
}; 

\addplot[densely dashed, color=red,mark=none,line width = 1.7,draw opacity=0.7] 
table[row sep=crcr]{%
0.5	0.205\\
1.5	0.205\\
};

\addplot [fill=black,mark=pentagon*,only marks,fill opacity=0.5,draw opacity=0.5, mark size=3,clip mode=individual]
table[row sep=crcr]{%
0.067	0.03811	\\
0.133	0.05610	\\
0.200	0.08081	\\
0.06429	0.016	\\
0.08571	0.026	\\
0.12857	0.049	\\
0.17143	0.073	\\
0.21429	0.078	\\
0.25714	0.097	\\
0.30000	0.170	\\
0.34286	0.152	\\
0.42857	0.171	\\
0.51429	0.194	\\
0.60000	0.182	\\
0.71429	0.216	\\
0.85714	0.192	\\
1.00000	0.189	\\
1.14286	0.193	\\
1.28571	0.210	\\
1.42857	0.180	\\
1.57143	0.198	\\
1.71429	0.179	\\
1.85714	0.168	\\
2.00000	0.151	\\
2.14286	0.135	\\
2.28571	0.138	\\
2.42857	0.151	\\
}; 
   
\addplot [fill=blue,mark=diamond*,only marks,fill opacity=0.5,draw opacity=0.5,draw=blue, mark size=3,clip mode=individual]
table[row sep=crcr]{%
0.011	0.001	\\
0.082	0.031	\\
0.152	0.072	\\
0.222	0.154	\\
0.293	0.174	\\
0.363	0.220	\\
0.434	0.220	\\
0.505	0.292	\\
0.576	0.127	\\
0.646	0.134	\\
0.717	0.108	\\
0.788	0.106	\\
0.858	0.076	\\
0.929	0.085	\\
0.082	0.099	\\
0.153	0.146	\\
0.224	0.218	\\
0.295	0.402	\\
0.366	0.313	\\
0.436	0.598	\\
0.507	0.468	\\
0.578	0.505	\\
0.649	0.489	\\
};

\end{groupplot}

\end{tikzpicture}